\documentclass[iop]{emulateapj}
\usepackage{natbib}
\usepackage{longtable}

\usepackage{savesym}
\savesymbol{tablenum}
\usepackage{siunitx}
\restoresymbol{SIX}{tablenum}

\begin{document}
\title{Variables in the Southern Polar region Evryscope 2016 dataset}

\author{Jeffrey~K.~Ratzloff\altaffilmark{1}, Henry~T.~Corbett\altaffilmark{1}, Nicholas~M.~Law\altaffilmark{1}, Brad~N.~Barlow\altaffilmark{2}, Amy~Glazier\altaffilmark{1}, \\ Ward~S.~Howard\altaffilmark{1}, Octavi~Fors\altaffilmark{1,3}, Daniel~del~Ser\altaffilmark{1,3}, and Trifon~Trifonov\altaffilmark{4}}

\altaffiltext{1}{Department of Physics and Astronomy, University of North Carolina at Chapel Hill, Chapel Hill, NC 27599-3255, USA}
\altaffiltext{2}{Department of Physics and Astronomy, High Point University, High Point, NC 27268, USA}
\altaffiltext{3}{Institut de Ci\`encies del Cosmos (ICCUB), Universitat de Barcelona, IEEC-UB, Mart\'{\i} i Franqu\`es 1, E08028 Barcelona, Spain}
\altaffiltext{4}{Max-Planck-Institut f\"{u}r Astronomie, K\"{o}nigstuhl  17, D-69117 Heidelberg, Germany}

\email[$\star$~E-mail:~]{jeff215@live.unc.edu}

%----------------------------------------------------------------------------------------
%	ABSTRACT
%----------------------------------------------------------------------------------------

\begin{abstract}
The regions around the celestial poles offer the ability to find and characterize long-term variables from ground-based observatories. We used multi-year Evryscope data to search for high-amplitude ($\approx$ 5\% or greater) variable objects among 160,000 bright stars (m\textsubscript{v} $<$ 14.5) near the South Celestial Pole. We developed a machine learning based spectral classifier to identify eclipse and transit candidates with M-dwarf or K-dwarf host stars - and potential low-mass secondary stars or gas giant planets. The large amplitude transit signals from low-mass companions of smaller dwarf host stars lessens the photometric precision and systematics removal requirements necessary for detection, and increases the discoveries from long-term observations with modest light curve precision among the faintest stars in the survey. The Evryscope is a robotic telescope array that observes the Southern sky continuously at 2-minute cadence, searching for stellar variability, transients, transits around exotic stars and other observationally challenging astrophysical variables. The multi-year photometric stability is better than 1\% for bright stars in uncrowded regions, with a 3-sigma limiting magnitude of g=16 in dark time. In this study, covering all stars 9 $<$ m\textsubscript{v} $<$ 14.5, in declinations \ang{-75} to \ang{-90}, and searching for high-amplitude variability, we recover 346 known variables and discover 303 new variables, including 168 eclipsing binaries. We characterize the discoveries and provide the amplitudes, periods, and variability type. A 1.7 R\textsubscript{J}  planet candidate with a late K-dwarf primary was found and the transit signal was verified with the PROMPT telescope network. Further followup revealed this object to be a likely grazing eclipsing binary system with nearly identical primary and secondary K5 stars. Radial velocity measurements from the Goodman Spectrograph on the 4.1 meter SOAR telescope of the likely-lowest-mass targets reveal that six of the eclipsing binary discoveries are low-mass (.06 - .37 \(M_\odot\)) secondaries with K-dwarf primaries, strong candidates for precision mass-radius measurements.\\

\end{abstract}

%----------------------------------------------------------------------------------------
%	INTRO
%----------------------------------------------------------------------------------------

\section{INTRODUCTION} \label{section_intro}

Variable star discoveries provide information on stellar properties, formation, and evolution, and are critical for determining distances and ages of astronomical objects. Eclipsing binaries allow the measurement of masses, radii, and temperatures, and can be used to test stellar formation theory predictions. Lower mass eclipsing binaries are observationally challenging due to the low intrinsic brightness of the star, and more systems are needed to properly characterize the mass/radius relationship in stellar models \citep{2010AJ....140.1158T, 2015ApJ...804...64M, 2017ApJ...845...72K}. Ground-based surveys such as the Palomar Transient Factory \citep{2009PASP..121.1395L}, ATLAS \citep{2011PASP..123...58T}, HAT \citep{2004PASP..116..266B}, HAT-South \citep{2018arXiv180100849B}, SuperWASP \citep{2006PASP..118.1407P}, KELT \citep{2007PASP..119..923P}, CSTAR \citep{2015ApJS..218...20W}, and many others are very successful in detecting variables (including transiting exoplanets) and adding to known variable star catalogs such as the Variable Star Index\footnote{http://www.aavso.org/vsx/} (VSX). These surveys either observe at day or longer time-scale cadences, or observe dedicated sky areas to reach fast cadence at the expense of all sky coverage. In contrast, the Evryscope is optimized for shorter-timescale observations with continuous all sky coverage and a multi-year period observation strategy. The continuous, fast-cadence, all-sky Evryscope light curves are sensitive to variations (including transits and eclipses) lasting only a few minutes, and provide fine sampling for ten minute level variations or longer.

The Evryscope is a robotic camera array mounted into a 6 ft-diameter hemisphere which tracks the sky \citep{2015PASP..127..234L}. The telescope is located at CTIO in Chile and observes continuously, covering 8150 sq. deg. in each 120s exposure. The Evryscope was deployed with 22 cameras and can accommodate 27 total cameras (with a corresponding increased field of view of 10,000 sq. deg). Each camera features a 29MPix CCD providing a plate scale of 13"/pixel. The Evryscope monitors the entire accessible Southern sky at 2-minute cadence, and the Evryscope database includes tens of thousands of epochs on 16 million sources. In this paper, we limited the search field to the region around the South Celestial Pole, and chose the brighter stars in order to maximize the number of epochs per source and minimize systematics.

The Southern Polar sky area is less explored than other parts of the sky, primarily due to the difficulty in reaching it. This is evidenced by the comparatively low number of planet, eclipsing binary, and variable star discoveries in this region. For example, the sky area in the declination region of \ang{-75} to \ang{-90} comprises 3.4\% of the southern sky's total area; however the VSX catalog of known variables in the same region accounts for only 1.2\% of the southern sky total. Surveys of the Southern Polar sky region typically either use a telescope located at a low latitude South American site or an instrument in the Antarctic. The former choice can be challenging depending on the airmass of the target region, while the second poses engineering difficulties due to the harsh environment \citep{1538-3881-145-3-58, 2015ApJS..218...20W}.

We use the Evryscope to explore the Southern Polar region (declinations \ang{-75} to \ang{-90}). While the airmass is non-optimal ($\sim$ 1.7 average), the Evryscope monitors the Southern Polar region continuously every night for the entire night at 2 minute cadence, with the same camera for multiple years. This long-term, same-camera coverage at short cadence results in many continuous data points with consistent airmass, and minimizes systematics. Targets in this region average over 60,000 epochs per year. Our observing strategy results in several hundred thousand light curves with targets ranging in brightness from 9 $<$ m\textsubscript{v} $<$ 15. The light curves have the precision necessary to potentially detect eclipsing binaries, variable stars, transiting gas-giant planets around small-cool host stars, and short-transit-time planets around small compact stellar remnants including white dwarfs and hot subdwarfs. With additional filtering, the light curves are precise enough to potentially detect gas-giant planets around bright solar type stars; we will address this in future work. These Evryscope light curves also facilitate searches with wide period ranges (for the Polar Search we searched from 3-720 hours), longer periods, and wide amplitude ranges. Long-period discoveries are typically non-interacting stars and are challenging to detect due to the low number of transits.

The primary target of this paper's search is eclipsing binaries, particularly low-mass and long-period systems. The secondary target of this paper's search is gas-giant planets around M-dwarf or late K-dwarf primaries. This survey relies on detection power to narrow the candidates and uses observations from mid 2016 to early 2017. The more challenging transiting exoplanet detections will be conducted with additional systematics removal steps, additional candidate filtering to push to lower power detections, and will use the full three plus year data set (Ratzloff et al., in prep). 

Eclipsing binaries are the best calibrators for determining relations between mass, radius, luminosity, and temperature. Relatively few low-mass (M-dwarf or late-K-dwarf secondary) eclipsing binaries have been discovered \citep{2011ApJ...731....8K, 2012ApJ...757..133L, 2018AJ....156...27C}, and many are too faint for easy radial-velocity followup measurements. This has limited our ability to measure the mass/radius relation at low masses, where many low-mass systems suggest larger radii than stellar models predict \citep{2010AJ....140.1158T, 2015ApJ...804...64M, 2017ApJ...845...72K}. This is particularly important for the determination of transiting planet radii around low-mass single stars, where some of the most exciting nearby planets are likely to be discovered \citep{2014SPIE.9143E..20R, 2018arXiv181002826C, 2018AJ....156..102S}.

In this paper, we report the discovery of 303 new variables including seven eclipsing binaries with low-mass secondary stars. We perform spectroscopic followup on select eclipsing binaries to confirm the stellar type and secondary size. Radial velocity measurements reveal that seven of the eclipsing binary discoveries are low-mass (.06 - .34 \(M_\odot\)) secondaries with K-dwarf primaries.

In \S~\ref{section_observations} we describe the Evryscope photometric observations that led to the discoveries as well as our analysis of the light curves and detection algorithms for identifying variables. In \S~\ref{section_followup_observations} we describe the followup observations performed for the low-mass eclipsing binaries including PROMPT \citep{2005NCimC..28..767R} followup photometry, identification spectra and radial velocity followup using the Goodman \citep{2004SPIE.5492..331C} spectrograph on the 4.1 meter SOAR telescope and the CHIRON \citep{2013PASP..125.1336T} echelle spectrometer on the CTIO/SMARTS 1.5 meter telescope. In \S~\ref{section_analysis} we present and characterize our discoveries. We also detail our analysis of the radial velocity followup work including the Monte-Carlo simulation to fit the masses, radii, and other parameters. We conclude in \S~\ref{section_summary}.

%----------------------------------------------------------------------------------------
%	SECTION
%----------------------------------------------------------------------------------------

\section{OBSERVATIONS AND VARIABILITY SEARCH} \label{section_observations}

%----------------------------------------------------------------------------------------
%	SUBSECTION
%----------------------------------------------------------------------------------------

\subsection{Evryscope Photometry}
All eclipsing binary and variable discoveries were detected in a transit search of the polar region (declinations \ang{-75} to \ang{-90}). The observations were taken from August 9, 2016 to April 4, 2017. The exposure time was 120s through a Sloan-g filter and each source typically had 16,000 epochs. We briefly describe the calibration and reduction of images and the construction of light curves; further details will be presented in an upcoming Evryscope instrumentation paper. Raw images are filtered with a quality check, calibrated with masterflats and masterdarks, and have large-scale backgrounds removed using the custom Evryscope pipeline. Forced photometry is performed using APASS-DR9 \citep{2015AAS...22533616H} as our master reference catalog. Aperture photometry is performed on all sources using multiple aperture sizes; the final aperture for each source is chosen to minimize light curve scatter. The primary systematics challenges are: background and airmass changes and the subsequent effects on stars of different magnitude and color, the ratchet observing cycle causing the targets to switch cameras and appear in different positions on the CCD chips over the observing season, daily aliases, source blending, PSF distortions, and vignetting. We use the quality filter, calibrations, aperture photometry, along with a custom implementation of the SysRem \citep{2005MNRAS.356.1466T} algorithm to remove the systematics challenges described above.

%----------------------------------------------------------------------------------------
%	SUBSECTION
%----------------------------------------------------------------------------------------

\subsection{Detection of Variables} \label{section_det_of_var}
\begin{figure}[h!]
\centering
\includegraphics[width=1.0\columnwidth]{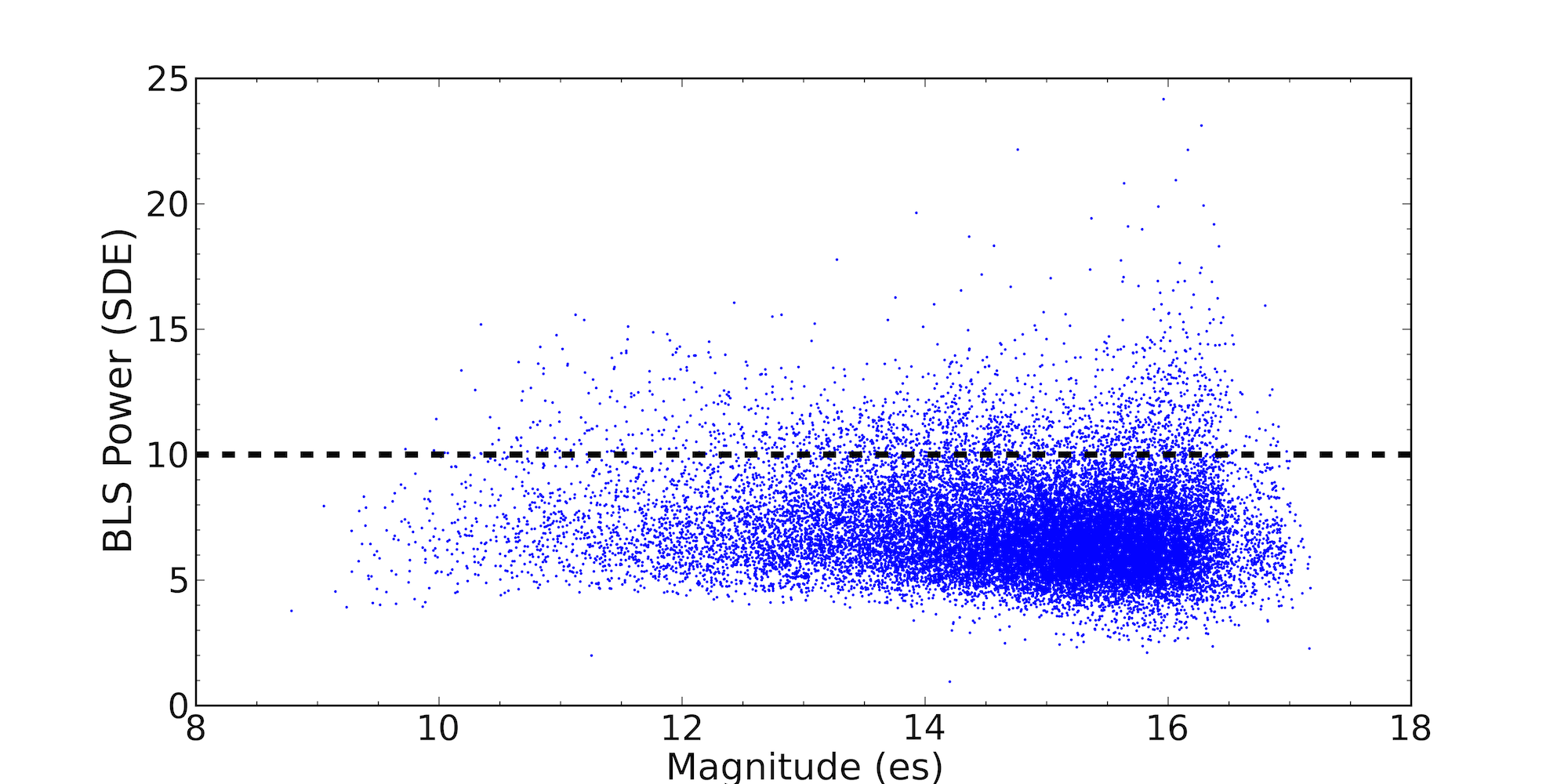}
\includegraphics[width=.45\columnwidth]{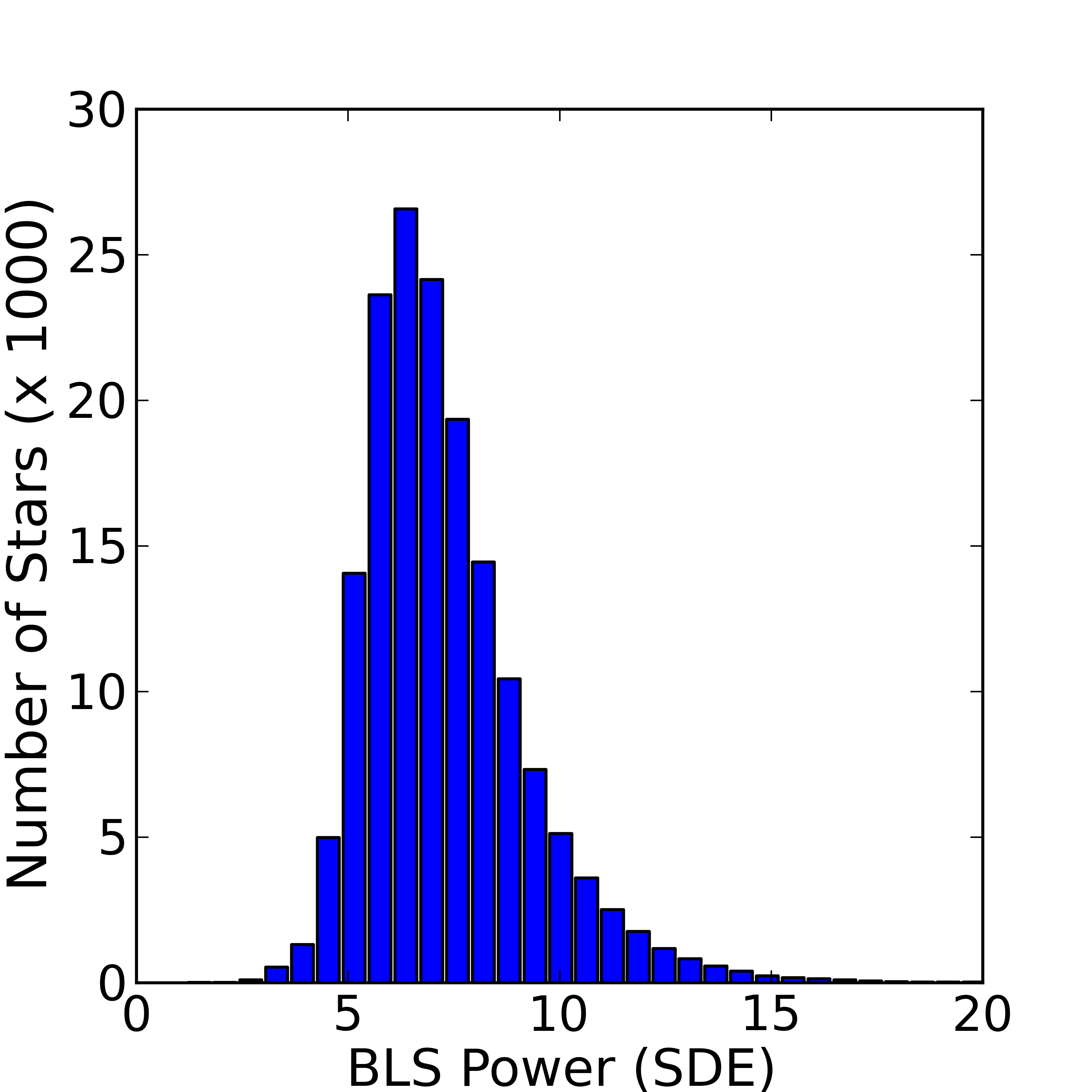}
\includegraphics[width=.45\columnwidth]{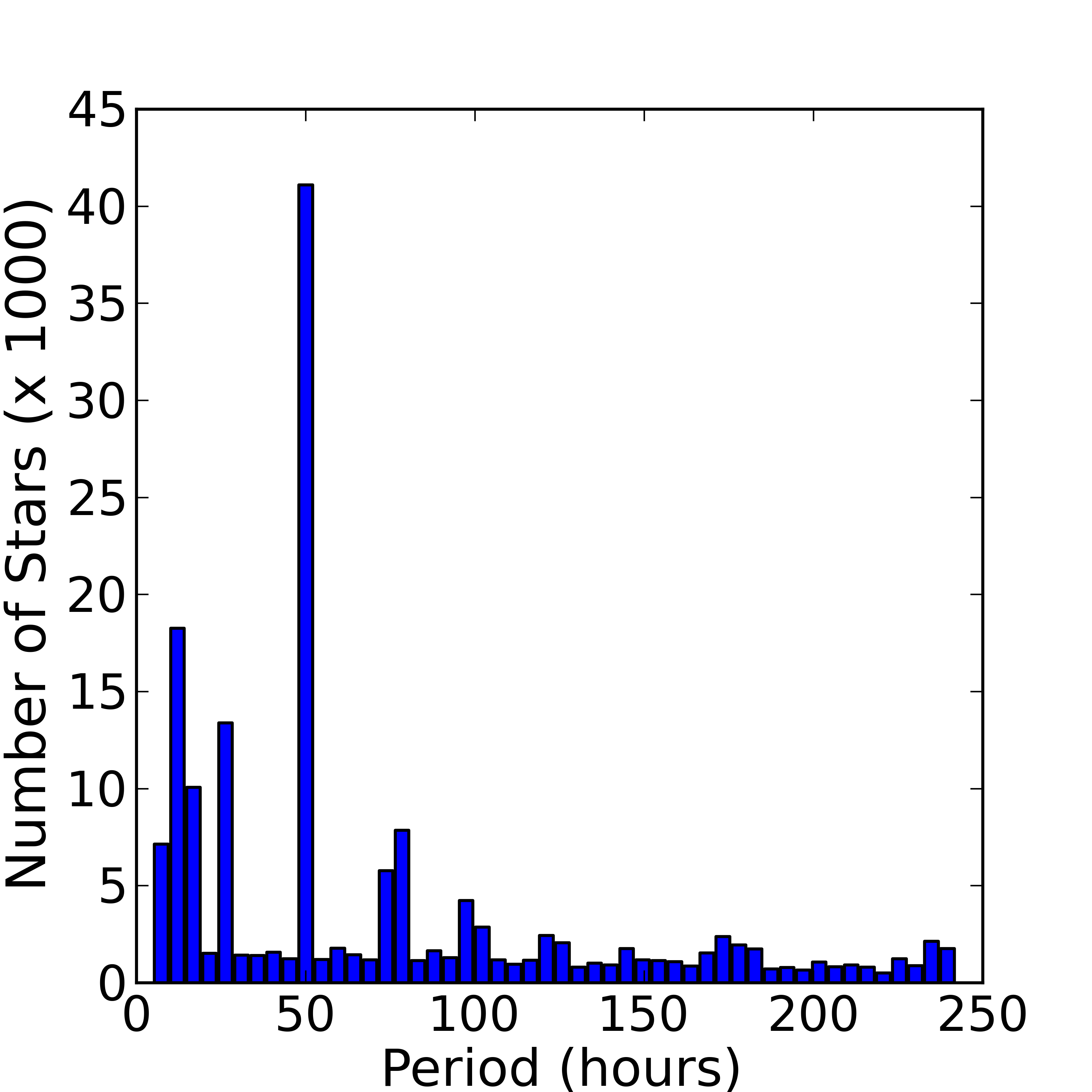}
\caption{Detection characteristics from the BLS results of the polar search. The top panel shows the BLS power in SDE vs. magnitude (15\% of the points are shown for better visualization), the lower left panel is the histogram of BLS power in SDE, the lower right is the histogram of periods found. Targets with an SDE $>$ 10 are selected for further inspection.}
\label{fig:bls_power}
\end{figure}

Filtering by declination and magnitude returns 239,991 initial targets from the Evryscope light curve database. 76,407 are eliminated by an additional quality filter based on non-blended sources. The remaining 163,584 are analyzed using Box Least Squares (BLS) \citep{Kovacs:2002gn, 2014A&A...561A.138O} with the pre-filtering, daily-alias masking, and settings described in \S~\ref{variability_algorithms}. The light curves are then sorted by BLS detection power, in terms of Signal Detection Efficiency (SDE) \citep{Kovacs:2002gn}. Figure \ref{fig:bls_power} shows the BLS SDE distribution for the targets along with the distribution of detected periods. Targets with an SDE $>$ 10 and with nearby reference stars gives 9104 suspects for further inspection. The 10-SED cutoff is chosen to: 1) limit the number of targets to an amount that is reasonable for human followup (in this case $\sim$ 10,000), 2) ensure a reasonable chance of detecting high-amplitude candidates without accumulating excessive false-positives, and 3) reach three percent level signal depths on bright stars to potentially detect low-mass secondaries and gas-giant planets around M-dwarfs or late K-dwarfs.

We compare the target light curve (both unfolded and folded to the best BLS period) to two nearby reference stars of similar magnitude looking for any signs that the detected variation is present in the references indicating systematics (see \S~\ref{false_positive}). The folded plots are colored by time to check how well-mixed the detection is, since a transit or eclipse with only a single or few occurrences is more likely to be an artifact of the detection algorithm. The light curves are also folded on the second and third best BLS periods to check for aliases, as well as the best Lomb-Scargle (LS) \citep{1975Ap&SS..39..447L, 1982Ap&SS..263..835S} period to check for sinusoidal variability. From visual inspection, we identify 649 variables from the machine filtered 9104 suspects. 346 are known variables and 303 are new discoveries.

%----------------------------------------------------------------------------------------
%	SUBSECTION
%----------------------------------------------------------------------------------------

\subsection{Machine-Learning Stellar Classification} \label{section_stellar_class}

We developed a machine-learning based classifier that uses publicly available catalog data to estimate stellar size from a B-V color/magnitude space, and to estimate spectral type from multiple color-differences. The discovery candidates were matched to APASS-DR9 \citep{2015AAS...22533616H} and PPMXL \citep{2010AJ....139.2440R} catalogs to obtain reduced proper motion (RPM) and color differences (B-V, V-K, J-H, H-K) for each target. Modifying the method in \cite{1972ApJ...173...xxx} with a two step machine learning process described below, we classify stars based on B-V and RPM to identify stellar size - main sequence, giants, white dwarfs, or sub dwarfs. The RPM and B-V combination provides a high return on our target catalog (99\% of our targets are classified as demonstrated below) and captures spectral information using available data. After the stellar size estimation is completed, the four color differences are used to approximate the spectral type.

\begin{figure}[h!]
\centering
\includegraphics[width=1.0\columnwidth]{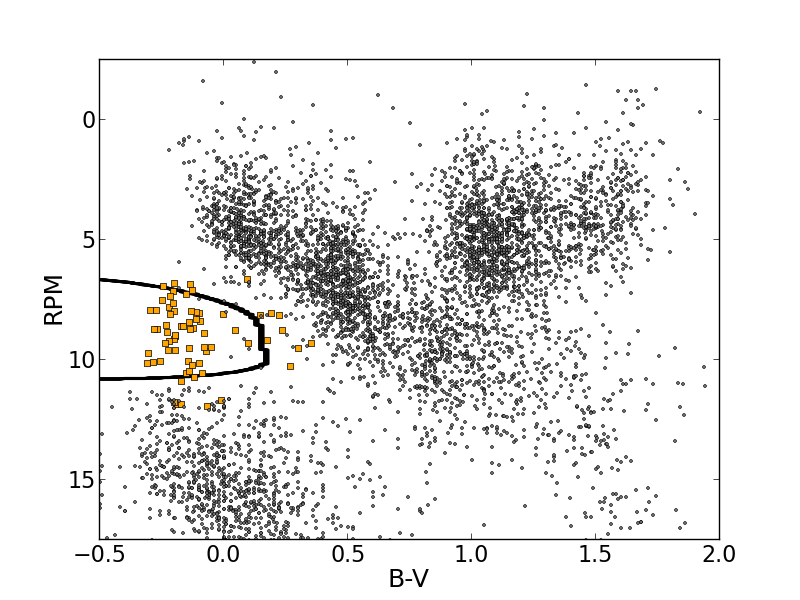}
\includegraphics[width=1.0\columnwidth]{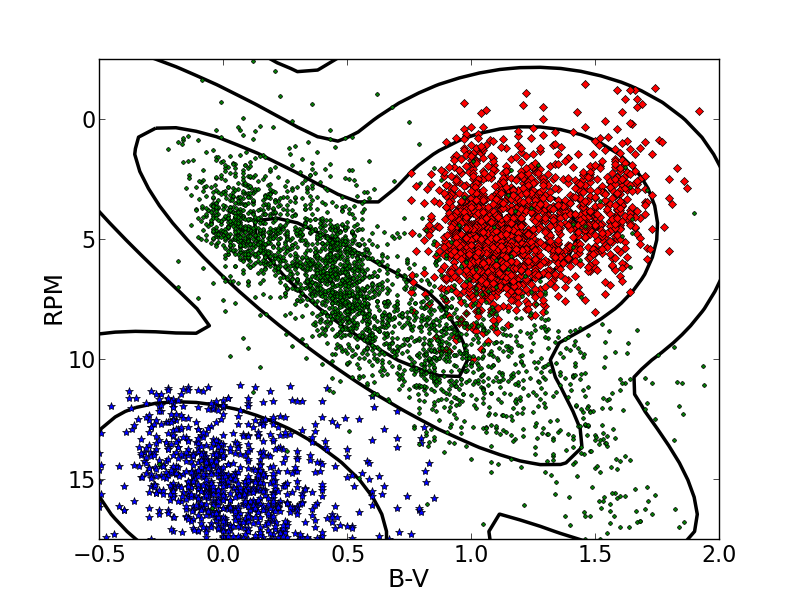}
\caption{The Evryscope Target Classification - We use B-V color differences and reduced proper motion (RPM) data with a two step machine learning algorithm to classify star size. Top: the training data (gold squares=hot subdwarfs, grey=all others) for the support vector machine (SVM) which returns the resulting hot subdwarf classification region (the area inside the black border). Bottom: the training data (blue stars=white dwarfs, green=main sequence, red diamonds=giants) for the Gaussian Mixture Model (GMM) which returns the resulting classification contours. Negative log likelihood plot-lines 1, 1.7, 2.8 are shown.}
\label{fig:classifier}
\end{figure}

In the first step of the machine learning process, we use a support vector machine (SVM) from the SKYLEARN python module \citep{scikit-learn} to identify likely hot subdwarfs (HSD) from all other stars. The HSD are challenging to separate since they can be close to main sequence O/A stars in this parameter space. We find the SVM to be an effective way to segregate the HSD, shown in the top panel of Figure \ref{fig:classifier} as the small confined area enclosed in the black border. This is done by using a training set of HSD from \citep{2017OAst...26..164G} and other types of stars from SIMBAD \citep{2000A&AS..143....9W}, filtering the outliers, then computing the contour boundaries. The SVM method is a non-probabilistic two-class classifier that computes a hard boundary (decision boundary) by minimizing the distance (or margin) between the points closest to the boundary. As with any classifier there are missed targets and contaminants, and there are physical reasons the results can be skewed (reddening for example). Our goal in this step is to separate the most challenging class (the HSD) from all the other classes while providing a boundary with a reasonable contingency space to the nearby white dwarf and main sequence regions.

Once the HSD are identified, all remaining objects are classified using a Gaussian Mixture Model (GMM) \citep{scikit-learn} with three classes to identify white dwarfs, main sequence, and giants. We again use an outlier filtered training set of stars of each type from SIMBAD (20,972 main sequence, 1515 white dwarfs (WD), and 10,000 giants). The GMM classifier results are shown in the bottom panel of Figure \ref{fig:classifier}. The GMM method is a best fit to 2-D Gaussian function (probability density function), using the training points to adjust the Gaussian centers, orientations, and elongations. Our application of this method uses three dimensions (WD, main sequence, and giants). Although more dimensions are possible, overlapping or poorly separated classes tend to give poor results (part of the motivation of using the SVM for the HSD step). The GMM produces contour lines with Negative-log-likelihood (NLL) values that can be converted ($LH=10^{-NLL}$) to give an estimate of the confidence level the data point belongs in the class.

\begin{figure}[h!]
\centering
\includegraphics[width=1.0\columnwidth]{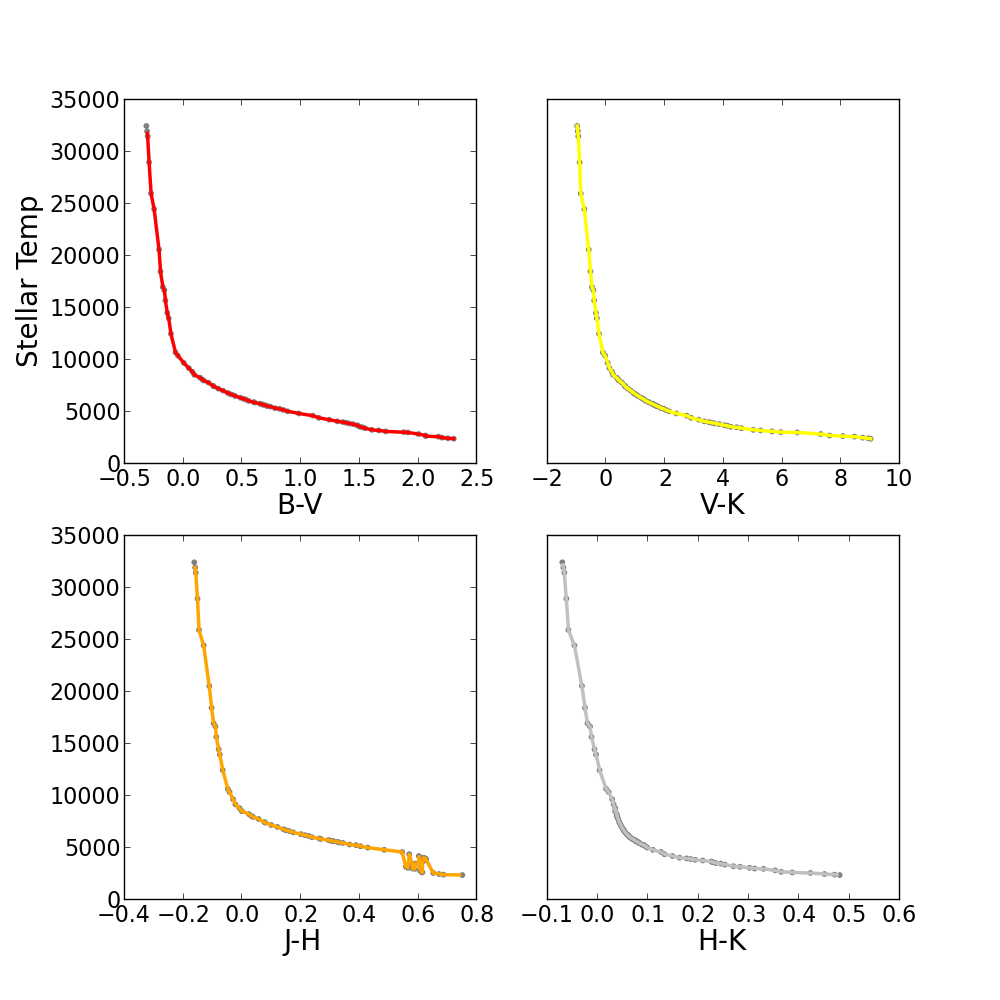}
\caption{The Evryscope Target Classification - We use (B-V, V-K, J-H, H-K) color differences to estimate temperature and spectral type using the data in \citep{2013ApJS..208....9P} to interpolate profiles for each color difference. The data are the grey points and the interpolations are the colored lines in the figures. We average the four results and pick the closest spectral type.}
\label{fig:temp_prof}
\end{figure}

We use the spectral type and temperature profiles in \citep{2013ApJS..208....9P} to derive a function (using 1-D interpolation) that uses available color differences to derive an estimate for spectral type (Figure \ref{fig:temp_prof}). If only B-V is available, we classify simply by the letter (O,B,A,F,G,K,M); if multiple colors are available we average the fits and choose the closest spectral type (G9, K4, M3 for example). For main sequence stars we add the luminosity class V. The code produces a function with RPM and color differences inputs and outputs the star size, star type, and NLL score for the GMM step. We used this to classify all of our discoveries, with the added requirement that the HSD also be apparent spectral type O or B and that the WD have a NLL score of less than 4.0. The added requirements help filter contaminants from main sequence A stars for the HSD, and borderline WD stars. Candidates identified as likely K or M-dwarfs with shallow (typically less than 10\%) eclipses or transits are identified as potentially high value targets and analyzed in more detail.

The Evryscope classifier is designed to: 1) facilitate identification of as many of the target light curves as practical, 2) identify targets to be included in Evryscope transit searches (white dwarfs, hot subdwarfs, K and M-dwarfs), and 3) classify variability discoveries helping to identify those as potentially interesting for further followup. For the Polar Search, 98.5\% of our targets have B-V and RPM data available, and 91.0\% have all four color differences and RPM data. Tests using 485,000 targets spread across the entire southern sky (all RA and declinations \ang{+10} to \ang{-90} ) have demonstrated very high returns - 99\% of Evryscope targets have all four color differences and RPM data available for classification. Once the catalogs were compiled and matched, the classifier took only a few minutes to classify the 485,000 test targets, making it practical for use on the full Evryscope database. All discoveries in this work are classified using the APASS-DR9 \citep{2015AAS...22533616H} and PPMXL \citep{2010AJ....139.2440R} catalogs as described above. A similar approach using the GAIA-DR2 \citep{2018A&A...616A...1G} catalog will be used as an additional target filter for the transiting exoplanet searches (Ratzloff et al., in prep).

We tested the Evryscope classifier in several ways. We chose known WD from APASS \citep{2017MNRAS.472.4173R} and high confidence ($>$.80) WD suspects from ATLAS \citep{2017arXiv170309714P} and SDSS \citep{2015arXiv150105309P} with m\textsubscript{v} $<$ 16.5, for a total of 211 classifier test targets. Using \cite{2017OAst...26..164G} with m\textsubscript{v} $<$ 15.0, we obtain 1560 HSD classifier test subjects (which may include WD due to the difficulty in separating the two groups). We use \cite{2011AJ....142..138L} to obtain 3764 high-confidence M-dwarfs. Using \cite{2001KFNT...17..409K} and filtering out the bright stars we have 999 main sequence, 452 giants, and 895 K-dwarfs for classifier testing.

Table \ref{tab:classifier_perf} shows the performance of the classifier to correctly determine star size (ms/giant/WD/HSD). Table \ref{tab:classifier_perf_2} shows the performance of the classifier to correctly determine letter spectral type (O,B,A,F,G,K,M). Table \ref{tab:classifier_perf_3} shows the performance of the classifier to correctly determine full spectral type (O0 - M9).

\begin{table}
\caption{Evryscope Classifier star size (ms/giant/WD/HSD) performance.}
\begin{tabular}{ l c c}
Test group & Star size & ES Classifier \% correct\\
 \hline
M-dwarfs & ms & 95.3\%\\
K-dwarfs & ms & 86.2\%\\
M-giants & giant & 98.7\%\\
Main Sequence & ms & 94.1\%\\
HSD & HSD & 54.5 (77.7 w/WD)\%\\
WD & WD & 87.0\%\\
 \hline
\end{tabular}
\label{tab:classifier_perf}
\end{table}

\begin{table}
\caption{Evryscope Classifier letter spectral type (O,B,A,F,G,K,M) performance.}
\begin{tabular}{ l c c}
Test group & letter spectral type & ES Classifier \% correct\\
 \hline
M-dwarfs & M & 95.2\%\\
K-dwarfs & K & 81.1\%\\
M-giants & M & 97.9\%\\
Main Sequence & O-M & 69.5\%\\
HSD & O,B & 76.5\%\\
 \hline
\end{tabular}
\label{tab:classifier_perf_2}
\end{table}

\begin{table}
\caption{Evryscope Classifier full spectral type (O0 - M9) performance.}
\begin{tabular}{ l c c c c}
  &   &   & ES Classifier & \\
Test group & spectral type & mean & variance & \% +/-3\\
 \hline
M-dwarfs & M0-M9 & -.50 & 1.9 & 95.3\%\\
K-dwarfs & K0-K9 & -.98 & 2.7 & 81.6\%\\
M-giants & M0-M9 & -2.0 & 1.7 & 78.0\%\\
Main Sequence & O0-M9 & -.87 & 3.7 & 63.1\%\\
 \hline
 \footnote{Shown is the mean difference and variance in classifier performance numerical class versus the known class. The last column shows the percent of the test group that is classified correctly to within 3 of the known numerical class.}
\end{tabular}
\label{tab:classifier_perf_3}
\end{table}

We also compared the classifier results to SOAR ID spectra taken for the low-mass eclipsing binaries (\S~\ref{section_lmeb}). 7 of the 8 were classified as the correct spectral type (K for example), and within +/- 1 numeric class (K5 or K6 for example) (Table \ref{tab:classifier_comp}).

\begin{table}
\caption{Comparison of the Evryscope Classifier to SOAR ID spectra.}
\begin{tabular}{ l c c}
ID (EVR+) & SOAR ID Sptp & ES Classifier Sptp\\
 \hline
J053513.22-774248.2 & G7V & K1V\\
J06456.10-823501.0 & G8V & G9V\\
J103938.18-872853.8 & K7V & K6V\\
J110815.96-870153.8 & K4V & K3V\\
J165050.23-843634.6 & K5V & K4V\\
J180826.26-842418.0 & G5V & G6V\\
J184114.02-843436.8 & K2V & K3V\\
J211905.47-865829.3 & K5V & K6V\\
 \hline
\end{tabular}
\label{tab:classifier_comp}
\end{table}

%----------------------------------------------------------------------------------------
%	SUBSECTION
%----------------------------------------------------------------------------------------

\subsection{Variability search algorithms} \label{variability_algorithms}
We selected sources in the polar region with m\textsubscript{v} $<$ 14.5 and with light curves that passed quality tests to eliminate sources with blending, narrow time coverage, or low number of epochs (\S~\ref{section_det_of_var}). Light curves (with MJD timestamps) were pre-filtered with a Gaussian smoother to remove variations on periods greater than 30 days, and a 3rd order polynomial fit was subtracted to remove long-term variations. Light curves were then searched for transit-like, eclipse-like, and stellar variability signals using the Box Least Squares (BLS) \citep{Kovacs:2002gn, 2014A&A...561A.138O} and Lomb-Scargle (LS) \citep{1975Ap&SS..39..447L, 1982Ap&SS..263..835S} algorithms.

We tested the recovery rates on Evryscope light curves with different BLS settings - with periods ranging from 2-720 hours, 10,000-100,000 periods tested, and transit fractions from .001 to 0.5. Recovery rate tests were run on known eclipsing binaries in our magnitude range with different transit depths ranging from .01 to .25, and on simulated few-percent level transit signals injected onto Evryscope light curves representative of low-mass secondaries. The tests showed that a very wide BLS test period range (2-720 hours) led to decreased detections as the periodogram becomes biased to long periods or spikes in longer periods arise from data gaps. This challenge combined with the survey 6-month time coverage (\S~\ref{section_intro}), shows too aggressive of a period range can detect fewer eclipsing binary candidates. Based on these tests, the final BLS settings used on the Evryscope Polar Search were a period range 3-250 hours with 25,000 periods tested and a transit fraction of .01 to .25.

Period detections of 24-hours and corresponding aliases (4, 6, 8, 16, 36, 48, and 72 hours) were masked in +/- .1 hour widths. The results were sorted by BLS signal detection strength - BLS periodogram peak power in terms of sigmas above the mean power. Targets with peak power greater than 10-sigma were verified visually with a panel detection plot. We use the Lomb-Scargle (LS) algorithm to identify sinusoidal variables. For LS, we used a period range 3-720 hours to include sensitivity to longer period variables. We recover slightly lower amplitude variables (minimum discovery amplitude in this work $= .008$) than eclipsing binaries (minimum discovery depth in this work $= .029$) as shown in the Appendix.

\begin{figure}[h!]
\centering
\includegraphics[width=1.0\columnwidth]{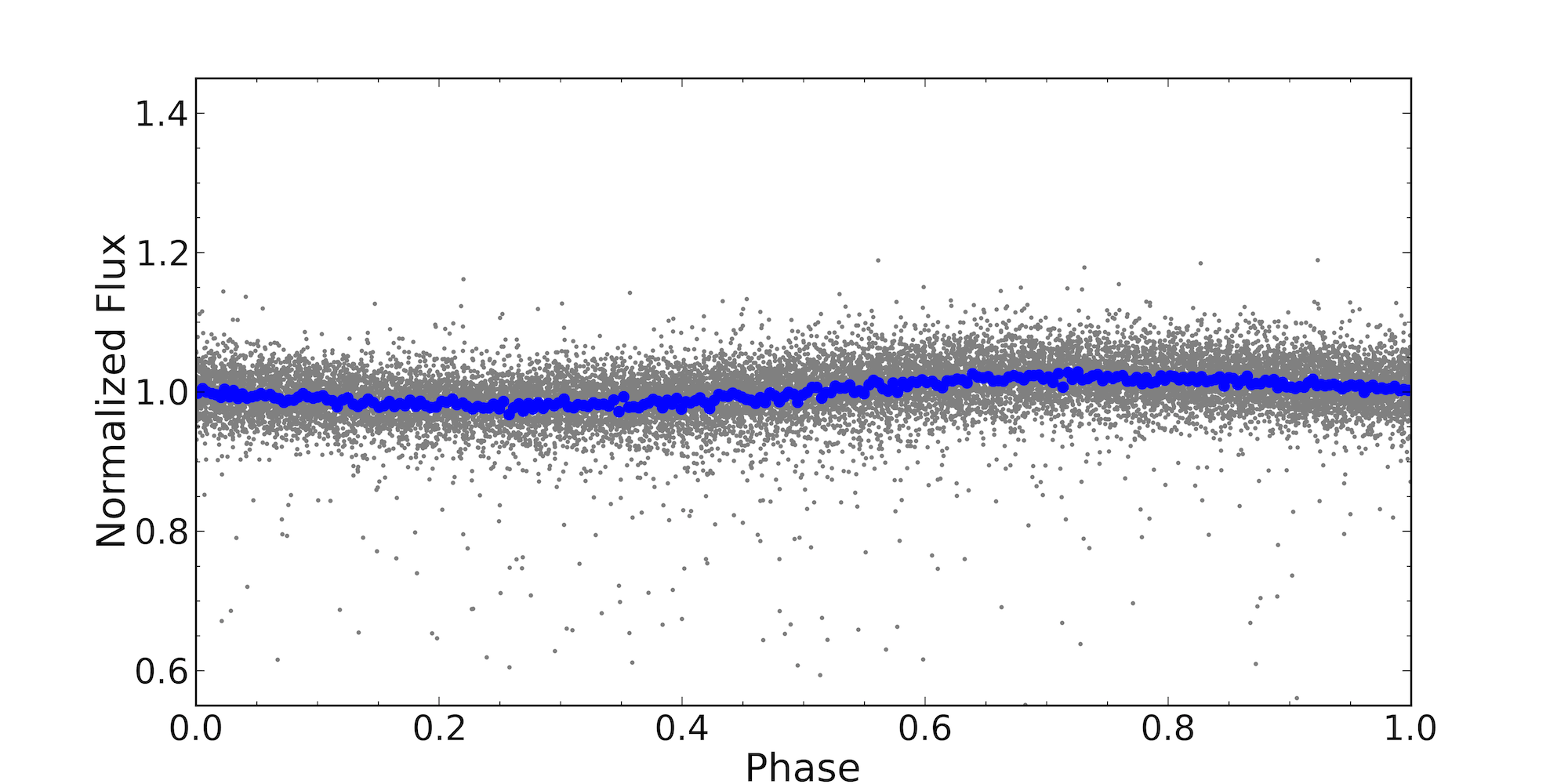}
\includegraphics[width=1.0\columnwidth]{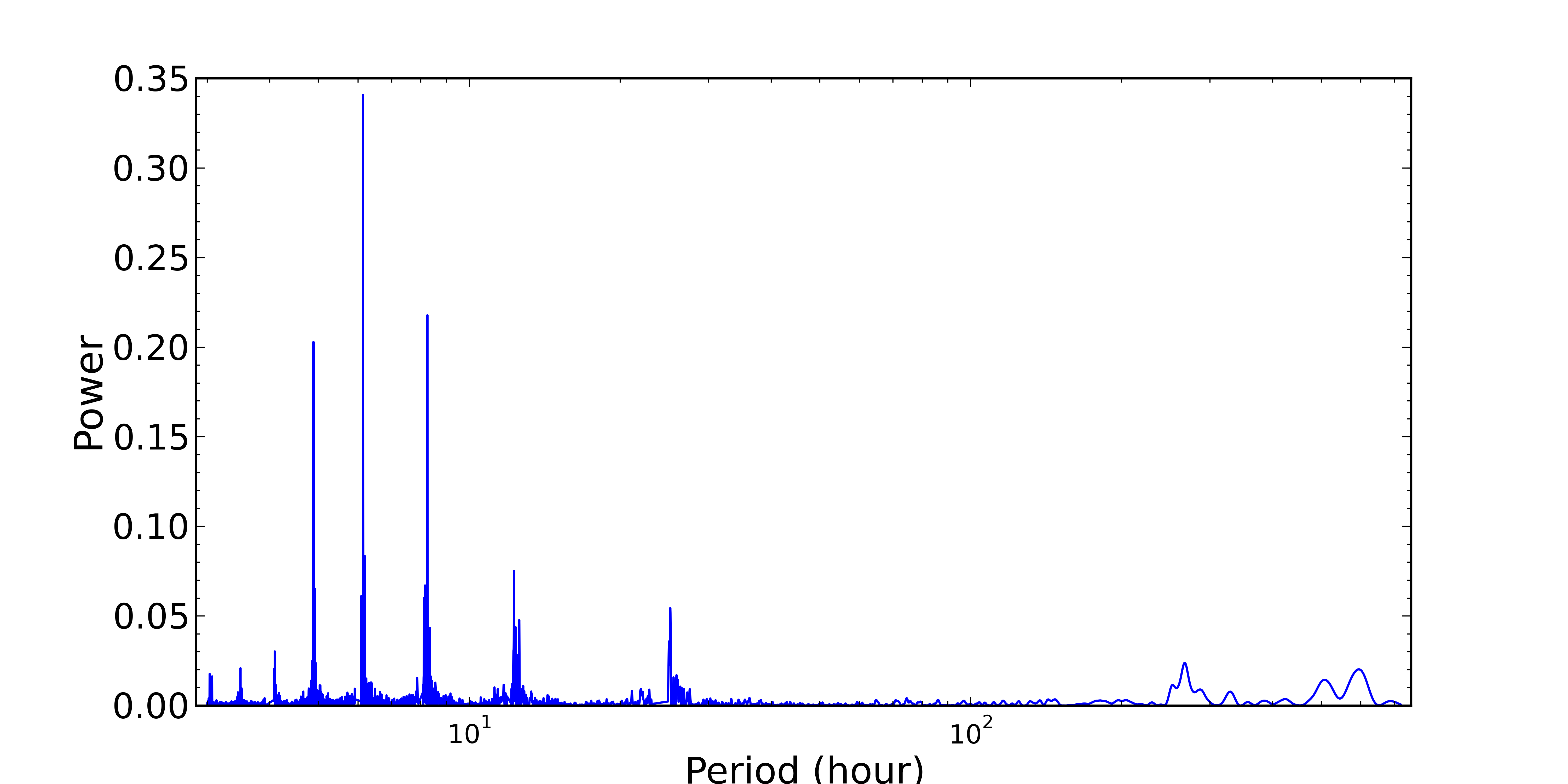}
\caption{An example variable discovery (EVRJ032442.50-780853.9) from this survey. The Evryscope light curve phased on its period of 4.676 hours is shown on the top panel. Grey points = 2 minute cadence, blue points = binned in phase. The bottom panel shows the LS power spectrum with the highest peak at the 4.676 hour detection.}
\label{fig:var_detection}
\end{figure}

Figure \ref{fig:eb_detection} shows the phase-folded Evryscope light curve for EVRJ110815.96-870153.8 a K4V primary and .21 \(M_\odot\) secondary with a BLS detected period of 12.28 hours. Figure \ref{fig:var_detection} shows the light curve for EVRJ032442.50-780853.9 a variable star with a LS detected period of 4.67 hours.

\begin{figure}[h!]
\centering
\includegraphics[width=1.0\columnwidth]{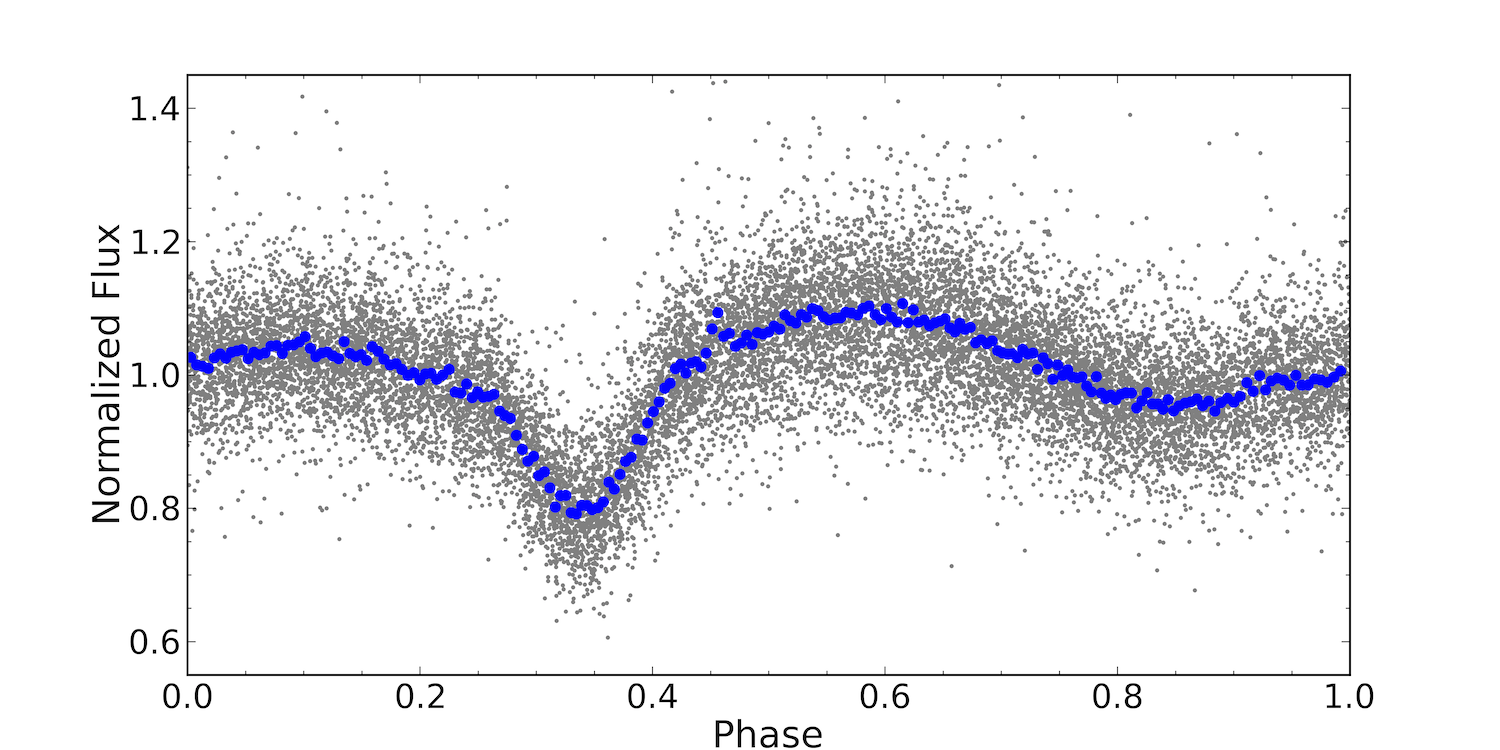}
\includegraphics[width=1.0\columnwidth]{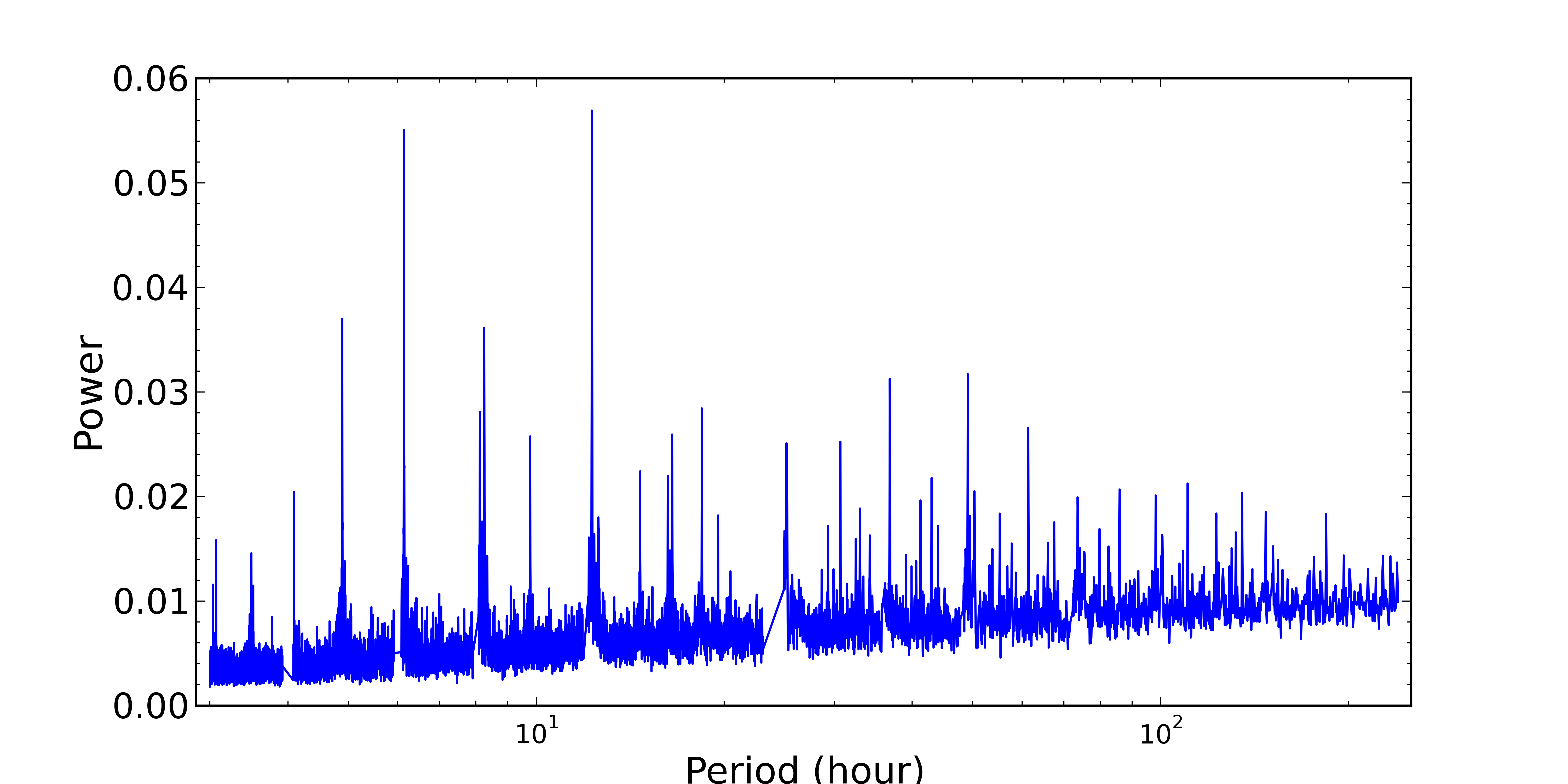}
\caption{An example low mass eclipsing binary discovery (EVRJ110815.96-870153.8) from this survey. The Evryscope light curve phased on its period of 12.277 hours is shown on the top panel. Grey points = 2 minute cadence, blue points = binned in phase. The bottom panel shows the BLS power spectrum with the highest peak at the 12.277 hour detection.}
\label{fig:eb_detection}
\end{figure}

\subsection{False Positive Tests} \label{false_positive}

We performed several tests to verify the variability signals were not false positives. First, we compared the candidate light curve with several nearby reference star light curves looking for similar variation to test for systematics or PSF blending. The nearest reference stars within 0.2 degrees of the reference star were filtered by magnitude and light curve coverage. The nearest three with magnitudes within 0.5 mag of the target star and with a light curve coverage and number of data points within 20\% of the target light curve are chosen for comparison. The references are folded at the same period as the detected period of the candidate, and are inspected visually for signs of similar signals, offsets, or outliers. Candidates with references showing similar variability are assumed to be systematics and thrown out.

Next we tested how well-mixed in phase the observations were, with poor mixing potentially indicating matched-filter fits to systematics or data gaps instead of astrophysical signals. This is performed by folding the candidate on the detected period and color coding the points by time (ranging from a blue-to-red scheme mapped to early-to-late times) and visually inspecting the resulting plot. For each discovery, we also compared the phased light curve of the first and the second half of the data looking for inconsistency. Candidates with marginal results from these tests were reviewed by an additional person and thrown out if both agreed the target is suspect.

Eclipsing binary light curves that did not reveal a secondary eclipse or out-of-transit ellipsoidal variation were tested further. For these candidates, we folded the light curves at twice the detected period, looking for differences in odd/even transit depths to rule out finding half of the actual period. Candidates passing these tests were then flagged as probable variable discoveries and analyzed further as detailed in \S~\ref{section_analysis}.

%----------------------------------------------------------------------------------------
%	SECTION
%----------------------------------------------------------------------------------------

\section{FOLLOWUP OBSERVATIONS} \label{section_followup_observations}

Followup observations for select eclipsing targets were made with the PROMPT telescopes \citep{2005NCimC..28..767R} in order to confirm the Evryscope detection. We used the SOAR Goodman spectrograph \citep{2004SPIE.5492..331C} for stellar classification and intermediate-resolution radial velocity measurements. We used the CHIRON \citep{2013PASP..125.1336T} spectrograph for high-resolution radial velocity measurements to measure the companion masses of select suspected low-mass secondaries.

%----------------------------------------------------------------------------------------
%	SUBSECTION
%----------------------------------------------------------------------------------------

\subsection{SOAR Goodman ID Spectroscopy} \label{section_ID_spec}

We observed the low mass candidates on April 29, 2018 on the SOAR 4.1 m telescope at Cerro Pachon, Chile with the Goodman spectrograph. We used the red camera with the 400 1/mm grating with a GG-455 filter in M1 and M2 preset mode with 2x2 binning and the 1" slit (R $\sim$ 825). The red camera \footnote{http://www.ctio.noao.edu/soar/content/goodman-red-camera} is optimized for the optical red part of the spectrum and when used with the M1 and M2 presets provides a wavelength coverage of 3500-9000 Angstroms. The Goodman spectra are 2-D, single order. We took eight consecutive 60s spectra for each of the targets and for the standard LTT3864. For calibrations, we took 3 x 60s FeAr lamps, 10 internal quartz flats using 50\% quartz power and 10s integrations, and 10 bias spectra.

We processed the spectra with a custom pipeline written in Python by the Evryscope team; this pipeline is described in detail here. The eight spectra for each target are median-combined, bias-subtracted, and flat-corrected. A 3rd-order polynomial is fit to the brightest pixels in each row; the spectra are then extracted in a 10-pixel range and background subtracted. We identify 8 prominent lamp emission lines for each preset (including 3749, 4806, 6965 Angstroms and many others spread across the entire wavelength range) and compare with the known lines of the Iron-Argon arc lamp using a Gaussian fit of each feature. We use a 4th-order polynomial to fit the Gaussian peaks and wavelength-calibrate each spectrum. We used the standard star LTT3864 to flux-calibrate by first removing prominent absorption features then fitting a 7th-order polynomial to the continuum. The resulting SOAR standard star spectra was visually matched to the template from the ESO library and verified to fit within the template precision. The spectra were normalized and the results from the M1 and M2 presets were combined for each target with a wavelength coverage of 3500-9000 Angstroms.

Errors in the SOAR spectra arise from instrumentation systematics, observational conditions, and the extraction pipeline. Instrumentation error sources are dominated by flexure, component alignment, and limitations in optical quality due to manufacturing constraints; see \cite{2004SPIE.5492..331C} for an elaborate discussion of these contributions. Observational sources of errors are primarily due to background noise, airmass, and atmospheric effects. Errors in the spectra from the extraction process are discussed in detail in \cite{2017ASPC..509..263F}; the chosen standard, normalization process, and resolution are the error sources relevant to this work. The Goodman spectrograph has been operating consistently for over 15 years, and we use the accumulated knowledge to minimize errors from instrumentation, observation, and processing sources. In \S~\ref{section_SOAR_ID_spectra_analysis} we compare the SOAR ID spectra to the spectra of stars with known stellar types. The known spectra are from different instruments, observational strategies, and pipelines; additionally the available known spectra are limited to an accuracy of $\approx$1-2 in the luminosity class. The combined errors in the high SNR SOAR ID spectra are less than this limitation. We demonstrate this in \S~\ref{section_SOAR_ID_spectra_analysis} by comparing the results from different stellar classification methods, which are consistent to $\approx$1-2 in the luminosity class.

%----------------------------------------------------------------------------------------
%	SUBSECTION
%----------------------------------------------------------------------------------------

\subsection{PROMPT Photometry}
EVRJ114225.51-793121.0, EVRJ06456.10-823501.0, EVRJ184114.02-843436.8, and EVRJ211905.47-865829.3 were observed with the PROMPT P8 60cm telescope located at CTIO Chile. All observations were taken with Johnson B and Johnson R filters, interleaved. Table \ref{tab:prompt_phot} summarizes the PROMPT followup work.

\begin{table}
\caption{PROMPT observations of select targets.}
\begin{tabular}{ l c c c}
ID (EVR+) & Date & Images & B/R(s)\\
 \hline
J06456.10-823501.0 & Dec 10, 2017 & 412 & 40/20\\
J114225.51-793121.0 & Oct 30, 2017 & 190 & 90/60\\
J114225.51-793121.0 & Feb 16, 2018 & 288 & 90/60\\
J184114.02-843436.8 & Dec 19, 2017 & 202 & 100/45\\
J211905.47-865829.3 & Nov 21, 2017 & 120 & 130/90\\
 \hline
\end{tabular}
\label{tab:prompt_phot}
\end{table}

\begin{figure}[h!]
\centering
\includegraphics[width=1.0\columnwidth]{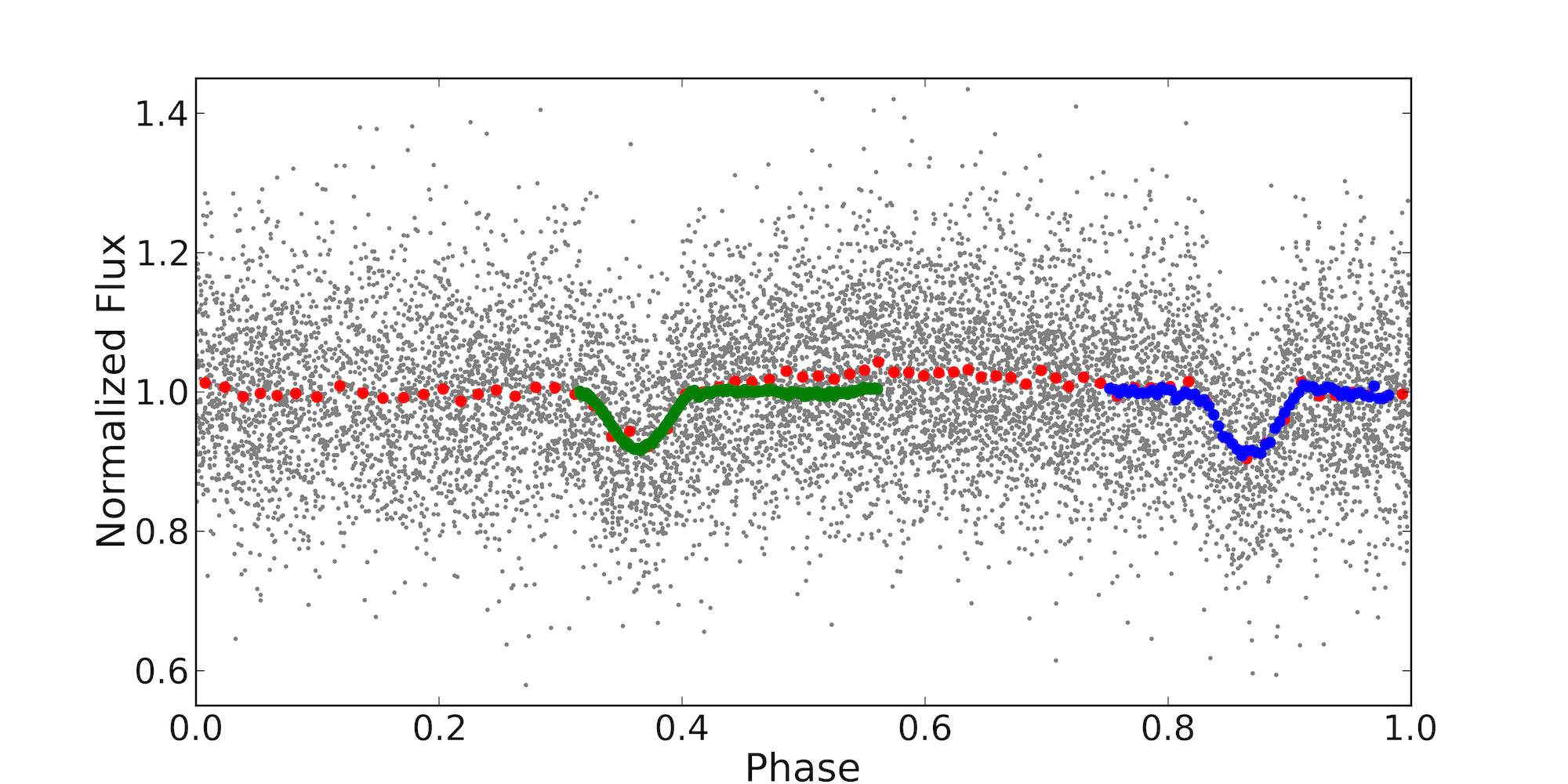}
\includegraphics[width=1.0\columnwidth]{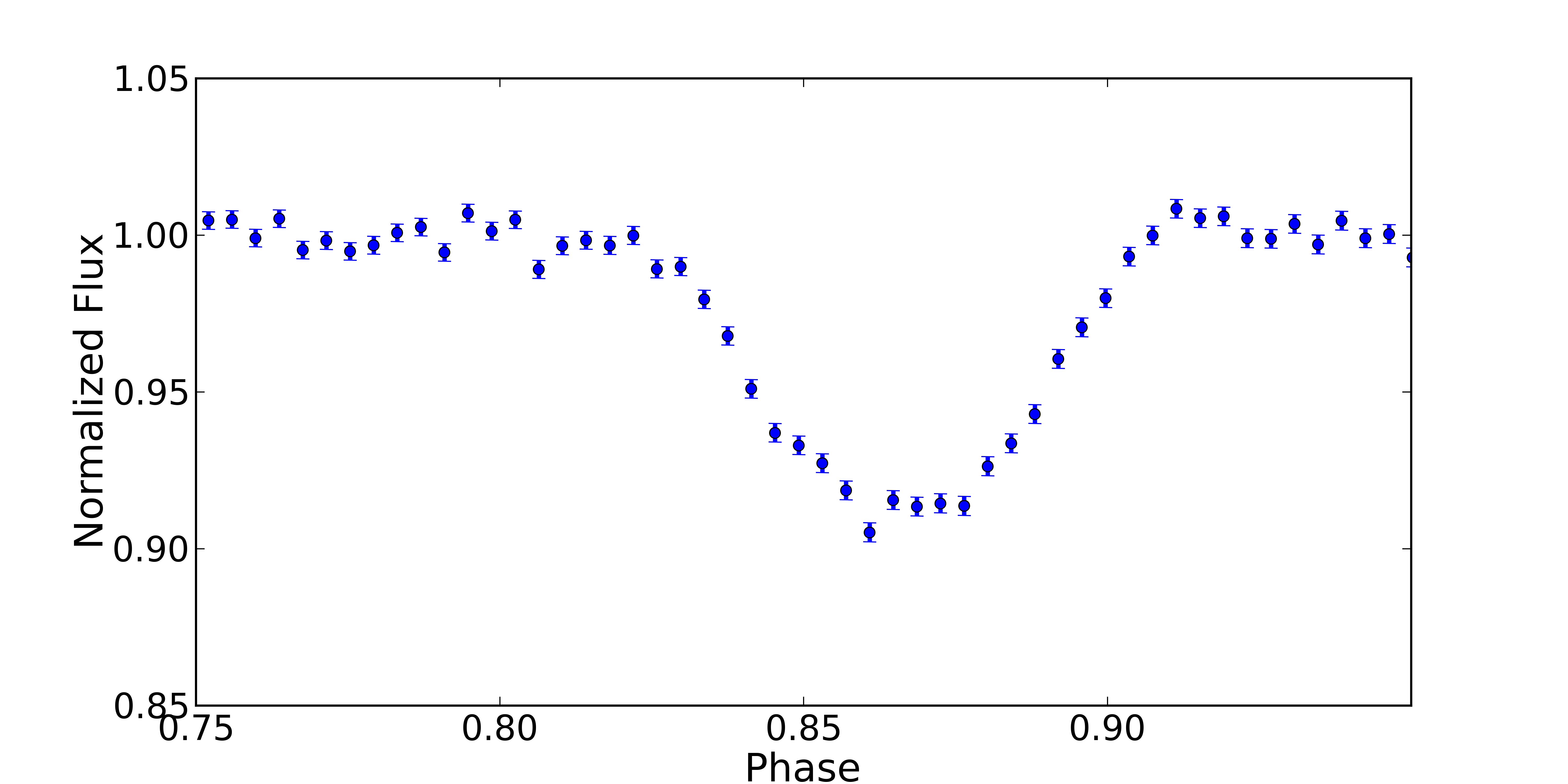}
\caption{\textit{Top:} Combined light curves of EVRJ211905.47-865829.3. This object was flagged as a potential 9.3 hour transiting gas giant planet as the transit depths are unchanged by color and in odd/even phase. There is a slight out of phase ellipsoidal variation when folded at the 18.6 hour period indicating it is most likely a grazing eclipsing binary with nearly identical primary and secondaries. \textit{Bottom:} A detailed view of the transit in the PROMPT light curve with 1$\sigma$ errors shown.}
\label{fig:combined_lc}
\end{figure}

The PROMPT followup observations confirm the candidate variability is astrophysical and not an Evryscope systematic by observing the Evryscope detection signal with a separate instrument and different eclipse time. The PROMPT telescopes have a 100 times larger aperture than the Evryscope cameras, giving the PROMPT light curves a lower root-mean-square (RMS) scatter and improved signal-to-noise-ratio (SNR) compared to the Evryscope discovery light curves. The amount of improvement depends on many factors including target brightness and sky background; here we show a representative example, EVRJ211905.47-865829.3 in Figure \ref{fig:combined_lc}. The light curve RMS (after removing the eclipse) for this target is .006 in PROMPT and .108 in Evryscope (unbinned 2-minute cadence). This corresponds to a SNR of $\approx{167}$ for the PROMPT single transit light curve and $\approx{9.5}$ for the Evryscope one year light curve. These results compare nicely to estimated theoretical SNR of 175 and 12 for PROMPT and Evryscope respectively, using reasonable values for sky background, throughput, and airmass for these telescopes observing an $m\textsubscript{v} =$ 14.0 magnitude target. We point out that the Evryscope binned light curve can reach the SNR of the PROMPT light curve, in this example with reduced sampling. In an upcoming white dwarf / hot subdwarf fast binary discovery paper (Ratzloff et al., in prep) we demonstrate the ability to reach higher than PROMPT SNR with multiple year binned Evryscope data. In this work, we use PROMPT to verify the Evryscope candidates and better characterize the eclipse depth and shape to reduce the error in the companion radii calculation. For this target, we also observed the secondary eclipse for comparison with the primary eclipse shown in Figure \ref{fig:combined_lc}. The PROMPT data also provides an additional eclipse time (several months past the latest Evryscope eclipse), and by phase-folding both light curves, the period accuracy is increased.

The PROMPT images were processed with a custom aperture photometry pipeline written in Python. The images were dark and bias-subtracted and flat-field-corrected using the master calibration frames. Five reference stars of similar magnitude are selected and aperture photometry is performed using a range of aperture sizes. The background is estimated using a sigma clipped annulus for each star scaled by the aperture size. A centroid step Gaussian fits the PSF to calculate the best center and ensures each aperture center is consistent regardless of pixel drift. The light curve rms variation is computed for the range of apertures, and the lowest-variation aperture size is chosen. A final detrending step using a 3rd order polynomial is applied to remove remaining systematics. Photometric errors are calculated per epoch using the estimated CCD aperture photometry noise in \cite{1995ExA.....6..163M} and the atmospheric scintillation noise approach in \cite{1967AJ.....72Q.328Y}. A detailed summary of the photometric error calculation is given in \cite{2017AJ....153...77C}. We combined the PROMPT and Evryscope light curves for final inspection. An example of a grazing eclipsing binary originally flagged as a 1.7 R\textsubscript{J} planet candidate is shown in Figure \ref{fig:combined_lc}. Radial velocity follow-up with the HARPS \citep{2003Msngr.114...20M} spectrograph on the ESO La Silla 3.6m telescope combined with the detailed light curve analysis confirms the candidate is a grazing eclipsing binary.

%----------------------------------------------------------------------------------------
%	SUBSECTION
%----------------------------------------------------------------------------------------

\subsection{Intermediate-resolution Spectroscopy and Radial Velocity}

EVRJ114225.51-793121.0, EVRJ06456.10-823501.0, EVRJ053513.22-774248.2, EVRJ184114.02-843436.8, and EVRJ211905.47-865829.3 were observed on November 15 and 19, 2017 and December 15 and 16, 2017 on the SOAR 4.1 m telescope at Cerro Pachon, Chile with the Goodman spectrograph. EVRJ110815.96-870153.8, EVRJ180826.26-842418.0, EVRJ165050.23-843634.6, and EVRJ103938.18-872853.8 were observed on February 12, 2018 and March 3, 2018. We used the blue camera with the 2100 1/mm grating in custom mode with 1x2 binning and the 1" slit (R $\sim$ 5500). We took four 300-360s spectra depending on the target and conditions. For all targets, we took 3 x 60s FeAr lamps after each group of science images. We took 10 internal quartz flats with 80\% quartz lamp power and 60s integration, and 10 bias spectra.\\
The spectra are processed using a modified version of the Python code described in \S~\ref{section_ID_spec} and radial velocity measurements are calculated (\S~\ref{section_SOAR_RV}). The SOAR spectra return radial velocity precision of $\approx$ 10 km/s for our targets, which allows us to characterize the secondary mass for small late M-dwarf stars. This also allowed us to rule out potential planetary-mass secondaries - the case in several of the grazing eclipses.\\

%----------------------------------------------------------------------------------------
%	SUBSECTION
%----------------------------------------------------------------------------------------

\subsection{High-resolution Radial Velocity}

EVRJ06456.10-823501.0 and EVRJ053513.22-774248.2 were observed between January 28, 2018 and March 25, 2018 on seven nights (one data point per night) with the SMARTS 1.5 m telescope at CTIO, Chile with the CHIRON spectrograph. EVRJ184114.02-843436.8 was observed on March 23, 2018. Spectra were taken in image slicer mode (R $\sim$ 80000). One 1500 to 1800 second spectrum was taken depending on the target and conditions. Spectra of RV standard HD131977 were taken to verify processing results.

\begin{figure}[h!]
\centering
\includegraphics[width=1.0\columnwidth]{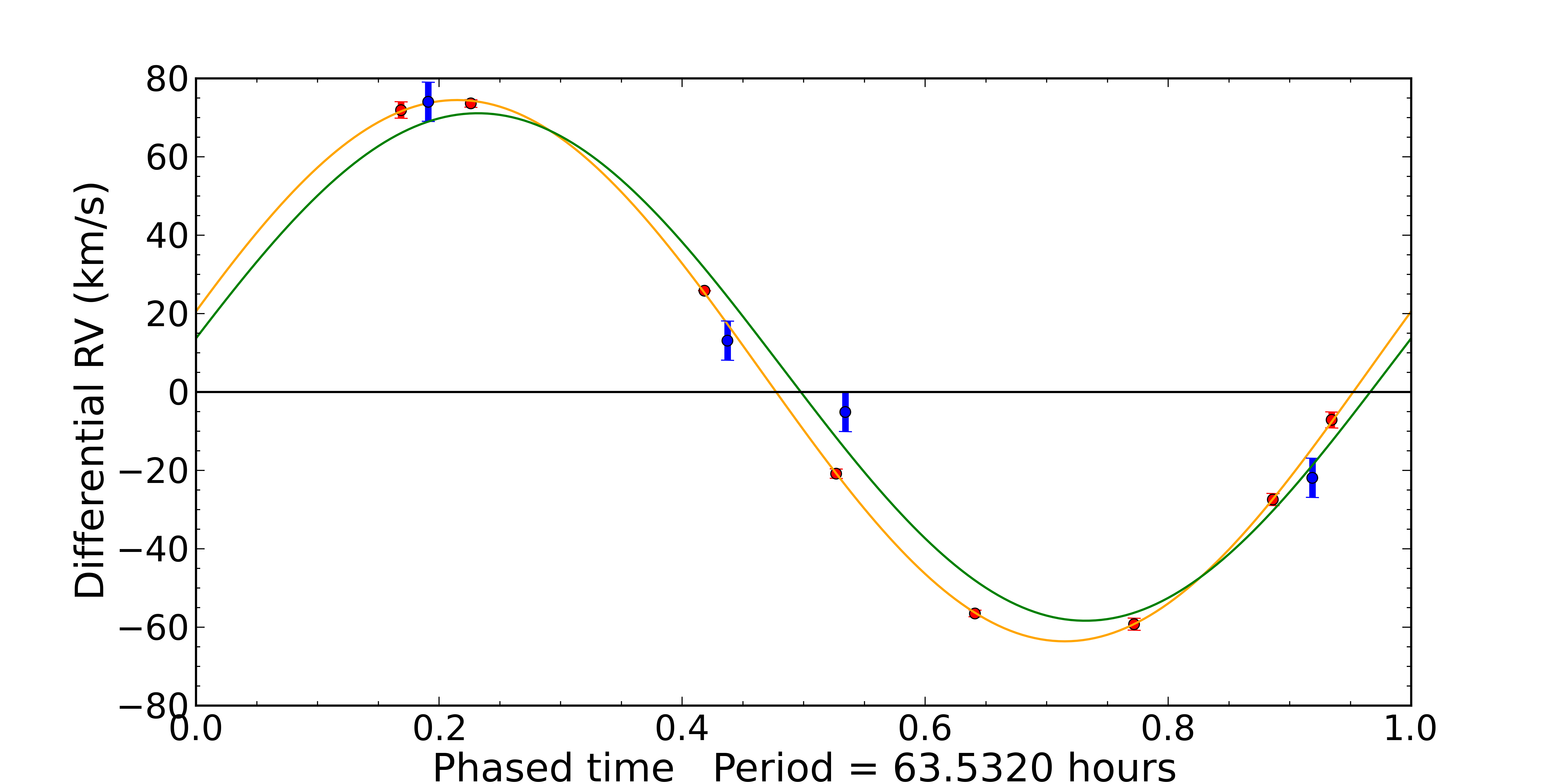}
\caption{Combined Radial Velocity curves for target EVRJ06456.10-823501.0. The red data points are from CHIRON RV data, and the blue points are SOAR data with the yellow and green curves of best fit.}
\label{fig:combined_CHIRON}
\end{figure}

Spectra were wavelength calibrated by the CHIRON pipeline, which we processed using a custom python code to measure radial velocity. We visually inspected the spectral orders and chose the top seven by SNR and with prominent atmospheric absorption features. The orders are spread throughout the wavelength range, and we select the most prominent atmospheric feature per order. Within each of the selected orders, for each observation, we clip a small section (typically 20 Angstroms) encompassing the best absorption feature. For example order nine uses the 4957 Angstrom feature, order fourteen uses the 5328 Angstrom feature, and order thirty-seven uses the 6563 Angstrom feature. We fit a Lorentzian to the absorption features and measure the wavelength shift of each observation in each order. For each observation, we sigma clip any outlier orders and use the average shift to calculate the velocity. Using the standard deviation of the measured shifts between the orders, we place error limits. The error in the RV standard is measured to $\approx$ 200 m/s, while the errors in the fainter targets are $\approx$ 1km/s. An example is shown in Figure \ref{fig:combined_CHIRON}; the best fit RV amplitude from the CHIRON data is 69.0 km/s and for the SOAR data is 64.7 km/s.

%----------------------------------------------------------------------------------------
%	SECTION
%----------------------------------------------------------------------------------------

\section{DISCOVERIES} \label{section_analysis}

In this section we present the discoveries beginning with the eclipsing binaries and variables. We measure the amplitudes of the variation, and for select targets we use the radial velocity measurement to estimate the companion mass. We show distributions of the periods, amplitudes, and magnitudes of the discoveries and summarize the important statistics of the search. All results are summarized in Tables 8-11.

%----------------------------------------------------------------------------------------
%	SUBSECTION
%----------------------------------------------------------------------------------------

\subsection{Discovery candidates parameter estimations}

\begin{figure}[h!]
\centering
\includegraphics[width=1.0\columnwidth]{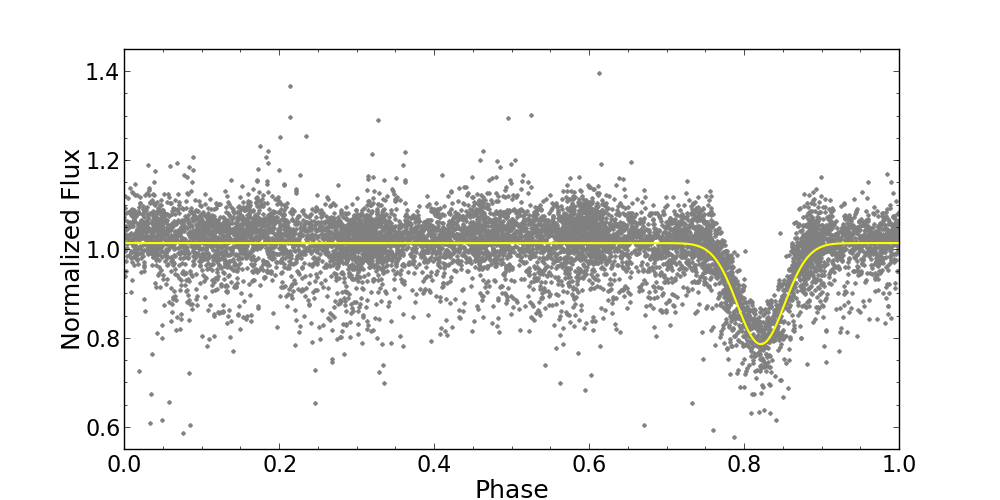}
\includegraphics[width=1.0\columnwidth]{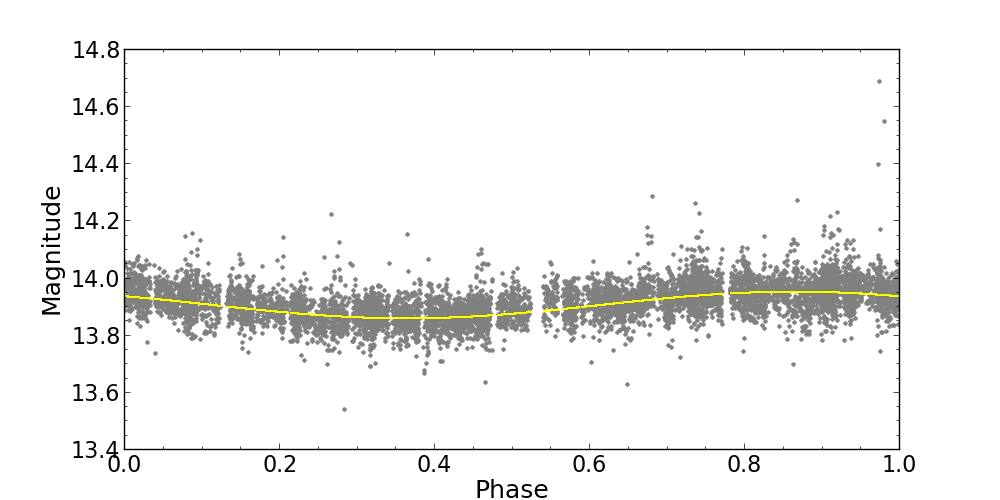}
\caption{Top: Eclipsing binary discovery EVRJ131324.31-792126.3 folded on its 33.7 hour period representative of 100's of Evryscope variable discoveries. Gray points are two minute cadence and yellow is the best Gaussian fit to measure depth. Bottom: variable star discovery EVRJ131228.85-782429.2 folded on its 136.665 hour period representative of 100's of Evryscope variable discoveries. Gray points are two minute cadence and yellow is the best LS fit to measure amplitude.}
\label{fig:best_fits}
\end{figure}

Candidates passing the false positive checks (\S~\ref{false_positive}) are separated by variation type (eclipse-like or sinusoidal variable-like) and measured. The eclipsing binary light curves are folded on the best period and fit with a Gaussian using the approximate phase and depth from the visual inspection plot as the prior. For the variable candidates, we use the best sinusoidal fit from the LS detection. Given the large number of candidates, fitting the light curve amplitude consistently and automatically is key. An additional challenge is the degeneracy due to orbital angle, limb darkening, and orbital eccentricity. We find the Gaussian (for eclipsing binaries) and best sinusoidal fit from the LS detection (for variables) methods to be effective and efficient to measure the variability of the discoveries, while select targets with followup data can be fit with more complicated tools (see \S~\ref{section_SOAR_RV}). Figure \ref{fig:best_fits} shows an example eclipsing binary and variable star fit.\\

%----------------------------------------------------------------------------------------
%	SUBSECTION
%----------------------------------------------------------------------------------------

\subsection{Identification Spectra} \label{section_SOAR_ID_spectra_analysis}

\begin{figure}[h!]
\centering
\includegraphics[width=1.0\columnwidth]{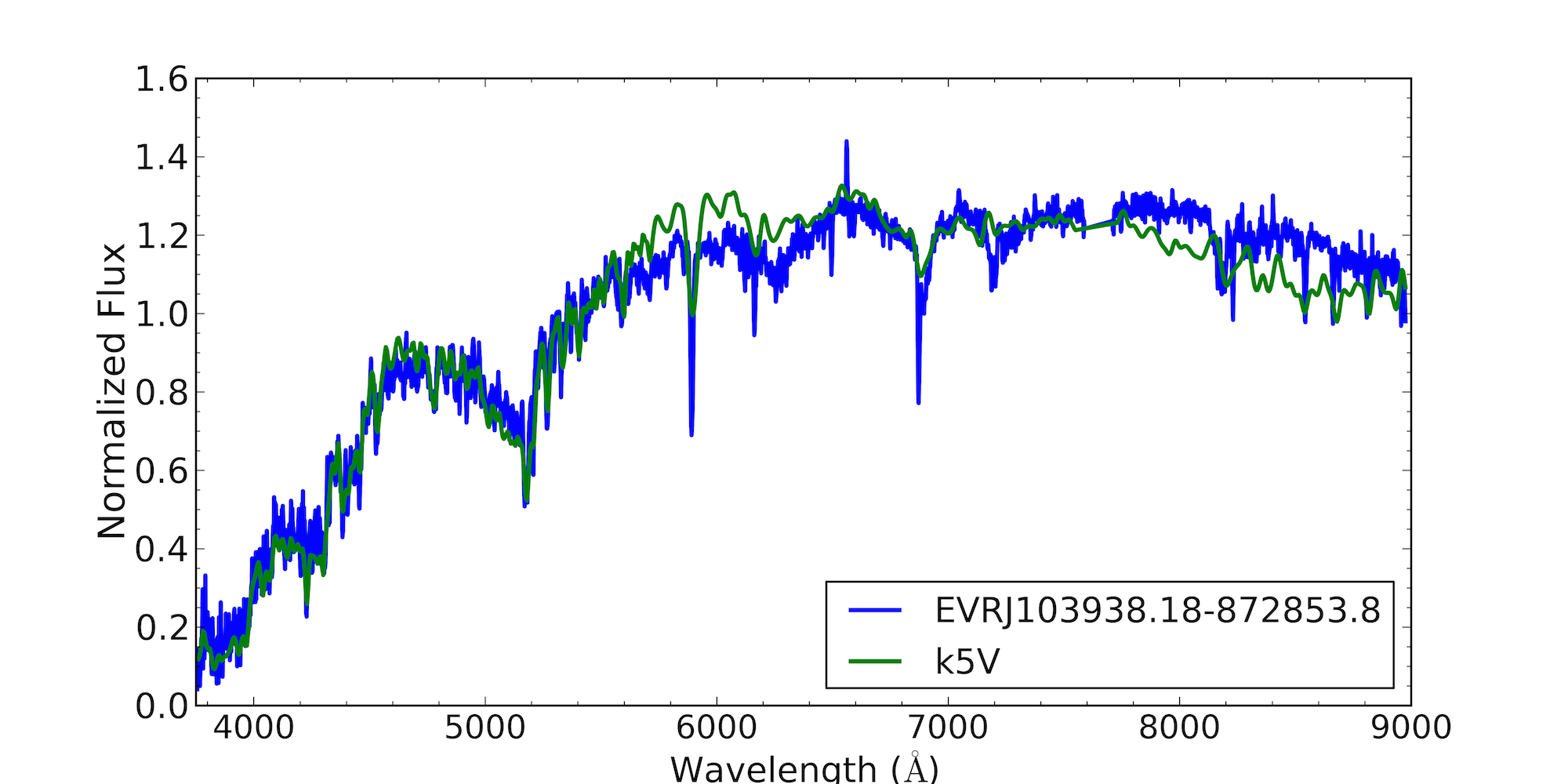}
\caption{An example low mass eclipsing binary discovery (EVRJ103938.18-872853.8) ID spectra taken with the Goodman Spectrograph on the 4.1m SOAR telescope at CTIO, Chile. The green line is a K5V template from the ESO library.}
\label{fig:id_spectra}
\end{figure}

For the discoveries with potential low-mass secondaries, we compare the SOAR ID spectra to ESO template spectra (available at www.eso.org), see Figure \ref{fig:id_spectra}. After finding the closest matching spectra, we compare the results from the color differences classifier described in the previous section. Finally, we use the PyHammer \citep{2007AJ....134.2398C} spectra fitting tool to confirm our fits. PyHammer uses empirical templates of known spectral types and performs a weighted least squares best fit to the input spectra and returns the estimated spectral type. For the the low-mass secondary eclipsing binaries, the results from the three methods are in agreement to within 1-2 in the luminosity class. The spectral types are shown in Table \ref{tab:low_mass_eb}.

%\clearpage

%----------------------------------------------------------------------------------------
%	SUBSECTION
%----------------------------------------------------------------------------------------

\subsection{Radial Velocity - SOAR Data} \label{section_SOAR_RV}

We cross-correlate the SOAR spectra and measure the velocity shift throughout the period found in the Evryscope photometry. Using the color differences in \S~\ref{section_stellar_class}, and the stellar type, radii, and mass profiles from \cite{2013ApJS..208....9P}, we derive functions (using 1-d interpolation) to estimate the primary radius and mass. The secondary radius and mass are then determined using Keplarian/Newtonian calculations described in the following section. For this step, we assume a circular orbit, zero inclination angle, and no limb-darkening. We run a Monte Carlo (MC) simulation to estimate the radius and mass ranges. Due to the simplifying zero-inclination angle assumption and the uncertainty in the SOAR RV measurements, our mass calculations for the secondaries are lower limits. More detailed modeling will be addressed in future work. We discuss our final solutions in \S~\ref{section_lmeb}. The results are listed in Table \ref{tab:low_mass_eb} and plots of the photometric and radial velocity light curves are shown in the appendix.

%----------------------------------------------------------------------------------------
%	SUBSUBSECTION
%----------------------------------------------------------------------------------------

\subsubsection{Secondary mass and radius determination}

\textbf{Photometry:} From visual inspection of the candidate light curves, an initial guess is made for the transit phase and depth and fit with a Gaussian (Figure \ref{fig:soar_rv}). The data is fit with a least squares minimization using scipy to measure the amplitude and phase.\\

\textbf{Radial Velocity:} 
An initial sine curve fit is made using a guess for the amplitude and zero point, while the phase and period are controlled by the transit time and the period found in the photometric light curve (Figure \ref{fig:soar_rv}). The amplitude and zero point are used as inputs to a sine fitting function that uses a sine curve with a fixed phase and period. The function fits the data with a least squares fit; this is the gold line and it returns an RV of 56 km/s for target EVRJ110815.96-870153.8. This assumes a circular orbit and edge on geometry. We leave more detailed analysis with additional variables to future work.

%----------------------------------------------------------------------------------------
%	SUBSUBSECTION
%----------------------------------------------------------------------------------------

\subsubsection{MC best fit of mass and radius}

Using the methods described in the previous section, we perform a Monte Carlo simulation (as described in \cite{Press:2007:NRE:1403886}) to determine the best fit and distribution of the primary and secondary mass and radius.

\begin{figure}[h!]
\centering
\includegraphics[width=1.0\columnwidth]{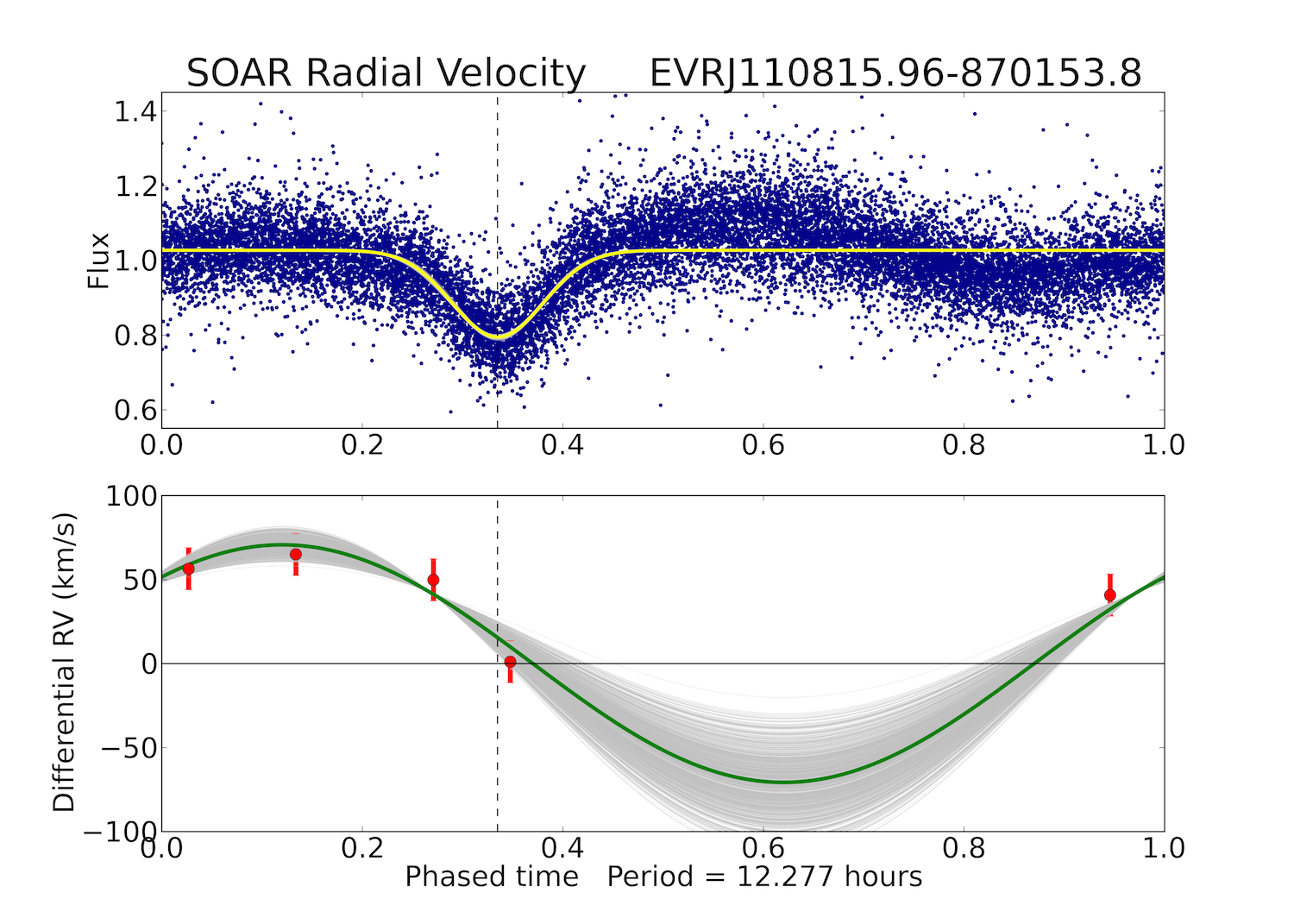}
\caption{EVRJ110815.96-870153.8 K-dwarf eclipsing binary eclipse and radial velocity fit. Top: The best fit (yellow) to the Evryscope photometry using a Gaussian with an initial guess to measure the depth and determine secondary radius. Bottom: The best fit (green) to the SOAR RV data (red points) using a sine curve with an initial guess to measure the velocity and determine the secondary mass. The silver lines are the MC simulation to determine the best fit and error range.}
\label{fig:soar_rv}
\end{figure}

\begin{figure}[h!]
\centering
\includegraphics[width=0.45\columnwidth]{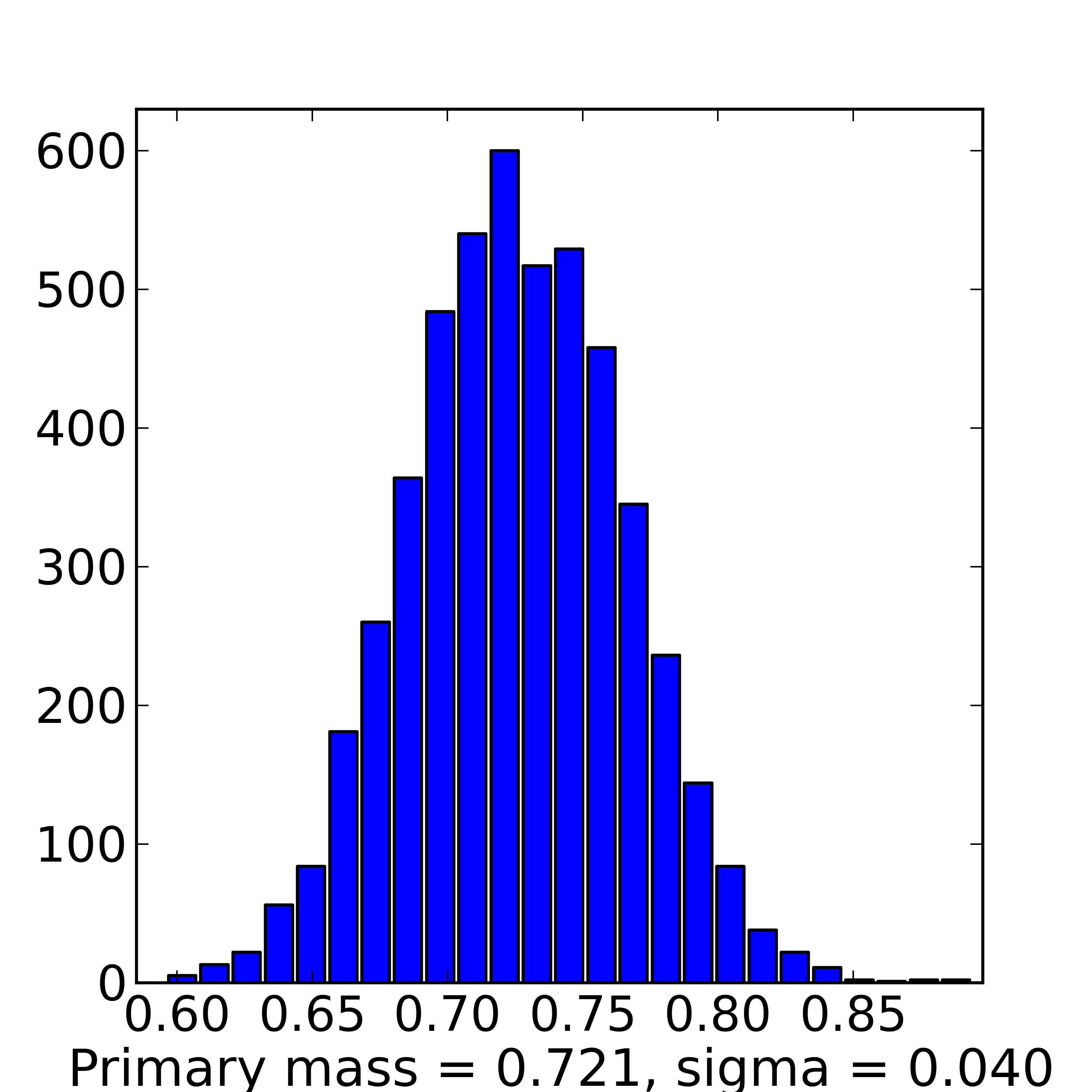}
\includegraphics[width=0.45\columnwidth]{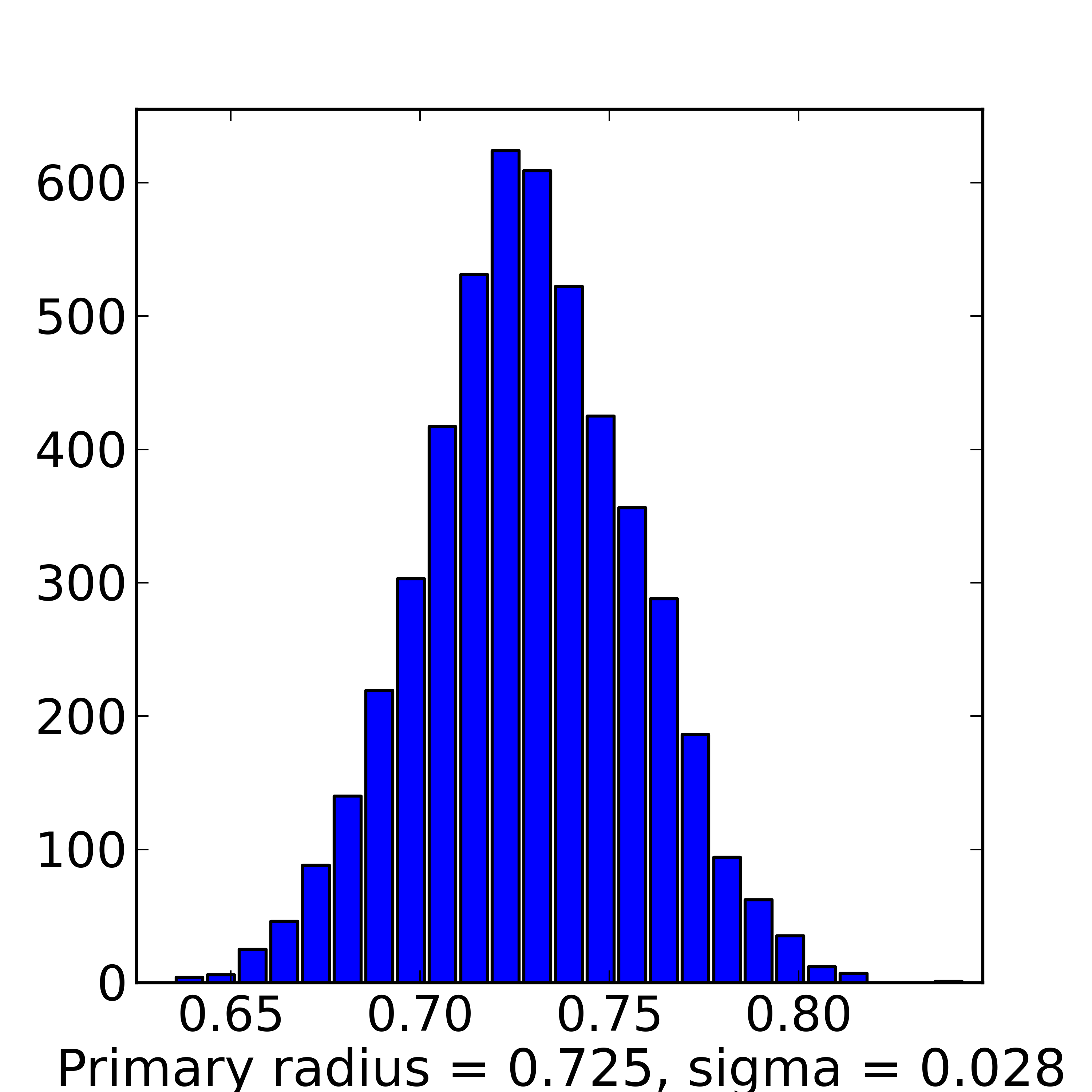}
\includegraphics[width=0.45\columnwidth]{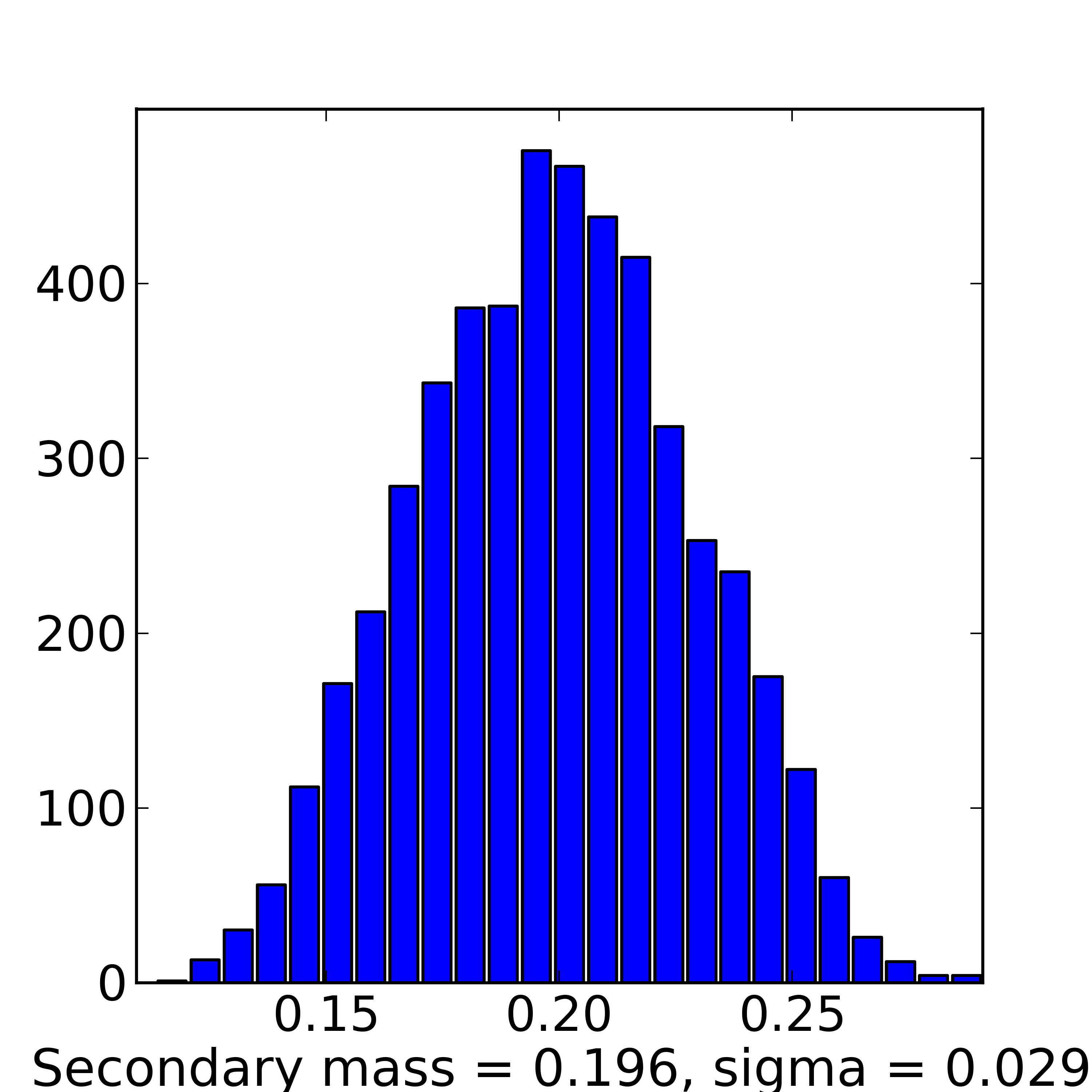}
\includegraphics[width=0.45\columnwidth]{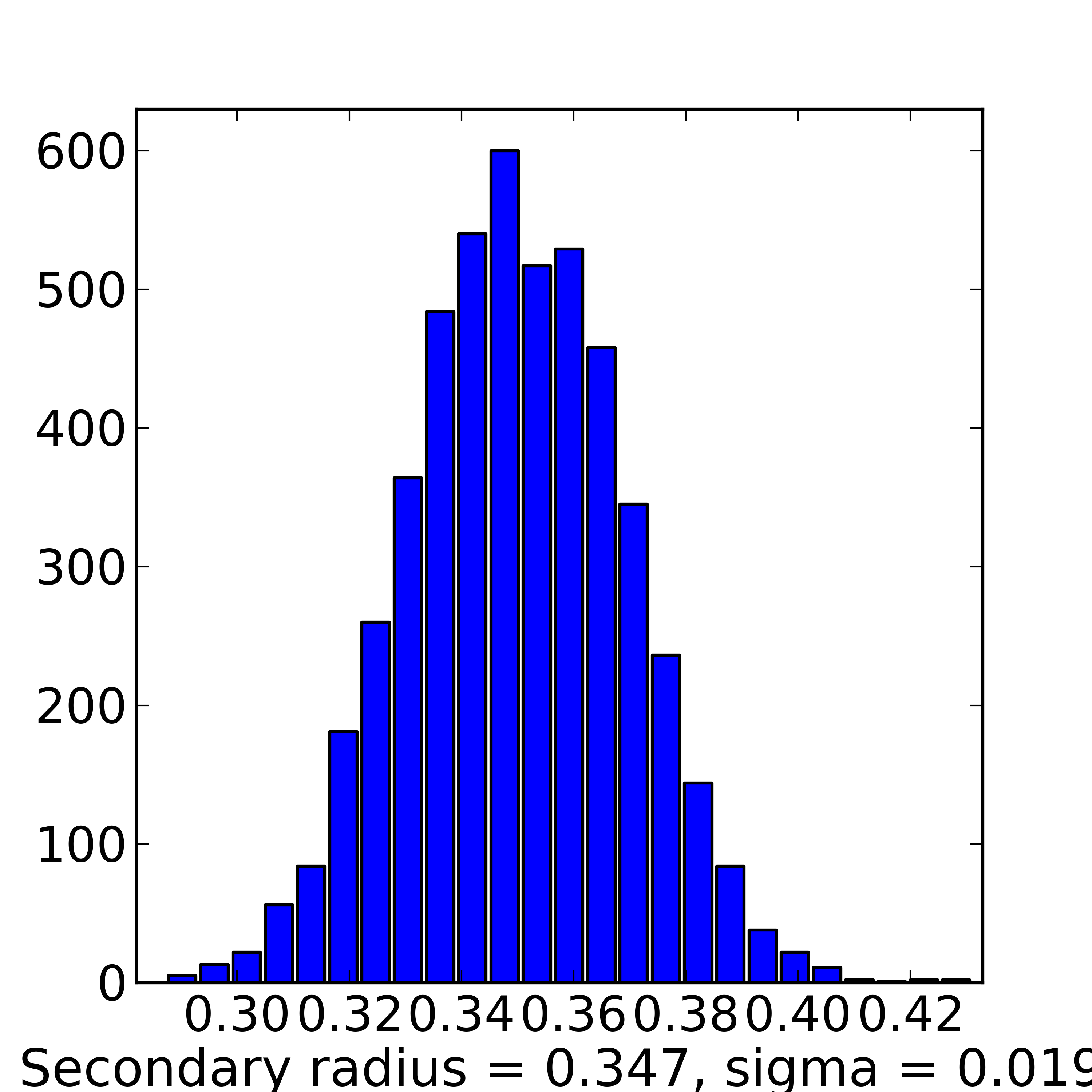}
\caption{Primary and secondary mass and radius determined from our MC simulation. The top panels are the mass and radius of the primary in solar units, the bottom panels are the mass and radius of the secondary. The y-axis is the counts from the MC simulation totaling 5000 trials.}
\label{fig:soar_rv_errors}
\end{figure}

From the Evryscope photometry, we use a bootstrap technique to leverage the very large number of epochs. We randomly choose half of the data points for each iteration with 5000 trials, and fit the data with a least squares minimization for each iteration. We also vary the radius of the primary for each trial by the range in \citep{2013ApJS..208....9P} (spanning +/- 1 in numeric class). From the radial velocity data, we choose a random number in the error bar range of each of the data points (red) and fit the best sine curve (the silver curves) shown in Figure \ref{fig:soar_rv}. We vary the mass of the primary for each trial by the error range in the estimated mass. The propagated results are shown in Figure \ref{fig:soar_rv_errors}.

%----------------------------------------------------------------------------------------
%	SUBSECTION
%----------------------------------------------------------------------------------------

\subsection{Search Statistics}
Sorting by BLS sigma power and choosing only the top candidates greater than 10 sigma narrows the candidates to 5.6\% (9104/163,584) of the filtered list. Visual inspection yields 7.3\% (649/9104) actual variables from the BLS 10 sigma power list. The fraction of all discoveries to all searched is .40\% (649/163,584). The false positive BLS rate is 5.2\% (8455/163,584). Of 649 total variables detected, 346 are known in VSX. The total known periodic variables listed in VSX for the same sky area as the Evryscope Polar Search is 1928, giving a return of 17.9\% (346/1928). There are 1050 known variables in the widest period ranges (3-720 hours) we searched, giving 33.0\% return. There are 858 known variables in the period ranges (3-240 hours) we searched with BLS, giving 40.3\% return. We add 303 new variables or 29\% to the known variables in the region.\\

%----------------------------------------------------------------------------------------
%	SUBSECTION
%----------------------------------------------------------------------------------------

\subsection{Eclipsing Binaries and Variables - Distribution of results}
\begin{figure}[h!]
\centering
\includegraphics[width=.45\columnwidth]{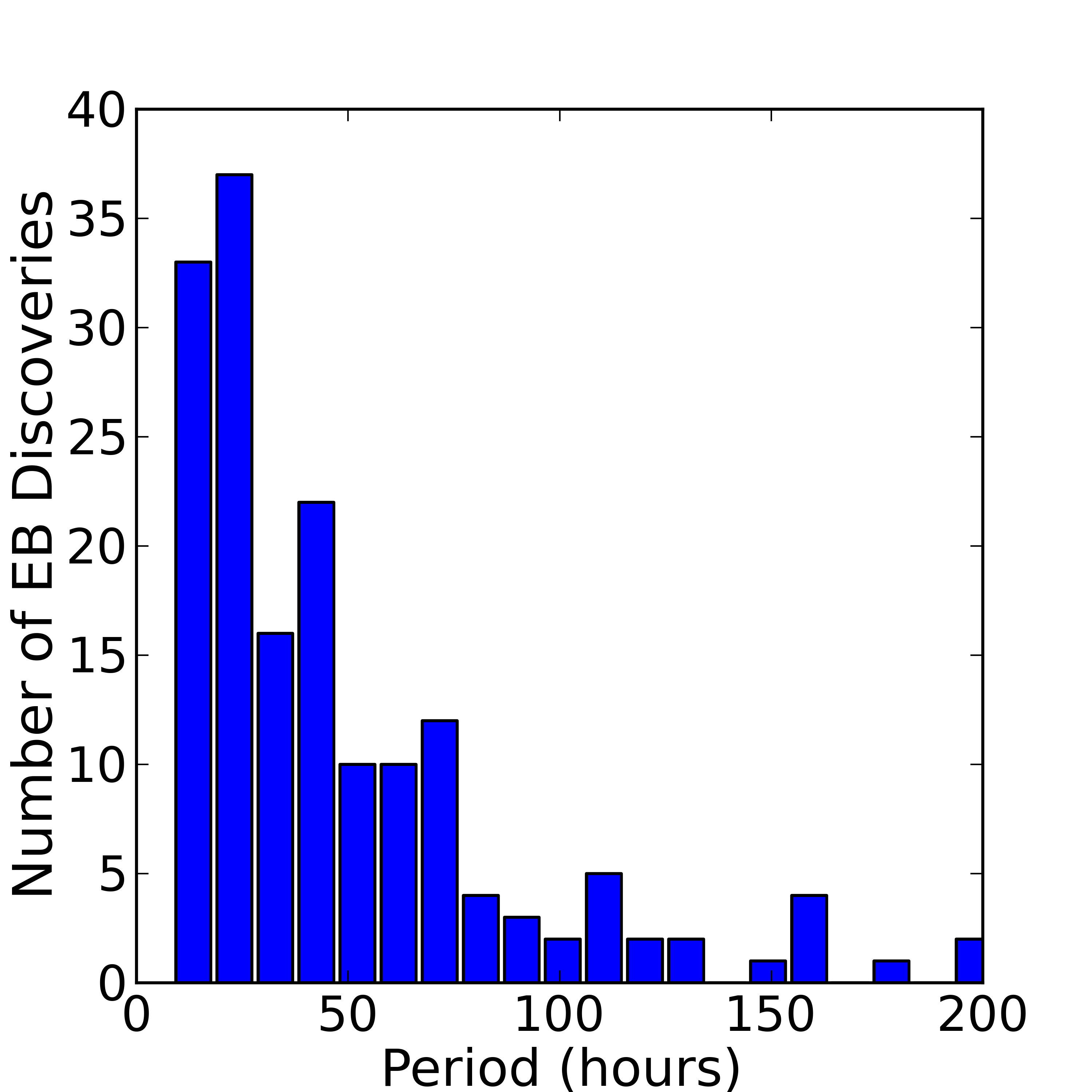}
\includegraphics[width=.45\columnwidth]{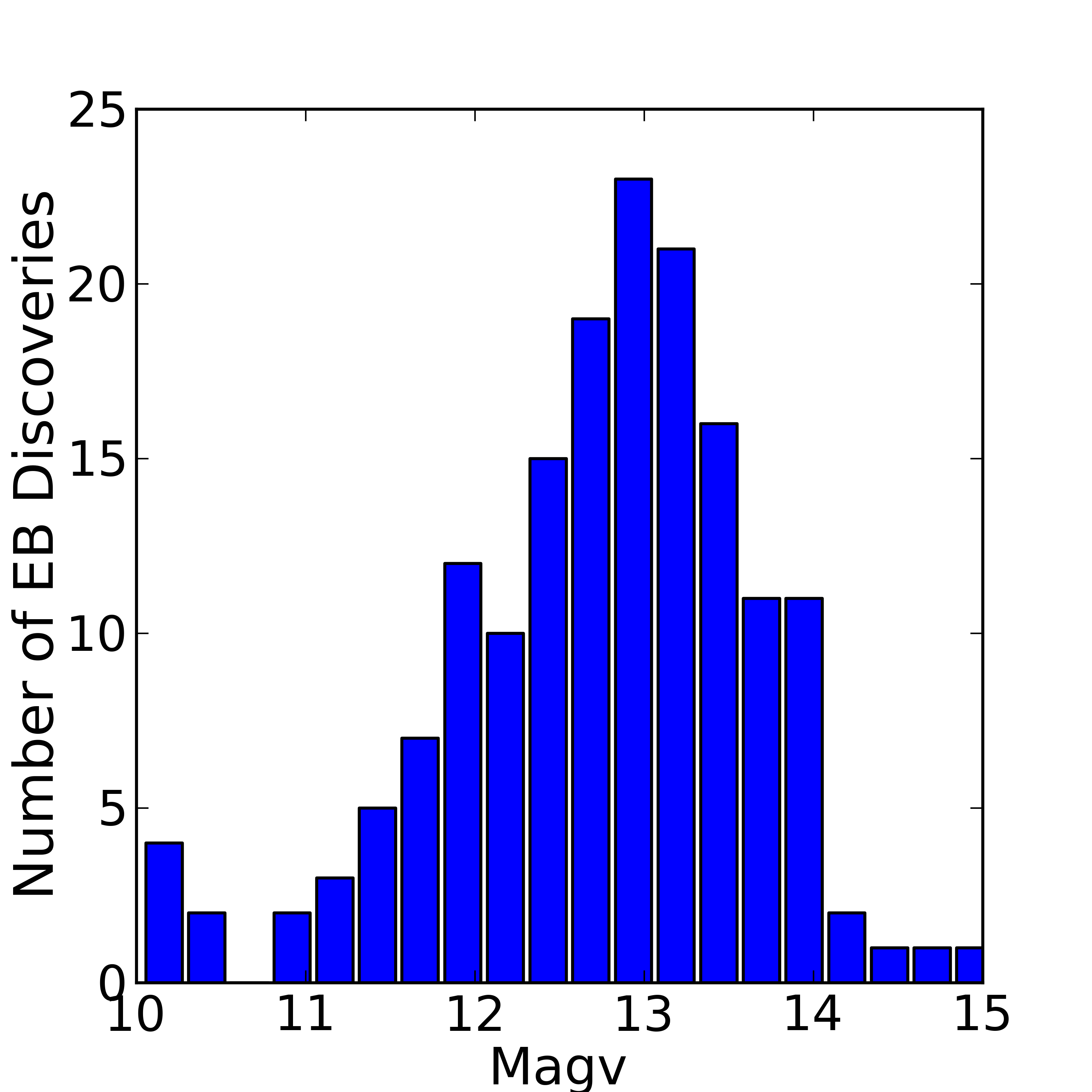}
\includegraphics[width=.45\columnwidth]{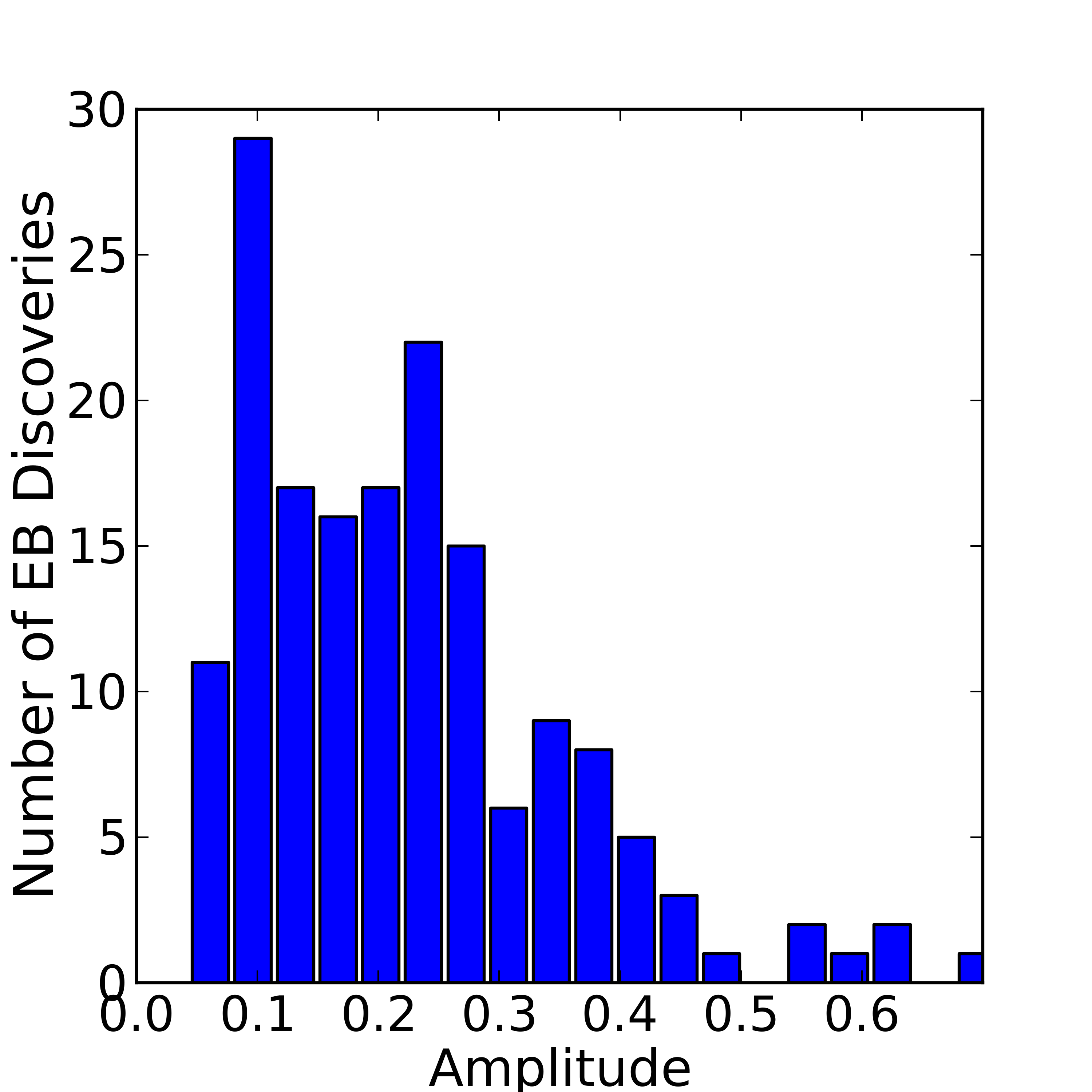}
\caption{Histogram plots summarizing the eclipsing binary discovery results. We are sensitive to periods of several hundred hours and a large fraction of our discoveries are greater than 10\% amplitude.}
\label{fig:discovery_dist}
\end{figure}

\begin{figure}[h!]
\centering
\includegraphics[width=.45\columnwidth]{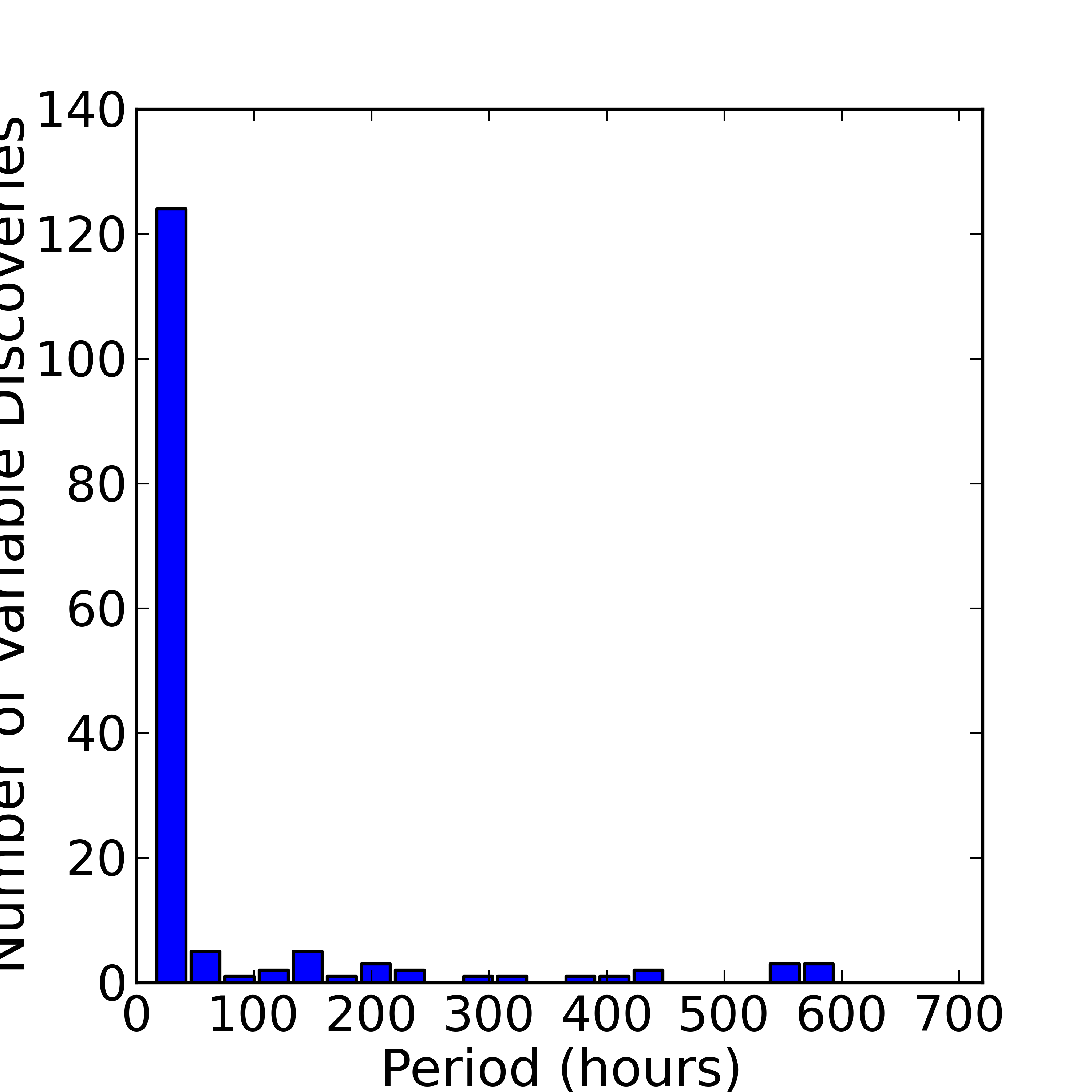}
\includegraphics[width=.45\columnwidth]{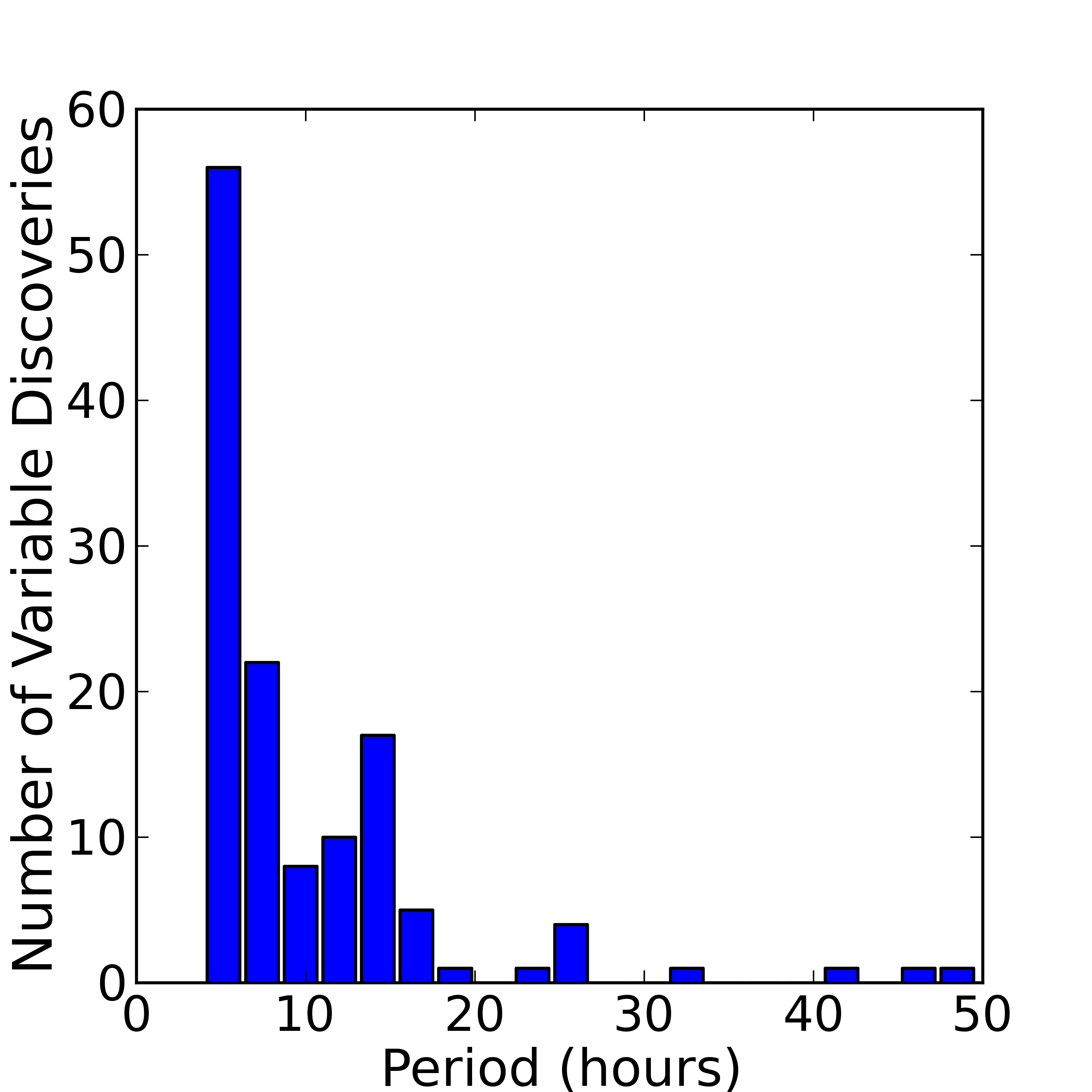}
\includegraphics[width=.45\columnwidth]{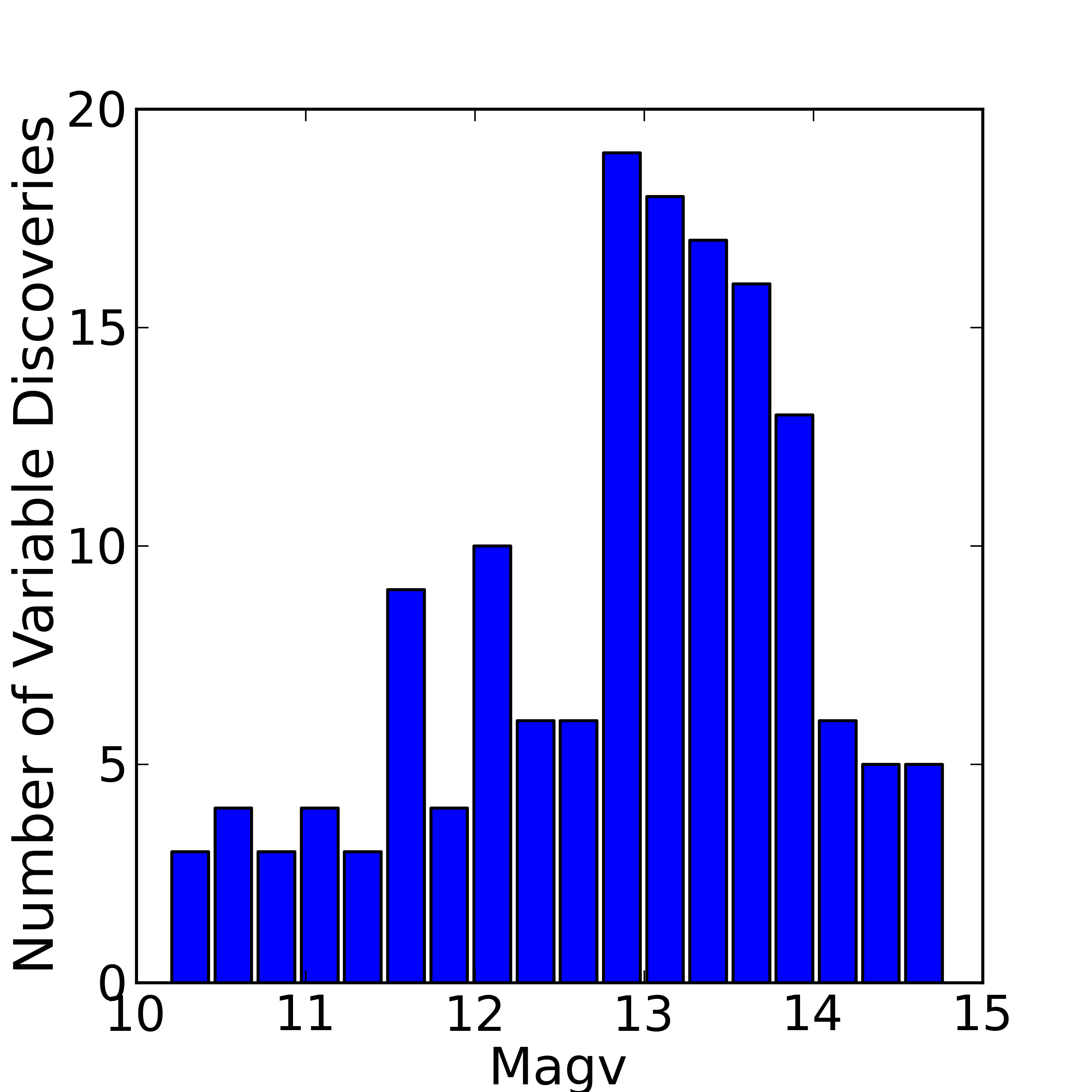}
\includegraphics[width=.45\columnwidth]{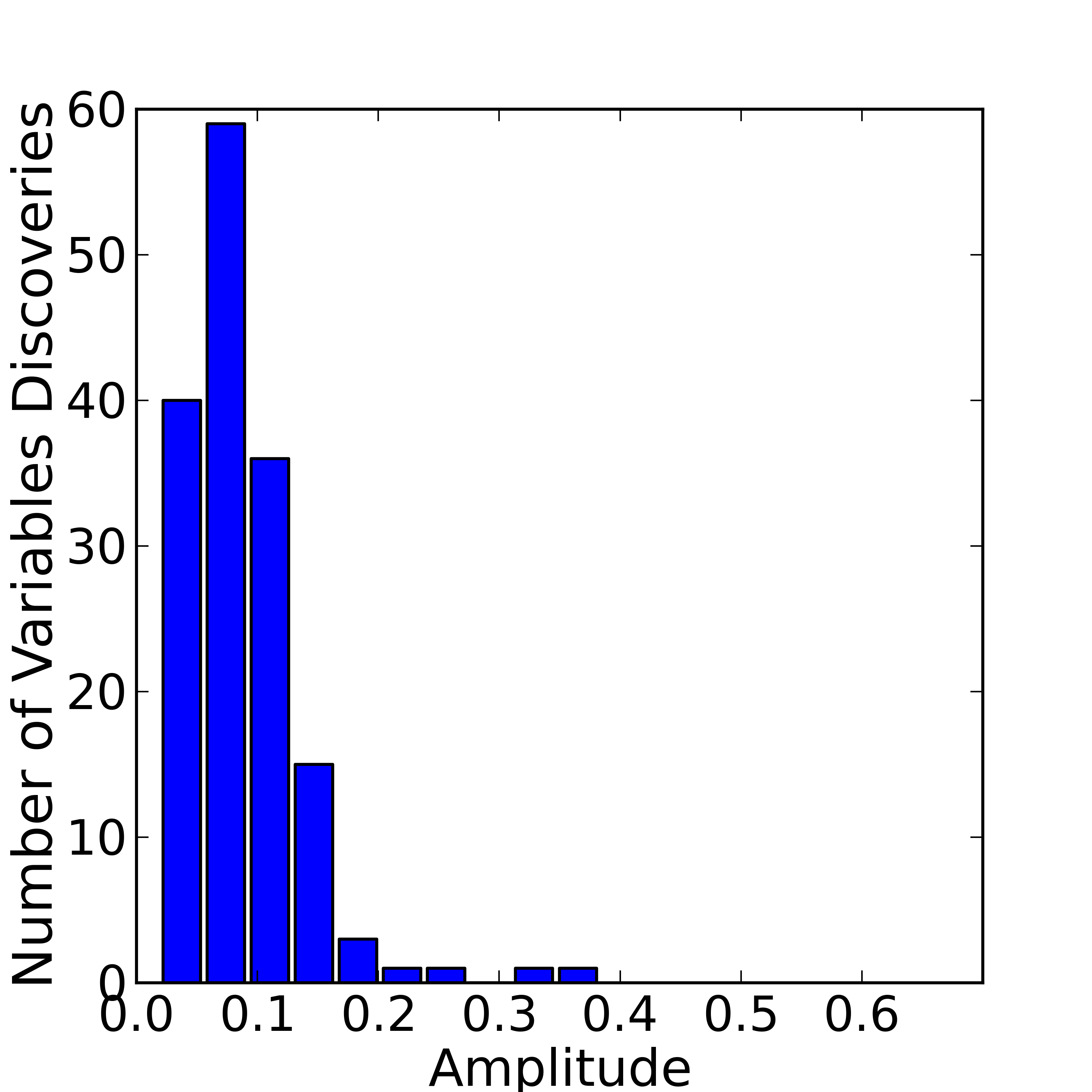}
\caption{Histogram plots summarizing the variable discovery results. A larger fraction of the variable star discoveries are small amplitude and short period.}
\label{fig:discovery_dist_var}
\end{figure}

Histograms of the eclipsing binary discoveries are shown in Figure \ref{fig:discovery_dist}. We discovered a total of 168 eclipsing binaries; most periods found are 75 hours or less, and most amplitudes found are 5-25\%. The results of the variables are shown in Figure \ref{fig:discovery_dist_var}, we found 135 total and most are smaller amplitudes and shorter periods.

%----------------------------------------------------------------------------------------
%	SUBSECTION
%----------------------------------------------------------------------------------------

\subsection{Classification}

\begin{figure}[h!]
\centering
\includegraphics[width=1.0\columnwidth]{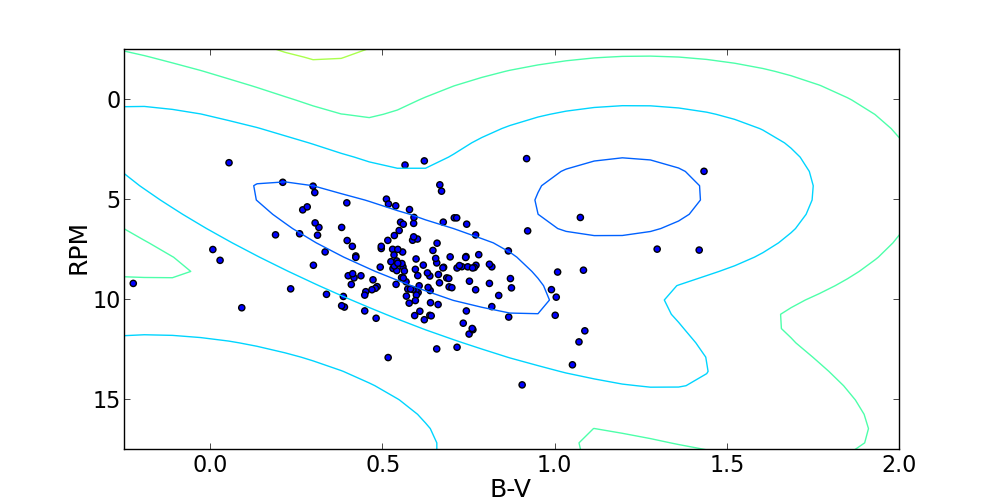}
\includegraphics[width=1.0\columnwidth]{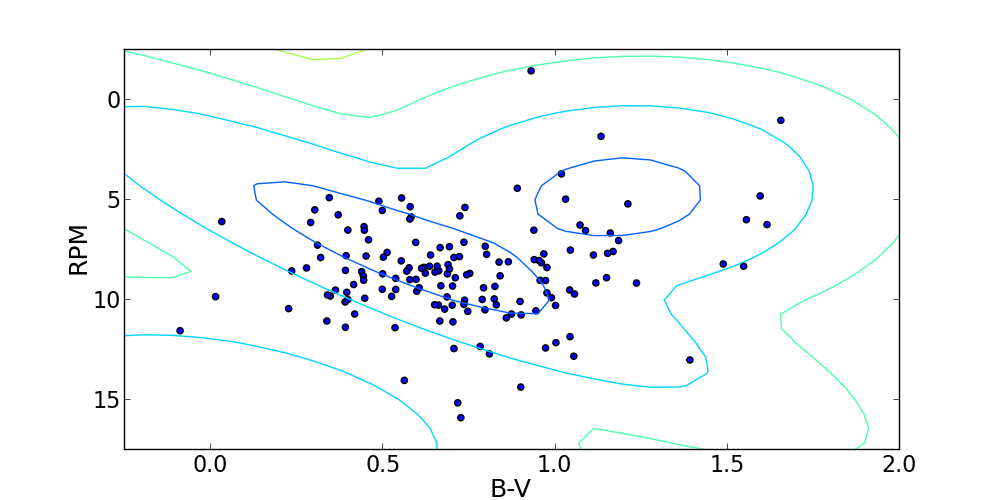}
\caption{Classification results of the eclipsing binary and variable discoveries - Negative log likelihood plot-lines 1, 1.7, 2.8 shown. Top: Eclipsing Binaries. Bottom: Variables.}
\label{fig:discovery_class}
\end{figure}

The discovery classification results are shown in Figure \ref{fig:discovery_class}. We find 267 are main sequence, 34 are giants, and two are not classified. Spectral type G is the most common, with the spectral types shown in Table \ref{tab:class_discovery}. We find more giant variables (24) than giant eclipsers (10) as shown in Figure \ref{fig:discovery_class}. Also shown are the discoveries by star size and spectral type compared to total targets searched (Table \ref{tab:class_discovery_total}).

\begin{table}
\caption{Classification discovery results - spectral type}
\begin{tabular}{ l c c }
Classifier Spectral Type & Number of Discoveries & Percent\\
 \hline
B & 2 & 0.7\\
A & 14 & 4.6\\
F & 89 & 29.4\\
G & 109 & 36.0\\
K & 76 & 25.1\\
M & 11 & 3.6\\
none & 2 & 0.7\\
Total & 303 & 100\\

 \hline
\end{tabular}
\label{tab:class_discovery}
\end{table}

\begin{table}
\caption{Classification discovery results - compared to total searched}
\begin{tabular}{ l c c c}
Classification & Total Searched & Number of Discoveries & Percent\\
 \hline
ms	&	114585	&	267	&	0.23\\
giant	&	40775	&	34	&	0.08\\
HSD	&	335	&	0	&	0.00\\
WD	&	21	&	0	&	0.00\\
O	&	20	&	0	&	0.00\\
B	&	331	&	2	&	0.60\\
A	&	4110	&	14	&	0.34\\
F	&	26102	&	89	&	0.34\\
G	&	49560	&	109	&	0.22\\
K	&	60964	&	76	&	0.12\\
M	&	14629	&	11	&	0.08\\

 \hline
\end{tabular}
\label{tab:class_discovery_total}
\end{table}

%----------------------------------------------------------------------------------------
%	SUBSECTION
%----------------------------------------------------------------------------------------

\subsection{Eclipsing Binaries with low-mass secondaries} \label{section_lmeb}

We identified seven of the eclipsing binary discoveries as hosting potential low-mass secondaries and found that four are less than .25 solar mass. Three of the systems are fully eclipsing binaries (p = 12.3 to 25.9 hours) with dwarf primaries (SpT\textsubscript{p} = G5V, K4V, K5V) and M-dwarf secondaries (mass = .06 - .20 \(M_\odot\)). The other three systems are grazing eclipses with (p = 20.8 to 137.1 hours) with dwarf primaries (SpT\textsubscript{p} = G8V, K2V, K7V), M-dwarf secondaries (mass = .24 - .37 \(M_\odot\)) and minimum radii (r = .20 to .26 \(r_\odot\)). Table \ref{tab:low_mass_eb} presents a list of all low-mass secondary targets. Also included is a likely visual binary EVRJ114225.51-793121.0 (separated in SOAR observations) and EVRJ211905.47-865829.3 a grazing EB with nearly identical primary and secondaries.

%----------------------------------------------------------------------------------------
%	SECTION
%----------------------------------------------------------------------------------------

\section{SUMMARY} \label{section_summary}
The Evryscope was deployed to CTIO in May 2015 and has been operational since that time. We conducted a variability search of the southern polar area using the first 6-months of available data and by selecting the brighter stars (m\textsubscript{v} $<$ 14.5) and limiting the declination range (\ang{-75} to \ang{-90}). We sorted by detection power and visually searched the top 5 \% for variability. We recovered 346 known variables and discovered 303 new variables, including 168 eclipsing binaries six of which we identify as low-mass (.06 - .37 \(M_\odot\)) secondaries with K-dwarf primaries. We encourage the community to followup further on these targets. We measured amplitudes, periods, and variability type and provide a catalog of all discoveries in the Appendix.

This research was supported by the NSF CAREER grant AST-1555175 and the Research Corporation Scialog grants 23782 and 23822. HC is supported by the NSFGRF grant DGE-1144081. BB is supported by the NSF grant AST-1812874. OF and DdS acknowledge support by the Spanish Ministerio de Econom\'ia y Competitividad (MINECO/FEDER, UE) under grants AYA2013-47447-C3-1-P, AYA2016-76012-C3-1-P, MDM-2014-0369 of ICCUB (Unidad de Excelencia 'Mar\'ia de Maeztu'). The Evryscope was constructed under NSF/ATI grant AST-1407589.

%----------------------------------------------------------------------------------------
%	BIBLIOGRAPHY
%----------------------------------------------------------------------------------------

%\newpage
\bibliographystyle{apj}
\bibliography{polar_search}

%----------------------------------------------------------------------------------------
%	APPENDIX
%----------------------------------------------------------------------------------------
\clearpage
\newpage

Note: Columns 1-4 are identification numbers, right ascension and declination, and magnitude. Columns 5-6 are the reduced proper motion (RPM) and color difference (B-V) which we use to estimate the star size and spectral type (see Section 4.2.1). Columns 7-8 are the spectral type from the classifier and from the SOAR ID spectra. Column is 9 the period found in hours, columns 10-17 are the mass of the primary and secondary derived from SOAR radial velocity, along with the one sigma error.

\appendix

\section{\\List of Eclipsing Binary discoveries with low-mass secondaries}
\begin{center}
\begin{longtable*}{l l l l l l l l l l l l l l l l l}
\caption{Eclipsing Binary discoveries with low-mass secondaries}\\
\hline
\noalign{\vskip 3pt}
\text{ESID} & \text{RA} & \text{Dec} & \text{M\textsubscript{v}} & \text{RPM} & \text{B-V} & \text{sptp} & \text{SOAR} & \text{period} & \text{mp} & \text{+/-} & \text{rp} & \text{+/-} &  \text{ms} & \text{+/-} & \text{rs} & \text{+/-} \\ [0.1ex]
\text{} & \text{(J2000)} & \text{(J2000)} & \text{} & \text{} & \text{} & \text{} & \text{} & \text{(hours)} & \text{$M_\odot$} & \text{} & \text{$R_\odot$} & \text{} & \text{$M_\odot$} & \text{} & \text{$R_\odot$} & \text{} \\ [0.1ex]
\hline
\noalign{\vskip 3pt}
\endfirsthead

\hline
\noalign{\vskip 3pt}
\text{ESID} & \text{RA} & \text{Dec} & \text{M\textsubscript{v}} & \text{RPM} & \text{B-V} & \text{sptp} & \text{SOAR} & \text{period} & \text{mp} & \text{+/-} & \text{rp} & \text{+/-} & \text{ms} & \text{+/-} & \text{rs} & \text{+/-} \\ [0.1ex]
\text{} & \text{(J2000)} & \text{(J2000)} & \text{} & \text{} & \text{} & \text{} & \text{} & \text{(hours)} & \text{$M_\odot$} & \text{} & \text{$R_\odot$} & \text{} & \text{$M_\odot$} & \text{} & \text{$R_\odot$} & \text{} \\ [0.1ex]
\hline
\noalign{\vskip 3pt}
\endhead
\endfoot
\hline
\endlastfoot

EVRJ06456.10-823501.0 & 101.2754 & -82.5836 & 11.80 & 8.19 & 1.12 & G9V & G8V & 63.53 & 0.90 & .04 & .876 & .059 & .37 & .02 & .26 & .01\\
EVRJ103938.18-872853.8 & 159.9091 & -87.4816 & 13.78 & 12.09 & 1.07 & K6V & K7V & 20.85 & 0.63 & .04 & .655 & .037 & .24 & .02 & .20 & .01\\
EVRJ110815.96-870153.8 & 167.0665 & -87.0316 & 12.68 & 9.77 & 0.84 & K3V & K4V & 12.28 & 0.72 & .04 & .725 & .028 & .20 & .03 & .35 & .02\\
EVRJ165050.23-843634.6 & 252.7093 & -84.6096 & 13.84 & 11.54 & 1.09 & K4V & K5V & 25.86 & 0.68 & .05 & .699 & .037 & .06 & .01 & .40 & .02\\
EVRJ180826.26-842418.0 & 272.1094 & -84.405 & 13.51 & 9.81 & 0.57 & G6V & G5V & 17.31 & 0.98 & .01 & .983 & .043 & .19 & .01 & .56 & .01\\
EVRJ184114.02-843436.8 & 280.3084 & -84.5769 & 13.34 & 10.19 & 0.91 & K3V & K2V & 137.16 & 0.78 & .07 & .764 & .052 & .28 & .03 & .24 & .02\\

\hline
%\noalign{\vskip 8pt}  
%\end{tabular}
\label{tab:low_mass_eb}
\end{longtable*}
%\end{longtable}
\end{center}

\begin{longtable*}{l l l l l l l l l l}
\caption{Peculiar Eclipsing Binary discoveries}\\
\hline
\noalign{\vskip 3pt}
\text{ESID} & \text{RA(J2000)} & \text{Dec(J2000)} & \text{M\textsubscript{v}} & \text{RPM} & \text{B-V} & \text{sptp} & \text{SOAR} & \text{period} & \text{note} \\ [0.1ex]
\hline
\noalign{\vskip 3pt}
\endfirsthead

\hline
\noalign{\vskip 3pt}
\text{ESID} & \text{RA(J2000)} & \text{Dec(J2000)} & \text{M\textsubscript{v}} & \text{RPM} & \text{B-V} & \text{sptp} & \text{SOAR} & \text{period} & \text{note} \\ [0.1ex]
\hline
\noalign{\vskip 3pt}
\endhead
\endfoot
\hline
\endlastfoot
EVRJ053513.22-774248.2 & 83.8051 & -77.7134 & 12.05 & 9.00 & 1.03 & K1V & G7V & 50.96 & Nearly identical primary and secondary\\
EVRJ114225.51-793121.0 & 175.6063 & -79.5225 & 12.81 & 12.66 & 0.95 & K3V & -- & 91.82 & Visual Double in SOAR image\\
EVRJ211905.47-865829.3 & 319.7728 & -86.9748 & 13.97 & 10.60 & 0.95 & K6V & K5V & 18.61 & Nearly identical primary and secondary\\

\hline
\label{tab:peculiar_eb}
\end{longtable*}

%----------------------------------------------------------------------------------------
%	APPENDIX other discoveries
%----------------------------------------------------------------------------------------
\clearpage
\newpage

\section{\\List of all variable discoveries}

Note: Columns 1-5 are identification numbers, right ascension and declination, and magnitude. Columns 6-9 are the reduced proper motion (RPM) and color difference (B-V) which we use to estimate the star size and spectral type (see Section 4.2.1). Columns 10 and 11 are the period found in hours, and the amplitude of the variability in magnitudes.

\begin{longtable*}{lllllllllll}
\caption{Variable Star discoveries}\\
\hline
\noalign{\vskip 3pt}
\text{ESID} & \text{APASSID} & \text{RA} & \text{Dec} & \text{M\textsubscript{v}} & \text{RPM} & \text{B-V} & \text{size} & \text{spec} & \text{period} & \text{amplitude} \\ [0.1ex]
\text{} & \text{} & \text{(J2000)} & \text{(J2000)} & \text{} & \text{} & \text{} & \text{} & \text{} & \text{(hours)} & \text{($\Delta$ mag)} \\ [0.1ex]
\hline
\noalign{\vskip 3pt}
\endfirsthead

\hline
\noalign{\vskip 3pt}
\text{ESID} & \text{APASSID} & \text{RA} & \text{Dec} & \text{M\textsubscript{v}} & \text{RPM} & \text{B-V} & \text{size} & \text{spec} & \text{period} & \text{amplitude} \\ [0.1ex]
\text{} & \text{} & \text{(J2000)} & \text{(J2000)} & \text{} & \text{} & \text{} & \text{} & \text{} & \text{(hours)} & \text{($\Delta$ mag)} \\ [0.1ex]
\hline
\noalign{\vskip 3pt}
\endhead
\endfoot
\hline
\endlastfoot

EVRJ000411.09-862200.5	&	36032704	&	1.0462	&	-86.3668	&	12.91	&	10.27	&	1.00	&	ms	&	K3V	&	3.7893	&	0.048	\\
EVRJ004033.86-852556.3	&	36034219	&	10.1411	&	-85.4323	&	13.88	&	8.56	&	0.57	&	ms	&	G0V	&	11.9673	&	0.108	\\
EVRJ010354.79-845024.4	&	36034985	&	15.9783	&	-84.8401	&	13.07	&	9.70	&	1.06	&	ms	&	K6V	&	66.0947	&	0.067	\\
EVRJ010428.68-752821.7	&	36397673	&	16.1195	&	-75.4727	&	12.74	&	8.80	&	0.44	&	ms	&	F1V	&	4.4661	&	0.056	\\
EVRJ010628.13-821135.5	&	36109108	&	16.6172	&	-82.1932	&	10.79	&	6.12	&	0.29	&	ms	&	A1V	&	11.5043	&	0.021	\\
EVRJ011610.92-853620.5	&	36033299	&	19.0455	&	-85.6057	&	13.56	&	9.95	&	0.82	&	ms	&	G8V	&	4.0396	&	0.061	\\
EVRJ013112.26-754727.2	&	36147371	&	22.8011	&	-75.7909	&	12.69	&	7.03	&	1.18	&	giant	&	K6	&	565.2369	&	0.057	\\
EVRJ014744.62-750722.8	&	36149487	&	26.9359	&	-75.1230	&	13.43	&	9.75	&	0.34	&	ms	&	F2V	&	7.9793	&	0.022	\\
EVRJ015507.25-842951.4	&	36036929	&	28.7802	&	-84.4976	&	12.03	&	1.82	&	1.13	&	giant	&	K4	&	207.4862	&	0.048	\\
EVRJ020014.59-824041.2	&	36042924	&	30.0608	&	-82.6781	&	13.68	&	11.05	&	0.66	&	ms	&	F3V	&	3.8633	&	0.048	\\
EVRJ020627.84-854259.4	&	36017206	&	31.6160	&	-85.7165	&	13.10	&	5.95	&	0.58	&	ms	&	F1V	&	4.2580	&	0.049	\\
EVRJ021301.18-850613.7	&	36035338	&	33.2549	&	-85.1038	&	14.18	&	8.74	&	0.74	&	ms	&	F9V	&	14.5239	&	0.044	\\
EVRJ022525.78-871212.6	&	36013743	&	36.3574	&	-87.2035	&	13.72	&	5.85	&	0.58	&	ms	&	F9V	&	5.4833	&	0.128	\\
EVRJ023633.12-841813.7	&	36036143	&	39.1380	&	-84.3038	&	12.99	&	8.31	&	1.55	&	giant	&	M	&	3.0339	&	0.026	\\
EVRJ024105.09-852056.8	&	36017369	&	40.2712	&	-85.3491	&	10.84	&	6.65	&	1.16	&	giant	&	M0	&	383.3562	&	0.031	\\
EVRJ031048.98-774610.2	&	36092849	&	47.7041	&	-77.7695	&	13.43	&	9.01	&	0.44	&	ms	&	F1V	&	5.5437	&	0.107	\\
EVRJ031437.25-812625.4	&	36050275	&	48.6552	&	-81.4404	&	11.83	&	10.24	&	0.65	&	ms	&	G6V	&	22.5360	&	0.038	\\
EVRJ032340.58-833811.4	&	36027122	&	50.9191	&	-83.6365	&	11.73	&	5.37	&	0.74	&	giant	&	G8	&	3.6540	&	0.115	\\
EVRJ032442.50-780853.9	&	36092101	&	51.1771	&	-78.1483	&	11.38	&	6.51	&	0.45	&	ms	&	F5V	&	4.6764	&	0.022	\\
EVRJ034027.74-771628.9	&	36093941	&	55.1156	&	-77.2747	&	11.26	&	8.02	&	0.95	&	ms	&	K4V	&	527.5689	&	0.041	\\
EVRJ035648.91-810628.4	&	36049584	&	59.2038	&	-81.1079	&	12.66	&	8.40	&	0.28	&	ms	&	A9V	&	10.7653	&	0.035	\\
EVRJ040145.72-794920.3	&	36086352	&	60.4405	&	-79.8223	&	13.50	&	7.80	&	0.45	&	ms	&	F1V	&	4.3002	&	0.063	\\
EVRJ041253.57-812919.7	&	36066969	&	63.2232	&	-81.4888	&	13.91	&	12.69	&	0.81	&	ms	&	G	&	5.5612	&	0.061	\\
EVRJ041719.75-803105.5	&	36068185	&	64.3323	&	-80.5182	&	11.38	&	6.23	&	1.61	&	giant	&	M	&	583.9396	&	0.066	\\
EVRJ042003.70-803034.9	&	36068172	&	65.0154	&	-80.5097	&	12.35	&	5.20	&	1.21	&	giant	&	M0	&	423.7755	&	0.158	\\
EVRJ042008.90-790539.1	&	36088224	&	65.0371	&	-79.0942	&	13.38	&	8.98	&	0.58	&	ms	&	G2V	&	5.0229	&	0.085	\\
EVRJ044109.58-774556.5	&	36081406	&	70.2899	&	-77.7657	&	10.19	&	5.99	&	1.55	&	giant	&	M	&	426.2952	&	0.046	\\
EVRJ044130.53-761227.0	&	36099120	&	70.3772	&	-76.2075	&	13.13	&	8.14	&	0.96	&	ms	&	K3V	&	3.3209	&	0.065	\\
EVRJ044735.35-775219.9	&	36081082	&	71.8973	&	-77.8722	&	11.30	&	8.69	&	0.50	&	ms	&	G1V	&	3.3276	&	0.026	\\
EVRJ050012.05-811743.4	&	36061636	&	75.0502	&	-81.2954	&	12.48	&	8.09	&	0.86	&	ms	&	G9V	&	3.4901	&	0.076	\\
EVRJ051213.46-761626.0	&	36201329	&	78.0561	&	-76.2739	&	12.99	&	9.39	&	0.60	&	ms	&	G1V	&	6.3173	&	0.074	\\
EVRJ051707.51-851934.3	&	36019388	&	79.2813	&	-85.3262	&	14.56	&	11.54	&	-0.09	&	ms	&	F0	&	17.4600	&	0.138	\\
EVRJ053531.56-792945.6	&	36073410	&	83.8815	&	-79.4960	&	13.45	&	10.57	&	0.75	&	ms	&	K0V	&	4.8621	&	0.061	\\
EVRJ053802.81-775651.0	&	36077655	&	84.5117	&	-77.9475	&	12.60	&	9.47	&	0.50	&	ms	&	G0V	&	6.8030	&	0.048	\\
EVRJ053857.96-840027.0	&	36021952	&	84.7415	&	-84.0075	&	13.18	&	8.45	&	0.69	&	ms	&	K2V	&	3.7923	&	0.120	\\
EVRJ053921.36-823350.0	&	36059589	&	84.8390	&	-82.5639	&	11.37	&	7.67	&	1.15	&	giant	&	K5	&	294.1877	&	0.042	\\
EVRJ055148.46-761552.6	&	36198063	&	87.9519	&	-76.2646	&	11.87	&	4.41	&	0.89	&	giant	&	K4	&	98.4013	&	0.048	\\
EVRJ055157.14-810831.9	&	36062821	&	87.9881	&	-81.1422	&	12.74	&	9.85	&	0.69	&	ms	&	F5V	&	4.9623	&	0.208	\\
EVRJ055856.16-785946.3	&	36075508	&	89.7340	&	-78.9962	&	13.06	&	9.65	&	0.98	&	ms	&	K4V	&	3.1286	&	0.078	\\
EVRJ060627.19-841156.4	&	37958830	&	91.6133	&	-84.1990	&	14.67	&	9.83	&	0.01	&	ms	&	F0	&	6.8412	&	0.142	\\
EVRJ061823.95-780812.1	&	38123485	&	94.5998	&	-78.1367	&	10.11	&	1.02	&	1.65	&	giant	&	M	&	544.3723	&	0.064	\\
EVRJ062714.64-794039.0	&	38117589	&	96.8110	&	-79.6775	&	13.19	&	8.31	&	0.66	&	ms	&	G2V	&	5.1073	&	0.080	\\
EVRJ064843.42-793349.3	&	38116489	&	102.1809	&	-79.5637	&	13.01	&	9.15	&	1.24	&	giant	&	M1	&	534.1532	&	0.095	\\
EVRJ070444.11-752812.7	&	39299363	&	106.1838	&	-75.4702	&	13.39	&	8.69	&	0.69	&	ms	&	G6V	&	4.9315	&	0.098	\\
EVRJ070553.98-813347.5	&	38085679	&	106.4749	&	-81.5632	&	13.28	&	12.32	&	0.78	&	ms	&	G	&	208.1137	&	0.080	\\
EVRJ070751.77-861600.8	&	37950806	&	106.9657	&	-86.2669	&	13.88	&	10.26	&	0.70	&	ms	&	G6V	&	4.4384	&	0.137	\\
EVRJ071040.15-854213.0	&	37951603	&	107.6673	&	-85.7036	&	13.28	&	10.46	&	0.68	&	ms	&	G7V	&	14.0116	&	0.042	\\
EVRJ071054.50-775214.5	&	38120922	&	107.7271	&	-77.8707	&	12.75	&	8.38	&	0.98	&	ms	&	K6V	&	119.7744	&	0.061	\\
EVRJ072150.93-814705.3	&	38084504	&	110.4622	&	-81.7848	&	12.48	&	3.69	&	1.02	&	giant	&	K4	&	569.9960	&	0.073	\\
EVRJ072411.33-865020.0	&	37932897	&	111.0472	&	-86.8389	&	12.06	&	7.72	&	0.80	&	ms	&	G8V	&	3.8106	&	0.036	\\
EVRJ074325.18-780127.5	&	38108272	&	115.8549	&	-78.0243	&	11.58	&	7.86	&	0.50	&	ms	&	F7V	&	3.3874	&	0.029	\\
EVRJ080401.49-824005.9	&	37961948	&	121.0062	&	-82.6683	&	11.55	&	7.70	&	0.97	&	giant	&	K6	&	4.2669	&	0.028	\\
EVRJ083403.17-811922.4	&	37975768	&	128.5132	&	-81.3229	&	13.54	&	-1.46	&	0.93	&	giant	&	K4	&	5.1269	&	0.047	\\
EVRJ084757.72-781627.1	&	38100581	&	131.9905	&	-78.2742	&	13.17	&	12.13	&	1.00	&	ms	&	K4V	&	266.6131	&	0.064	\\
EVRJ085910.42-813844.9	&	37967866	&	134.7934	&	-81.6458	&	13.69	&	9.23	&	0.41	&	ms	&	A4V	&	6.5406	&	0.073	\\
EVRJ090816.25-840058.0	&	37945248	&	137.0677	&	-84.0161	&	13.73	&	9.56	&	0.60	&	ms	&	G1V	&	4.1451	&	0.159	\\
EVRJ092130.62-780552.1	&	38047442	&	140.3776	&	-78.0978	&	12.06	&	7.58	&	1.17	&	giant	&	K8	&	14.2823	&	0.036	\\
EVRJ092934.58-883002.5	&	37925391	&	142.3941	&	-88.5007	&	12.70	&	11.83	&	1.04	&	ms	&	K4V	&	14.9276	&	0.072	\\
EVRJ094546.82-845901.7	&	37943780	&	146.4451	&	-84.9838	&	11.40	&	4.88	&	0.34	&	ms	&	F4V	&	10.5364	&	0.023	\\
EVRJ094914.47-765810.9	&	38050263	&	147.3103	&	-76.9697	&	11.94	&	7.38	&	0.67	&	ms	&	G1V	&	4.7866	&	0.047	\\
EVRJ095103.38-775148.6	&	38042145	&	147.7641	&	-77.8635	&	12.64	&	8.52	&	0.39	&	ms	&	A9V	&	3.3413	&	0.052	\\
EVRJ103805.95-823919.8	&	37947694	&	159.5248	&	-82.6555	&	12.25	&	9.15	&	1.12	&	ms	&	K5V	&	188.7851	&	0.037	\\
EVRJ103843.68-841342.6	&	37941994	&	159.6820	&	-84.2285	&	14.34	&	9.32	&	0.82	&	ms	&	K1V	&	3.4514	&	0.095	\\
EVRJ104338.88-812945.6	&	37997111	&	160.9120	&	-81.4960	&	13.10	&	12.81	&	1.05	&	ms	&	K	&	9.2306	&	0.071	\\
EVRJ104928.30-840709.1	&	37941961	&	162.3679	&	-84.1192	&	14.41	&	10.25	&	0.66	&	ms	&	G5V	&	4.3652	&	0.071	\\
EVRJ110736.10-801214.8	&	37995054	&	166.9004	&	-80.2041	&	13.20	&	8.60	&	0.65	&	ms	&	G6V	&	5.2520	&	0.103	\\
EVRJ112926.09-790251.7	&	38014623	&	172.3587	&	-79.0477	&	12.82	&	10.01	&	0.74	&	ms	&	K3V	&	3.2616	&	0.032	\\
EVRJ113221.60-773934.2	&	38021372	&	173.0900	&	-77.6595	&	12.62	&	7.52	&	0.64	&	ms	&	F6V	&	2.2946	&	0.089	\\
EVRJ113648.94-770820.8	&	38274528	&	174.2039	&	-77.1391	&	13.07	&	10.70	&	0.87	&	ms	&	K2V	&	40.9161	&	0.049	\\
EVRJ113950.33-823313.7	&	37986809	&	174.9597	&	-82.5538	&	12.70	&	8.23	&	0.69	&	ms	&	G9V	&	12.2413	&	0.045	\\
EVRJ121247.35-782517.8	&	47439342	&	183.1973	&	-78.4216	&	12.77	&	8.67	&	0.75	&	ms	&	G8V	&	7.1346	&	0.071	\\
EVRJ124042.22-852021.1	&	47199888	&	190.1759	&	-85.3392	&	12.33	&	7.11	&	0.73	&	ms	&	K1V	&	51.6660	&	0.093	\\
EVRJ124521.96-772053.5	&	47445563	&	191.3415	&	-77.3482	&	13.16	&	8.11	&	0.84	&	ms	&	G8V	&	5.0777	&	0.089	\\
EVRJ124614.88-851715.4	&	47199917	&	191.5620	&	-85.2876	&	13.75	&	12.43	&	0.71	&	ms	&	G8V	&	3.6741	&	0.109	\\
EVRJ124711.98-784313.8	&	--	&	191.7999	&	-78.7205	&	--	&	--	&	--	&	--	&	--	&	4.0535	&	0.100	\\
EVRJ125757.58-773925.9	&	47442656	&	194.4899	&	-77.6572	&	12.88	&	9.03	&	0.97	&	ms	&	K2V	&	13.6397	&	0.065	\\
EVRJ130148.17-831417.9	&	47209493	&	195.4507	&	-83.2383	&	12.92	&	8.20	&	1.49	&	giant	&	M3	&	121.8445	&	0.068	\\
EVRJ130916.78-775637.7	&	47433337	&	197.3199	&	-77.9438	&	13.61	&	9.88	&	0.99	&	ms	&	K4V	&	7.8242	&	0.153	\\
EVRJ131216.99-803701.6	&	47370231	&	198.0708	&	-80.6171	&	11.08	&	5.79	&	0.72	&	ms	&	G9V	&	7.9778	&	0.008	\\
EVRJ131228.85-782429.2	&	--	&	198.1202	&	-78.4081	&	--	&	--	&	--	&	--	&	--	&	136.7678	&	0.047	\\
EVRJ131248.91-794104.9	&	47374655	&	198.2038	&	-79.6847	&	10.48	&	7.50	&	1.04	&	giant	&	K4	&	5.5061	&	0.015	\\
EVRJ134036.82-810805.3	&	47361846	&	205.1534	&	-81.1348	&	13.08	&	10.74	&	0.90	&	ms	&	K0V	&	3.4605	&	0.038	\\
EVRJ134909.43-795116.2	&	47384377	&	207.2893	&	-79.8545	&	12.69	&	9.02	&	0.96	&	ms	&	K1V	&	4.4504	&	0.032	\\
EVRJ135636.96-852904.9	&	47198519	&	209.1540	&	-85.4847	&	13.04	&	9.29	&	0.67	&	ms	&	G8V	&	5.5375	&	0.039	\\
EVRJ135847.35-753616.9	&	47466750	&	209.6973	&	-75.6047	&	12.64	&	4.90	&	0.55	&	ms	&	F8V	&	13.0043	&	0.081	\\
EVRJ135948.38-773732.9	&	47408549	&	209.9516	&	-77.6258	&	12.25	&	7.12	&	0.59	&	ms	&	G0V	&	6.4798	&	0.063	\\
EVRJ142647.47-774451.4	&	47396223	&	216.6978	&	-77.7476	&	13.36	&	11.36	&	0.39	&	ms	&	F0	&	3.0402	&	0.110	\\
EVRJ143601.82-832825.3	&	47204308	&	219.0076	&	-83.4737	&	13.62	&	9.82	&	0.52	&	ms	&	F2V	&	6.8145	&	0.117	\\
EVRJ150519.75-753054.0	&	47509935	&	226.3323	&	-75.5150	&	10.53	&	4.96	&	1.03	&	giant	&	K3	&	6.7618	&	0.036	\\
EVRJ151320.26-833642.1	&	47215664	&	228.3344	&	-83.6117	&	11.52	&	7.98	&	0.94	&	ms	&	K3V	&	202.2009	&	0.027	\\
EVRJ152213.03-850853.2	&	47177228	&	230.5543	&	-85.1481	&	12.59	&	7.32	&	0.80	&	ms	&	K0V	&	44.3809	&	0.081	\\
EVRJ152422.85-765355.7	&	47316062	&	231.0952	&	-76.8988	&	10.58	&	5.74	&	0.37	&	ms	&	A7V	&	3.3395	&	0.042	\\
EVRJ154618.22-824405.3	&	47194254	&	236.5759	&	-82.7348	&	13.77	&	11.09	&	0.70	&	ms	&	F6V	&	4.5592	&	0.091	\\
EVRJ155137.18-824810.8	&	47194265	&	237.9049	&	-82.8030	&	12.70	&	8.58	&	0.44	&	ms	&	F5V	&	8.0204	&	0.052	\\
EVRJ155320.98-824654.8	&	47193653	&	238.3374	&	-82.7819	&	12.98	&	5.06	&	0.49	&	ms	&	F8V	&	10.0283	&	0.044	\\
EVRJ155645.07-802819.2	&	47288797	&	239.1878	&	-80.4720	&	11.75	&	6.08	&	0.03	&	ms	&	B9V	&	116.9858	&	0.032	\\
EVRJ155942.19-824514.8	&	47193923	&	239.9258	&	-82.7541	&	12.12	&	5.33	&	0.58	&	ms	&	G0V	&	14.6855	&	0.038	\\
EVRJ162259.47-805820.3	&	47254401	&	245.7478	&	-80.9723	&	11.73	&	8.04	&	0.55	&	ms	&	F1V	&	4.8019	&	0.048	\\
EVRJ162554.07-854342.2	&	47181390	&	246.4753	&	-85.7284	&	13.37	&	11.38	&	0.53	&	ms	&	F5V	&	5.0865	&	0.083	\\
EVRJ163031.56-834849.7	&	47191010	&	247.6315	&	-83.8138	&	13.35	&	8.31	&	0.63	&	ms	&	G0V	&	12.2352	&	0.089	\\
EVRJ163252.27-832748.2	&	47191300	&	248.2178	&	-83.4634	&	12.51	&	6.51	&	0.94	&	giant	&	K6	&	150.4504	&	0.092	\\
EVRJ163852.99-843744.0	&	47182121	&	249.7208	&	-84.6289	&	13.21	&	10.49	&	0.80	&	ms	&	K2V	&	32.2767	&	0.048	\\
EVRJ165216.22-825416.6	&	47186184	&	253.0676	&	-82.9046	&	13.94	&	9.91	&	0.45	&	ms	&	F8V	&	15.1777	&	0.367	\\
EVRJ165457.58-775615.7	&	47277017	&	253.7399	&	-77.9377	&	12.87	&	5.49	&	0.30	&	ms	&	F1V	&	7.4973	&	0.069	\\
EVRJ171344.76-825649.9	&	47243917	&	258.4365	&	-82.9472	&	13.44	&	7.62	&	0.51	&	ms	&	F3V	&	5.2027	&	0.075	\\
EVRJ171929.86-800819.0	&	47250893	&	259.8744	&	-80.1386	&	12.85	&	8.54	&	0.66	&	ms	&	F3V	&	4.5325	&	0.071	\\
EVRJ171939.84-852016.4	&	47180802	&	259.9160	&	-85.3379	&	14.67	&	13.00	&	1.39	&	ms	&	K6V	&	3.2852	&	0.072	\\
EVRJ172044.45-811413.2	&	47249474	&	260.1852	&	-81.2370	&	10.80	&	7.87	&	0.71	&	ms	&	G1V	&	12.2459	&	0.019	\\
EVRJ173256.64-833249.9	&	47242117	&	263.2360	&	-83.5472	&	14.11	&	10.70	&	0.42	&	ms	&	F8V	&	7.1155	&	0.135	\\
EVRJ173854.53-825617.2	&	47243150	&	264.7272	&	-82.9381	&	13.40	&	11.05	&	0.34	&	ms	&	F3	&	6.2448	&	0.087	\\
EVRJ174904.92-853647.9	&	47180335	&	267.2705	&	-85.6133	&	13.89	&	7.33	&	0.69	&	ms	&	G1V	&	4.3006	&	0.093	\\
EVRJ175053.26-794941.5	&	47264410	&	267.7219	&	-79.8282	&	11.96	&	8.79	&	0.84	&	ms	&	K0V	&	134.4649	&	0.051	\\
EVRJ175230.22-785048.8	&	47268452	&	268.1259	&	-78.8469	&	12.01	&	7.88	&	0.32	&	ms	&	F1V	&	8.3917	&	0.041	\\
EVRJ175340.68-753831.9	&	47710653	&	268.4195	&	-75.6422	&	12.35	&	7.83	&	0.72	&	ms	&	G8V	&	5.0187	&	0.089	\\
EVRJ175347.33-854135.5	&	47180309	&	268.4472	&	-85.6932	&	12.81	&	8.92	&	0.54	&	ms	&	F9V	&	4.4685	&	0.051	\\
EVRJ180742.10-824651.6	&	57161419	&	271.9254	&	-82.7810	&	14.02	&	10.10	&	0.39	&	ms	&	F0V	&	7.0683	&	0.224	\\
EVRJ181807.06-800525.4	&	57185282	&	274.5294	&	-80.0904	&	11.04	&	4.80	&	1.59	&	giant	&	M	&	370.8165	&	0.025	\\
EVRJ182044.62-754759.6	&	57415585	&	275.1859	&	-75.7999	&	11.50	&	5.52	&	0.50	&	ms	&	F7V	&	3.4102	&	0.026	\\
EVRJ182608.74-864925.3	&	57090707	&	276.5364	&	-86.8237	&	14.41	&	10.07	&	0.90	&	ms	&	K2V	&	4.2908	&	0.130	\\
EVRJ183802.93-811827.4	&	57166241	&	279.5122	&	-81.3076	&	10.90	&	10.24	&	0.83	&	ms	&	K1V	&	48.6808	&	0.051	\\
EVRJ192029.45-860534.4	&	57091324	&	290.1227	&	-86.0929	&	13.27	&	9.80	&	0.35	&	ms	&	A7V	&	6.0642	&	0.025	\\
EVRJ195449.01-840216.4	&	57097667	&	298.7042	&	-84.0379	&	12.73	&	10.22	&	0.73	&	ms	&	G9V	&	5.5082	&	0.063	\\
EVRJ200551.67-820620.5	&	57105858	&	301.4653	&	-82.1057	&	13.63	&	8.88	&	0.71	&	ms	&	G9V	&	5.5203	&	0.087	\\
EVRJ203404.49-871549.0	&	57072262	&	308.5187	&	-87.2636	&	13.87	&	9.50	&	0.36	&	ms	&	F1V	&	4.8050	&	0.095	\\
EVRJ204931.97-845034.4	&	57076995	&	312.3832	&	-84.8429	&	14.21	&	15.14	&	0.72	&	ms	&	G	&	4.4090	&	0.078	\\
EVRJ205037.49-774637.2	&	57175393	&	312.6562	&	-77.7770	&	11.00	&	6.99	&	0.46	&	ms	&	F8V	&	3.3040	&	0.023	\\
EVRJ205225.92-855742.5	&	57074554	&	313.1080	&	-85.9618	&	14.07	&	12.40	&	0.97	&	ms	&	K3V	&	133.5587	&	0.062	\\
EVRJ205802.14-793349.7	&	57118550	&	314.5089	&	-79.5638	&	12.97	&	8.41	&	0.61	&	ms	&	F7V	&	6.2271	&	0.048	\\
EVRJ210729.26-763906.5	&	57215979	&	316.8719	&	-76.6518	&	13.58	&	10.43	&	0.23	&	ms	&	F3	&	4.7894	&	0.024	\\
EVRJ210937.03-785828.2	&	57118801	&	317.4043	&	-78.9745	&	12.99	&	6.25	&	1.07	&	giant	&	K4	&	180.8028	&	0.051	\\
EVRJ213403.58-865953.5	&	57070601	&	323.5149	&	-86.9982	&	12.69	&	8.97	&	0.60	&	ms	&	F9V	&	4.7182	&	0.082	\\
EVRJ215744.06-790828.7	&	57141740	&	329.4336	&	-79.1413	&	13.51	&	7.75	&	1.11	&	giant	&	K5	&	10.6690	&	0.148	\\
EVRJ220737.90-813510.0	&	57090539	&	331.9079	&	-81.5861	&	11.93	&	14.35	&	0.90	&	ms	&	G	&	14.9246	&	0.016	\\
EVRJ223616.97-773616.2	&	57136977	&	339.0707	&	-77.6045	&	13.55	&	8.89	&	1.15	&	giant	&	K6	&	2074.7965	&	0.299	\\
EVRJ235019.03-840248.8	&	57081420	&	357.5793	&	-84.0469	&	11.40	&	7.79	&	0.39	&	ms	&	F5V	&	5.8564	&	0.020	\\

\hline
%\noalign{\vskip 8pt}  
%\end{tabular}
\end{longtable*}
%\end{longtable}

\clearpage
\newpage

Note: Columns 1-5 are identification numbers, right ascension and declination, and magnitude. Columns 6-9 are the reduced proper motion (RPM) and color difference (B-V) which we use to estimate the star size and spectral type (see Section 4.2.1). Columns 10 and 11 are the period found in hours, and the fractional eclipse depth from normalized flux.

\begin{longtable*}{lllllllllll}
\caption{Eclipsing Binary discoveries}\\
\hline
\noalign{\vskip 3pt}
\text{ESID} & \text{APASSID} & \text{RA} & \text{Dec} & \text{M\textsubscript{v}} & \text{RPM} & \text{B-V} & \text{size} & \text{spec} & \text{period} & \text{depth} \\ [0.1ex]
\text{ } & \text{ } & \text{(J2000)} & \text{(J2000)} & \text{ } & \text{} & \text{} & \text{} & \text{} & \text{(hours)} & \text{(fractional)} \\ [0.1ex]

\hline
\noalign{\vskip 3pt}
\endfirsthead

\hline
\noalign{\vskip 3pt}
\text{ESID} & \text{APASSID} & \text{RA} & \text{Dec} & \text{M\textsubscript{v}} & \text{RPM} & \text{B-V} & \text{size} & \text{spec} & \text{period} & \text{depth} \\ [0.1ex]
\text{ } & \text{ } & \text{(J2000)} & \text{(J2000)} & \text{ } & \text{} & \text{} & \text{} & \text{} & \text{(hours)} & \text{(fractional)} \\ [0.1ex]

\hline
\noalign{\vskip 3pt}
\endhead
\endfoot
\hline
\endlastfoot

EVRJ002445.62-784031.1	&	36136315	&	6.1901	&	-78.6753	&	11.78	&	6.21	&	0.56	&	ms	&	G9V	&	109.3650	&	0.118	\\
EVRJ004748.46-754942.6	&	36397363	&	11.9519	&	-75.8285	&	11.79	&	5.20	&	0.52	&	ms	&	F9V	&	157.4220	&	0.229	\\
EVRJ005637.85-782127.0	&	36134521	&	14.1577	&	-78.3575	&	12.39	&	8.65	&	0.63	&	ms	&	G5V	&	74.0640	&	0.273	\\
EVRJ010726.33-774753.2	&	36135199	&	16.8597	&	-77.7981	&	12.96	&	8.26	&	0.30	&	ms	&	A5V	&	15.7654	&	0.148	\\
EVRJ012740.66-841645.1	&	36035232	&	21.9194	&	-84.2792	&	13.25	&	7.92	&	0.65	&	ms	&	G1V	&	20.0793	&	0.109	\\
EVRJ013849.70-842426.6	&	36037250	&	24.7071	&	-84.4074	&	13.24	&	14.24	&	0.90	&	ms	&	K	&	198.5830	&	0.172	\\
EVRJ014115.60-800737.9	&	36113485	&	25.3150	&	-80.1272	&	14.97	&	9.44	&	0.23	&	ms	&	G2V	&	22.1010	&	0.187	\\
EVRJ023605.38-852430.6	&	36017318	&	39.0224	&	-85.4085	&	12.75	&	7.47	&	0.54	&	ms	&	F1V	&	118.7710	&	0.225	\\
EVRJ024203.26-750224.0	&	36161150	&	40.5136	&	-75.0400	&	11.98	&	6.76	&	0.31	&	ms	&	F4V	&	30.9563	&	0.243	\\
EVRJ024438.52-835122.7	&	36036709	&	41.1605	&	-83.8563	&	12.50	&	8.28	&	0.72	&	ms	&	G8V	&	25.9882	&	0.303	\\
EVRJ030147.71-761211.5	&	36155585	&	45.4488	&	-76.2032	&	10.14	&	2.93	&	0.92	&	giant	&	K3	&	68.3947	&	0.065	\\
EVRJ032000.70-760821.5	&	36156333	&	50.0029	&	-76.1393	&	13.15	&	10.03	&	0.59	&	ms	&	F9V	&	79.5734	&	0.202	\\
EVRJ032206.70-752842.6	&	36157171	&	50.5279	&	-75.4785	&	12.91	&	9.86	&	1.00	&	ms	&	K4V	&	7.3698	&	0.147	\\
EVRJ032355.42-783922.7	&	36091552	&	50.9809	&	-78.6563	&	11.40	&	8.26	&	0.62	&	ms	&	F8V	&	22.0561	&	0.215	\\
EVRJ033317.16-792812.7	&	36087183	&	53.3215	&	-79.4702	&	13.30	&	9.82	&	0.38	&	ms	&	A8V	&	21.1649	&	0.095	\\
EVRJ043634.42-863132.9	&	36012543	&	69.1434	&	-86.5258	&	14.30	&	11.43	&	0.76	&	ms	&	F1V	&	8.1408	&	0.221	\\
EVRJ043913.51-855448.6	&	36019521	&	69.8063	&	-85.9135	&	10.23	&	8.34	&	0.82	&	ms	&	G3V	&	36.8604	&	0.054	\\
EVRJ043932.02-794339.0	&	36070451	&	69.8834	&	-79.7275	&	11.53	&	6.84	&	0.59	&	ms	&	F9V	&	61.6244	&	0.117	\\
EVRJ044501.22-771324.6	&	36081663	&	71.2551	&	-77.2235	&	11.33	&	10.78	&	0.59	&	ms	&	F9V	&	85.9685	&	0.217	\\
EVRJ044545.98-770625.6	&	36081683	&	71.4416	&	-77.1071	&	13.16	&	4.63	&	0.30	&	ms	&	F1V	&	17.5505	&	0.387	\\
EVRJ045203.19-853702.6	&	36019627	&	73.0133	&	-85.6174	&	12.49	&	10.39	&	0.09	&	ms	&	F9	&	30.2652	&	0.092	\\
EVRJ045807.78-772037.7	&	36082386	&	74.5324	&	-77.3438	&	12.58	&	8.92	&	0.69	&	ms	&	G6V	&	38.1567	&	0.379	\\
EVRJ050731.01-760919.8	&	36201646	&	76.8792	&	-76.1555	&	13.38	&	10.14	&	0.64	&	ms	&	G2V	&	10.7099	&	0.065	\\
EVRJ052042.43-753131.8	&	36202184	&	80.1768	&	-75.5255	&	11.06	&	3.14	&	0.05	&	ms	&	A1V	&	64.8150	&	0.322	\\
EVRJ053006.26-811232.4	&	36062995	&	82.5261	&	-81.2090	&	11.95	&	7.32	&	0.49	&	ms	&	F8V	&	45.1688	&	0.217	\\
EVRJ053504.90-834045.5	&	36058479	&	83.7704	&	-83.6793	&	12.40	&	9.23	&	0.41	&	ms	&	F9V	&	35.7837	&	0.130	\\
EVRJ053541.69-753728.6	&	36199669	&	83.9237	&	-75.6246	&	12.75	&	8.21	&	0.81	&	ms	&	K2V	&	31.2876	&	0.113	\\
EVRJ054814.83-772912.5	&	36195270	&	87.0618	&	-77.4868	&	13.85	&	9.49	&	0.47	&	ms	&	G3V	&	17.3382	&	0.611	\\
EVRJ055000.48-780018.7	&	36076437	&	87.5020	&	-78.0052	&	13.60	&	8.79	&	0.44	&	ms	&	G0V	&	22.1010	&	0.243	\\
EVRJ055918.00-861604.1	&	36018864	&	89.8250	&	-86.2678	&	12.51	&	8.16	&	0.54	&	ms	&	G6V	&	16.0350	&	0.101	\\
EVRJ060300.82-763227.2	&	39301921	&	90.7534	&	-76.5409	&	13.01	&	9.37	&	0.63	&	ms	&	G6V	&	35.3178	&	0.368	\\
EVRJ060956.86-842635.9	&	37958393	&	92.4869	&	-84.4433	&	12.32	&	8.02	&	0.03	&	ms	&	F1	&	64.7228	&	0.106	\\
EVRJ061730.58-853507.8	&	37952210	&	94.3774	&	-85.5855	&	13.76	&	9.17	&	-0.23	&	ms	&	A0	&	19.5420	&	0.064	\\
EVRJ061942.89-872037.7	&	37932270	&	94.9287	&	-87.3438	&	12.77	&	8.30	&	0.53	&	ms	&	G0V	&	11.7925	&	0.176	\\
EVRJ062614.11-812323.6	&	38088026	&	96.5588	&	-81.3899	&	12.84	&	9.76	&	0.45	&	ms	&	F8V	&	22.1397	&	0.057	\\
EVRJ065136.84-775609.6	&	38124833	&	102.9035	&	-77.9360	&	11.32	&	4.24	&	0.66	&	giant	&	G5	&	154.2250	&	0.158	\\
EVRJ065350.93-840014.0	&	37957964	&	103.4622	&	-84.0039	&	12.91	&	9.75	&	0.60	&	ms	&	K0V	&	45.7400	&	0.212	\\
EVRJ065609.43-810853.5	&	38088326	&	104.0393	&	-81.1482	&	13.49	&	12.36	&	0.71	&	ms	&	K0V	&	13.4842	&	0.116	\\
EVRJ070327.70-813323.4	&	38085967	&	105.8654	&	-81.5565	&	13.29	&	9.59	&	0.45	&	ms	&	A9V	&	22.2891	&	0.357	\\
EVRJ071503.55-792949.6	&	38091930	&	108.7648	&	-79.4971	&	13.08	&	7.32	&	0.41	&	ms	&	G1V	&	35.9922	&	0.599	\\
EVRJ071744.33-854505.4	&	37951607	&	109.4347	&	-85.7515	&	13.09	&	6.75	&	0.77	&	ms	&	K3V	&	101.6850	&	0.314	\\
EVRJ071748.29-844104.6	&	37953026	&	109.4512	&	-84.6846	&	14.52	&	10.35	&	0.39	&	ms	&	G3V	&	7.6682	&	0.184	\\
EVRJ071938.74-794442.4	&	38091354	&	109.9114	&	-79.7451	&	12.18	&	5.14	&	0.39	&	ms	&	F2V	&	20.8352	&	0.107	\\
EVRJ072710.08-815757.2	&	38084205	&	111.7920	&	-81.9659	&	13.73	&	5.29	&	0.54	&	ms	&	G9V	&	41.0657	&	1.000	\\
EVRJ073157.14-815943.4	&	38084148	&	112.9881	&	-81.9954	&	9.93	&	6.74	&	0.19	&	ms	&	F1V	&	60.0174	&	0.074	\\
EVRJ074851.14-844938.3	&	37953692	&	117.2131	&	-84.8273	&	13.18	&	8.90	&	0.56	&	ms	&	F	&	23.3998	&	0.115	\\
EVRJ075512.70-831036.1	&	37960834	&	118.8029	&	-83.1767	&	12.40	&	8.05	&	0.54	&	ms	&	F8V	&	68.9536	&	0.237	\\
EVRJ080959.06-765721.2	&	38113121	&	122.4961	&	-76.9559	&	12.33	&	8.08	&	0.52	&	ms	&	F8V	&	109.3650	&	0.257	\\
EVRJ082431.85-771708.5	&	38110818	&	126.1327	&	-77.2857	&	10.99	&	6.53	&	0.55	&	ms	&	G1V	&	46.0730	&	0.162	\\
EVRJ083235.69-814208.3	&	37974452	&	128.1487	&	-81.7023	&	13.27	&	10.91	&	0.48	&	ms	&	G7V	&	22.5004	&	0.273	\\
EVRJ083610.66-822751.1	&	37966550	&	129.0444	&	-82.4642	&	12.08	&	7.46	&	1.30	&	giant	&	K	&	15.4668	&	0.066	\\
EVRJ084853.45-755536.1	&	38163662	&	132.2227	&	-75.9267	&	10.01	&	5.34	&	0.28	&	ms	&	F8V	&	59.1296	&	0.081	\\
EVRJ085629.66-833101.6	&	37964619	&	134.1236	&	-83.5171	&	12.47	&	4.96	&	0.51	&	ms	&	F1V	&	59.1383	&	0.250	\\
EVRJ090851.91-835702.5	&	37945252	&	137.2163	&	-83.9507	&	13.29	&	8.89	&	0.42	&	ms	&	F1V	&	10.8306	&	0.337	\\
EVRJ091345.72-822820.3	&	37965615	&	138.4405	&	-82.4723	&	11.75	&	6.95	&	0.60	&	ms	&	G5V	&	43.7381	&	0.182	\\
EVRJ092241.74-833802.0	&	37945406	&	140.6739	&	-83.6339	&	13.77	&	8.53	&	0.54	&	ms	&	G0V	&	4.4717	&	0.145	\\
EVRJ093342.00-865534.0	&	37929506	&	143.4250	&	-86.9261	&	13.03	&	8.34	&	0.75	&	ms	&	G9V	&	106.1730	&	0.347	\\
EVRJ093554.48-763543.8	&	38051623	&	143.9770	&	-76.5955	&	13.59	&	4.56	&	0.67	&	giant	&	F6	&	35.6159	&	0.248	\\
EVRJ093619.37-811153.2	&	37970069	&	144.0807	&	-81.1981	&	13.13	&	10.33	&	0.82	&	ms	&	K1V	&	129.2850	&	0.239	\\
EVRJ094641.04-781309.8	&	38041815	&	146.6710	&	-78.2194	&	13.00	&	7.60	&	0.56	&	ms	&	G2V	&	33.1735	&	0.231	\\
EVRJ095515.41-830705.9	&	37948497	&	148.8142	&	-83.1183	&	12.88	&	9.48	&	0.59	&	ms	&	G5V	&	151.7600	&	0.345	\\
EVRJ100205.04-814503.2	&	37950302	&	150.5210	&	-81.7509	&	12.79	&	6.15	&	0.30	&	ms	&	F5V	&	67.4233	&	0.196	\\
EVRJ100426.40-803846.0	&	38032563	&	151.1100	&	-80.6461	&	13.09	&	7.01	&	0.59	&	ms	&	F6V	&	28.3768	&	0.095	\\
EVRJ100649.61-801046.9	&	38033725	&	151.7067	&	-80.1797	&	12.65	&	12.88	&	0.51	&	ms	&	F4	&	44.7694	&	0.205	\\
EVRJ101423.47-774932.5	&	38040775	&	153.5978	&	-77.8257	&	13.29	&	13.24	&	1.05	&	ms	&	K	&	35.1543	&	0.083	\\
EVRJ103443.51-775813.1	&	38026295	&	158.6813	&	-77.9703	&	10.88	&	10.79	&	0.64	&	ms	&	G4V	&	20.9396	&	0.029	\\
EVRJ105421.24-782234.7	&	38024299	&	163.5885	&	-78.3763	&	12.41	&	10.76	&	1.00	&	ms	&	K6V	&	20.3265	&	0.081	\\
EVRJ105445.86-785351.4	&	38023795	&	163.6911	&	-78.8976	&	12.69	&	10.56	&	0.61	&	ms	&	G9V	&	62.0270	&	0.169	\\
EVRJ110105.30-864038.6	&	37928838	&	165.2721	&	-86.6774	&	13.73	&	9.45	&	0.58	&	ms	&	G7V	&	42.5378	&	0.096	\\
EVRJ110815.96-870153.8	&	37928658	&	167.0665	&	-87.0316	&	12.68	&	9.77	&	0.84	&	ms	&	K3V	&	12.2767	&	0.230	\\
EVRJ111244.66-830219.7	&	37987113	&	168.1861	&	-83.0388	&	13.18	&	7.43	&	0.49	&	ms	&	G7V	&	17.3342	&	0.277	\\
EVRJ111447.02-811836.7	&	37992322	&	168.6959	&	-81.3102	&	12.41	&	10.99	&	0.62	&	ms	&	G6V	&	35.0246	&	0.171	\\
EVRJ112755.49-842109.7	&	37940976	&	171.9812	&	-84.3527	&	13.81	&	9.72	&	0.34	&	ms	&	F8V	&	8.4533	&	0.252	\\
EVRJ114502.30-771447.0	&	38272180	&	176.2596	&	-77.2464	&	12.29	&	7.73	&	0.78	&	ms	&	G4V	&	39.2816	&	0.209	\\
EVRJ114706.02-835834.7	&	37940896	&	176.7751	&	-83.9763	&	12.90	&	9.39	&	0.87	&	ms	&	K3V	&	54.3094	&	0.091	\\
EVRJ120501.68-852738.9	&	47205316	&	181.2570	&	-85.4608	&	12.73	&	9.41	&	0.48	&	ms	&	F6V	&	103.6910	&	0.144	\\
EVRJ120856.86-770450.9	&	48482910	&	182.2369	&	-77.0808	&	13.21	&	6.55	&	0.92	&	giant	&	K2	&	40.7350	&	0.275	\\
EVRJ122230.55-772324.4	&	48480777	&	185.6273	&	-77.3901	&	11.60	&	8.26	&	0.77	&	ms	&	K0V	&	100.8010	&	0.248	\\
EVRJ125134.08-790133.2	&	47428800	&	192.8920	&	-79.0259	&	10.32	&	9.09	&	0.57	&	ms	&	G8V	&	114.7710	&	0.157	\\
EVRJ125505.76-851321.7	&	47199929	&	193.7740	&	-85.2227	&	11.73	&	6.37	&	0.38	&	ms	&	F3V	&	30.5570	&	0.083	\\
EVRJ131324.31-792126.3	&	47375486	&	198.3513	&	-79.3573	&	12.28	&	6.37	&	0.31	&	ms	&	F7V	&	33.7030	&	0.228	\\
EVRJ131504.46-763140.1	&	47445061	&	198.7686	&	-76.5278	&	12.08	&	7.84	&	0.69	&	ms	&	G9V	&	21.5120	&	0.055	\\
EVRJ131906.34-840040.3	&	47206948	&	199.7764	&	-84.0112	&	11.58	&	5.87	&	1.07	&	giant	&	K3	&	23.8310	&	0.086	\\
EVRJ131909.89-834711.0	&	47207423	&	199.7912	&	-83.7864	&	12.74	&	8.79	&	0.40	&	ms	&	F0V	&	15.7910	&	0.368	\\
EVRJ132210.78-790543.1	&	47376460	&	200.5449	&	-79.0953	&	13.29	&	9.33	&	0.48	&	ms	&	F8V	&	37.6460	&	0.108	\\
EVRJ132915.46-763040.0	&	47457701	&	202.3144	&	-76.5111	&	12.65	&	8.42	&	0.53	&	ms	&	F	&	19.3470	&	0.205	\\
EVRJ133026.14-852532.2	&	47199044	&	202.6089	&	-85.4256	&	12.55	&	8.51	&	1.08	&	giant	&	K5	&	12.6020	&	0.038	\\
EVRJ133347.26-833757.7	&	47210054	&	203.4469	&	-83.6327	&	13.39	&	8.35	&	0.56	&	ms	&	G3V	&	17.8470	&	0.057	\\
EVRJ133848.86-834425.4	&	47209861	&	204.7036	&	-83.7404	&	13.58	&	10.16	&	0.58	&	ms	&	G7V	&	71.4170	&	0.455	\\
EVRJ134321.74-845650.6	&	47199233	&	205.8406	&	-84.9474	&	12.78	&	8.18	&	0.55	&	ms	&	G5V	&	9.7290	&	0.365	\\
EVRJ135211.09-844337.9	&	47201544	&	208.0462	&	-84.7272	&	13.73	&	9.53	&	0.60	&	ms	&	G8V	&	27.8460	&	0.382	\\
EVRJ135212.36-785333.7	&	47399995	&	208.0515	&	-78.8927	&	12.07	&	6.17	&	0.59	&	ms	&	F9V	&	61.0300	&	0.443	\\
EVRJ135431.18-815457.2	&	47214874	&	208.6299	&	-81.9159	&	11.19	&	6.69	&	0.26	&	ms	&	A4V	&	79.1580	&	0.050	\\
EVRJ135907.01-842606.7	&	47203065	&	209.7792	&	-84.4352	&	13.05	&	7.80	&	0.42	&	ms	&	F7V	&	14.1710	&	0.324	\\
EVRJ140102.93-823656.9	&	47212126	&	210.2622	&	-82.6158	&	10.79	&	9.06	&	0.75	&	ms	&	K0V	&	43.2894	&	0.188	\\
EVRJ140546.42-835919.0	&	47203290	&	211.4434	&	-83.9886	&	11.29	&	8.38	&	0.76	&	ms	&	G9V	&	33.0580	&	0.044	\\
EVRJ141018.94-830332.8	&	47211232	&	212.5789	&	-83.0591	&	11.89	&	4.11	&	0.21	&	ms	&	F1V	&	18.8112	&	0.123	\\
EVRJ141043.54-834026.4	&	47204533	&	212.6814	&	-83.6740	&	12.19	&	10.85	&	0.86	&	ms	&	G3V	&	13.7081	&	0.045	\\
EVRJ142633.19-825021.5	&	47205241	&	216.6383	&	-82.8393	&	12.92	&	11.70	&	0.75	&	ms	&	G9V	&	13.7050	&	0.176	\\
EVRJ143328.56-833750.5	&	47202745	&	218.3690	&	-83.6307	&	12.77	&	8.89	&	0.68	&	ms	&	G5V	&	129.9480	&	0.420	\\
EVRJ143842.46-813049.0	&	47228767	&	219.6769	&	-81.5136	&	13.07	&	8.39	&	0.67	&	ms	&	G1V	&	16.4483	&	0.153	\\
EVRJ143916.73-844909.5	&	47178809	&	219.8197	&	-84.8193	&	12.18	&	6.11	&	0.55	&	ms	&	F	&	26.6630	&	0.082	\\
EVRJ144001.37-842946.7	&	47201869	&	220.0057	&	-84.4963	&	12.67	&	7.87	&	0.42	&	ms	&	F6V	&	18.0120	&	0.126	\\
EVRJ144437.08-775109.4	&	47397018	&	221.1545	&	-77.8526	&	12.23	&	4.31	&	0.30	&	ms	&	F1V	&	40.6480	&	0.151	\\
EVRJ144607.85-835647.8	&	47202209	&	221.5327	&	-83.9466	&	12.20	&	8.33	&	0.73	&	ms	&	G6V	&	16.1201	&	0.081	\\
EVRJ145302.21-823117.0	&	47219133	&	223.2592	&	-82.5214	&	12.80	&	8.93	&	0.87	&	ms	&	K0V	&	52.0750	&	0.236	\\
EVRJ145519.22-761940.8	&	47504010	&	223.8301	&	-76.3280	&	13.50	&	5.89	&	0.71	&	ms	&	G7V	&	20.7599	&	0.242	\\
EVRJ150051.41-823800.6	&	47219888	&	225.2142	&	-82.6335	&	11.69	&	7.50	&	1.42	&	giant	&	M2	&	50.1160	&	0.071	\\
EVRJ150534.08-873605.8	&	47165613	&	226.3920	&	-87.6016	&	12.70	&	8.88	&	0.55	&	ms	&	F7V	&	178.9330	&	0.127	\\
EVRJ151657.05-782900.6	&	47306697	&	229.2377	&	-78.4835	&	11.51	&	8.78	&	0.60	&	ms	&	G7V	&	155.4880	&	0.187	\\
EVRJ152229.16-855039.1	&	47176529	&	230.6215	&	-85.8442	&	12.70	&	11.49	&	0.76	&	ms	&	K0V	&	55.5406	&	0.319	\\
EVRJ152858.46-781119.0	&	47305454	&	232.2436	&	-78.1886	&	12.08	&	7.47	&	0.53	&	ms	&	F2V	&	37.9090	&	0.124	\\
EVRJ152933.41-755721.2	&	47507129	&	232.3892	&	-75.9559	&	9.95	&	5.49	&	0.27	&	ms	&	A6V	&	27.9790	&	0.099	\\
EVRJ153920.90-820531.2	&	47222149	&	234.8371	&	-82.0920	&	13.86	&	8.80	&	0.64	&	ms	&	G2V	&	21.6105	&	0.093	\\
EVRJ154507.37-793351.1	&	47291578	&	236.2807	&	-79.5642	&	12.75	&	7.16	&	0.66	&	ms	&	G5V	&	12.6404	&	0.244	\\
EVRJ160954.46-851858.3	&	47181489	&	242.4769	&	-85.3162	&	12.76	&	7.60	&	0.33	&	ms	&	F4V	&	44.7430	&	0.189	\\
EVRJ161508.66-835940.6	&	47187461	&	243.7861	&	-83.9946	&	13.18	&	11.16	&	0.73	&	ms	&	K0V	&	144.5400	&	0.403	\\
EVRJ162303.89-810023.4	&	47253636	&	245.7662	&	-81.0065	&	12.34	&	3.25	&	0.56	&	ms	&	G7V	&	62.7963	&	0.313	\\
EVRJ163423.28-841700.6	&	47182229	&	248.5970	&	-84.2835	&	13.63	&	9.48	&	0.99	&	ms	&	K2V	&	4.7610	&	0.397	\\
EVRJ163736.60-853222.2	&	47181298	&	249.4025	&	-85.5395	&	12.70	&	6.11	&	0.67	&	ms	&	G2V	&	25.4995	&	0.076	\\
EVRJ164003.19-835758.7	&	47185009	&	250.0133	&	-83.9663	&	12.91	&	9.45	&	0.57	&	ms	&	G6V	&	41.2660	&	0.247	\\
EVRJ164326.83-861143.8	&	47180996	&	250.8618	&	-86.1955	&	14.13	&	7.55	&	0.86	&	ms	&	G	&	9.7203	&	0.336	\\
EVRJ164433.50-860655.4	&	47181009	&	251.1396	&	-86.1154	&	12.22	&	5.87	&	0.59	&	ms	&	F1V	&	63.6382	&	0.219	\\
EVRJ165236.34-775955.0	&	47279936	&	253.1514	&	-77.9986	&	12.57	&	6.78	&	0.53	&	ms	&	F9V	&	45.8782	&	0.215	\\
EVRJ170835.06-804042.2	&	47250036	&	257.1461	&	-80.6784	&	13.36	&	10.77	&	0.64	&	ms	&	G6V	&	11.7251	&	0.185	\\
EVRJ171020.11-794022.8	&	47266105	&	257.5838	&	-79.6730	&	13.38	&	9.62	&	0.60	&	ms	&	G0V	&	34.9601	&	0.224	\\
EVRJ173306.41-841023.5	&	47184009	&	263.2767	&	-84.1732	&	11.94	&	7.48	&	0.01	&	ms	&	B8V	&	62.0777	&	0.107	\\
EVRJ175021.89-833451.2	&	47241898	&	267.5912	&	-83.5809	&	12.90	&	10.55	&	0.45	&	ms	&	G1V	&	97.8154	&	0.670	\\
EVRJ175151.98-770054.7	&	47705441	&	267.9666	&	-77.0152	&	12.99	&	7.95	&	0.60	&	ms	&	F1V	&	14.0583	&	0.312	\\
EVRJ175821.86-823023.4	&	47246340	&	269.5911	&	-82.5065	&	13.32	&	8.36	&	0.68	&	ms	&	G6V	&	32.6439	&	0.272	\\
EVRJ181441.14-755531.1	&	57414649	&	273.6714	&	-75.9253	&	11.71	&	9.14	&	0.66	&	ms	&	G3V	&	70.3589	&	0.235	\\
EVRJ181713.66-870538.4	&	57073210	&	274.3069	&	-87.0940	&	11.79	&	12.45	&	0.66	&	ms	&	K0V	&	8.4675	&	0.138	\\
EVRJ181714.18-773047.2	&	57406134	&	274.3091	&	-77.5131	&	13.02	&	7.02	&	0.51	&	ms	&	G1V	&	36.9321	&	0.603	\\
EVRJ181841.30-822733.5	&	57161826	&	274.6721	&	-82.4593	&	12.83	&	3.05	&	0.62	&	giant	&	G2	&	76.6038	&	0.070	\\
EVRJ184428.49-823046.8	&	57160536	&	281.1187	&	-82.5130	&	13.66	&	9.22	&	0.54	&	ms	&	G8V	&	71.0168	&	0.238	\\
EVRJ184600.48-775012.5	&	57197233	&	281.5020	&	-77.8368	&	12.93	&	8.55	&	0.56	&	ms	&	G1V	&	37.1489	&	0.192	\\
EVRJ185649.78-802355.0	&	57169339	&	284.2074	&	-80.3986	&	12.77	&	8.40	&	0.71	&	ms	&	K1V	&	11.4201	&	0.128	\\
EVRJ190936.96-825733.8	&	57102741	&	287.4040	&	-82.9594	&	12.21	&	9.17	&	0.81	&	ms	&	K1V	&	20.9770	&	0.097	\\
EVRJ191030.65-852500.5	&	57093523	&	287.6277	&	-85.4168	&	13.72	&	7.03	&	0.40	&	ms	&	F8V	&	69.7130	&	0.580	\\
EVRJ191957.17-815827.8	&	57163432	&	289.9882	&	-81.9744	&	11.82	&	8.40	&	0.76	&	ms	&	G9V	&	49.7553	&	0.190	\\
EVRJ195938.52-800854.6	&	57171413	&	299.9105	&	-80.1485	&	13.38	&	8.73	&	0.66	&	ms	&	G6V	&	11.8792	&	0.362	\\
EVRJ200844.69-841100.6	&	57097190	&	302.1862	&	-84.1835	&	12.36	&	6.21	&	0.74	&	ms	&	G8V	&	25.0280	&	0.183	\\
EVRJ204029.23-792542.2	&	57173726	&	310.1218	&	-79.4284	&	12.04	&	7.83	&	0.74	&	ms	&	G8V	&	27.8260	&	0.149	\\
EVRJ211629.93-755719.1	&	57218408	&	319.1247	&	-75.9553	&	11.50	&	8.69	&	0.41	&	ms	&	F4V	&	15.3512	&	0.213	\\
EVRJ213512.74-762221.4	&	57148511	&	323.8031	&	-76.3726	&	13.06	&	9.38	&	0.70	&	ms	&	G4V	&	21.3854	&	0.100	\\
EVRJ214611.09-783816.4	&	57142128	&	326.5462	&	-78.6379	&	13.01	&	7.87	&	0.74	&	ms	&	K1V	&	58.0007	&	0.138	\\
EVRJ215221.58-830052.9	&	57089418	&	328.0899	&	-83.0147	&	12.85	&	3.57	&	1.43	&	giant	&	M2	&	14.9228	&	0.069	\\
EVRJ215337.54-775041.6	&	57143831	&	328.4064	&	-77.8449	&	12.46	&	10.23	&	0.66	&	ms	&	G4V	&	4.7476	&	0.086	\\
EVRJ215538.66-830823.6	&	57089350	&	328.9111	&	-83.1399	&	12.97	&	9.49	&	0.77	&	ms	&	G9V	&	14.9228	&	0.095	\\
EVRJ223125.73-803027.7	&	57127400	&	337.8572	&	-80.5077	&	11.82	&	5.48	&	0.58	&	ms	&	F9V	&	41.7888	&	0.068	\\
EVRJ223812.41-780109.1	&	57136882	&	339.5517	&	-78.0192	&	13.50	&	10.29	&	0.38	&	ms	&	A8V	&	15.0893	&	0.256	\\
EVRJ225743.22-782429.2	&	57136245	&	344.4301	&	-78.4081	&	11.86	&	8.36	&	0.49	&	ms	&	G6V	&	19.8070	&	0.071	\\
EVRJ225936.22-830757.0	&	57083769	&	344.9009	&	-83.1325	&	11.22	&	9.34	&	0.69	&	ms	&	K0V	&	12.8218	&	0.064	\\
EVRJ230355.18-773000.7	&	57137831	&	345.9799	&	-77.5002	&	12.58	&	8.47	&	0.59	&	ms	&	G2V	&	34.7999	&	0.138	\\
EVRJ231712.07-790607.6	&	57132830	&	349.3003	&	-79.1021	&	11.51	&	8.15	&	0.66	&	ms	&	G6V	&	195.0140	&	0.154	\\
EVRJ232322.32-771336.5	&	57135188	&	350.8430	&	-77.2268	&	12.43	&	5.89	&	0.71	&	ms	&	G6V	&	66.0742	&	0.196	\\
EVRJ233038.69-760536.6	&	57261482	&	352.6612	&	-76.0935	&	12.56	&	9.29	&	0.60	&	ms	&	G0V	&	84.7218	&	0.390	\\
EVRJ234021.77-780703.4	&	57134048	&	355.0907	&	-78.1176	&	13.66	&	8.99	&	0.47	&	ms	&	F8V	&	11.6461	&	0.136	\\
EVRJ234130.91-790731.8	&	57133414	&	355.3788	&	-79.1255	&	13.36	&	7.75	&	0.53	&	ms	&	F5V	&	90.6906	&	0.291	\\

\hline
%\end{tabular}
%\end{table*}
%\begin{tablenotes}
%\item Columns 1-5 are identification numbers, right ascension and declination, and magnitude. Columns 6-9 are the reduced proper motion (RPM) and color difference (B-V) which we use to estimate the star size and spectral type (see Section 4.2.1). Columns 10 and 11 are the period found in hours, and the fractional eclipse depth from normalized flux.
%\end{tablenotes}
\end{longtable*}

\newpage
\begin{figure*}[ht]
\epsscale{1}
\begin{centering}
\includegraphics[width=.95\textwidth]{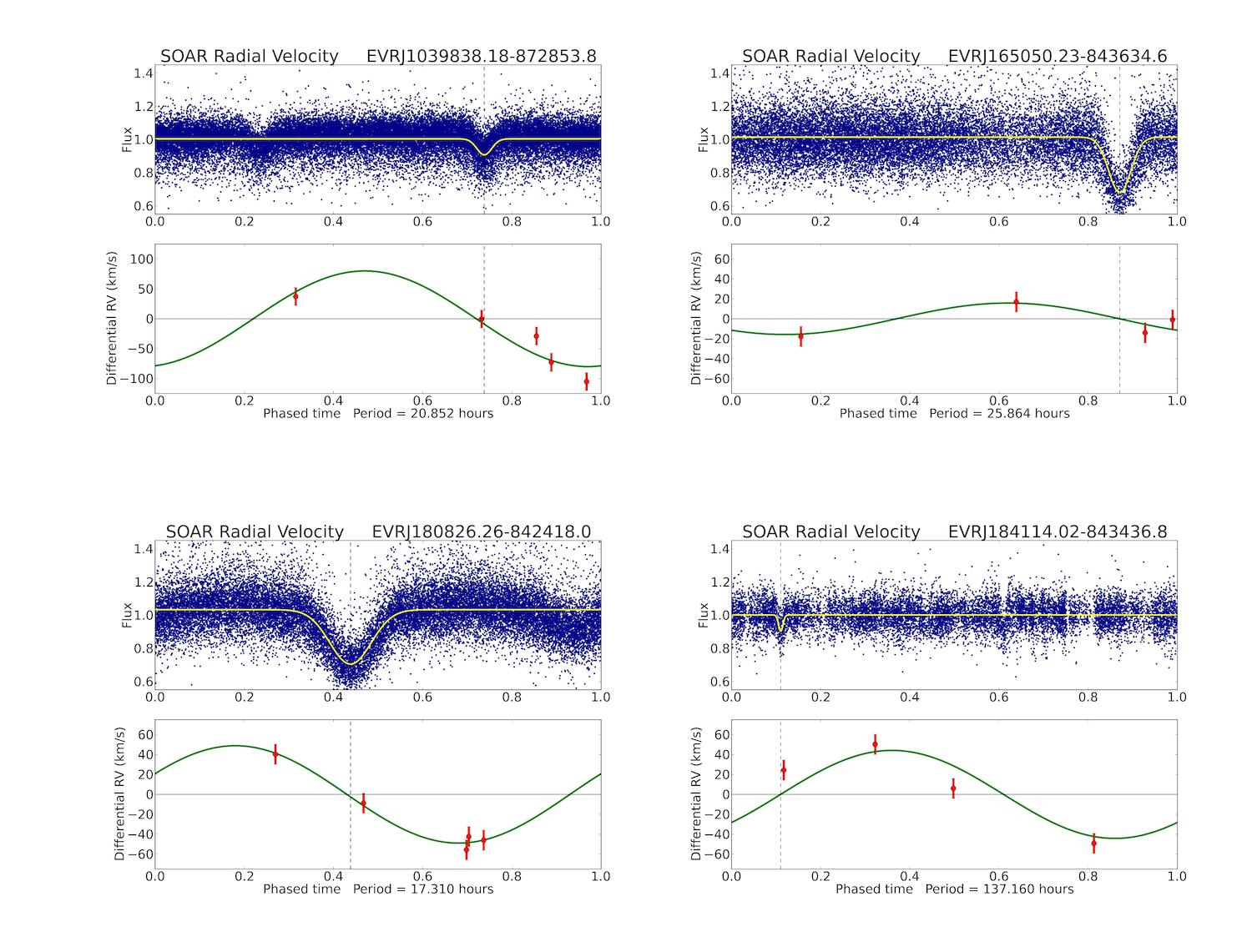}
\includegraphics[width=.45\textwidth]{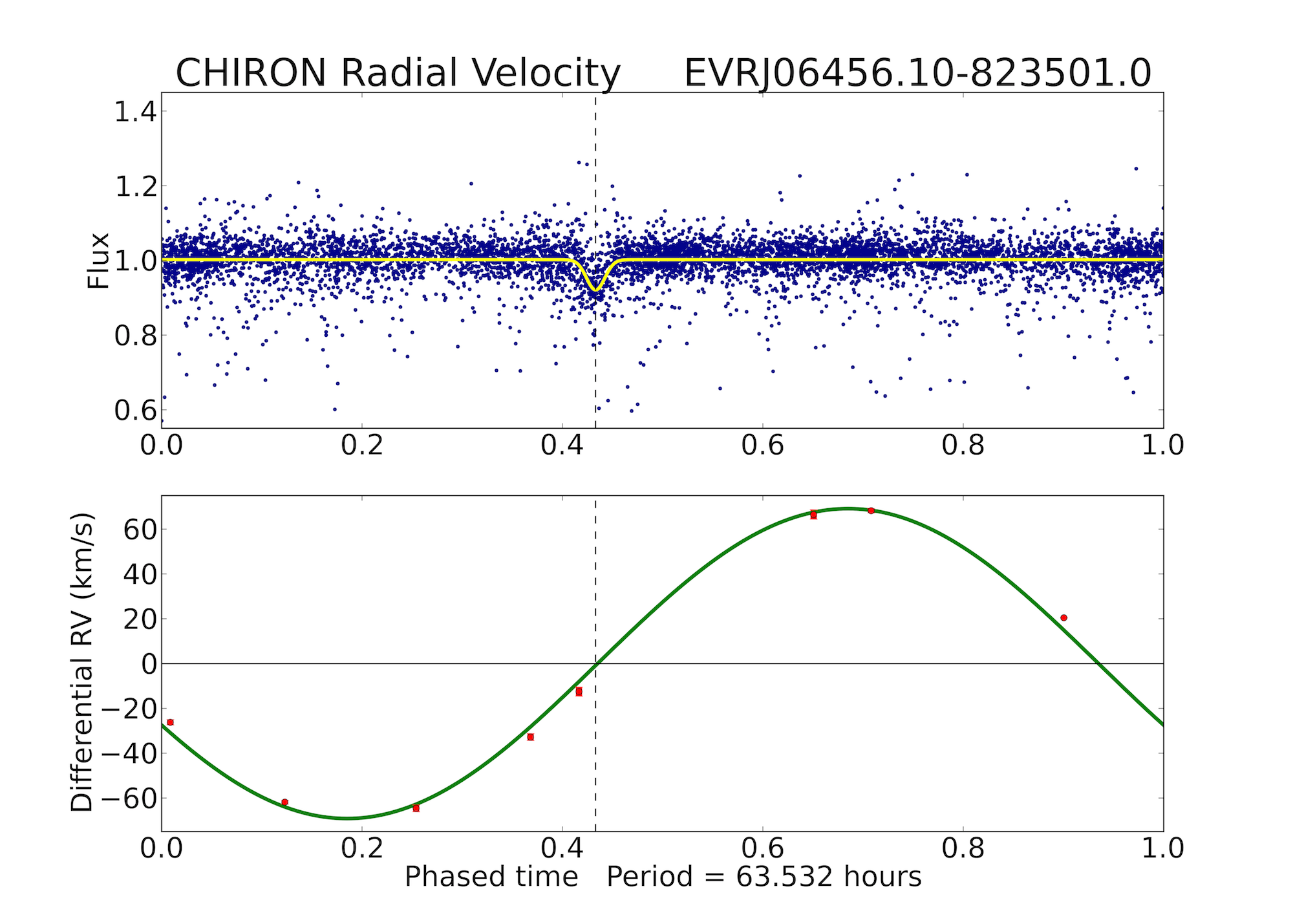}
\caption{Low mass secondary discoveries. Top panels are the Evryscope light curves with the best transit fit. Bottom panels are the SOAR RV points (red) with the best sinusoidal fit. Primary and secondary mass and radius values are shown in Table 2.}
\end{centering}
\label{fig:rv_combined}
\end{figure*}

\newpage
\begin{figure*}[ht]
\figurenum{16}
\epsscale{1}
\includegraphics[width=.95\textwidth]{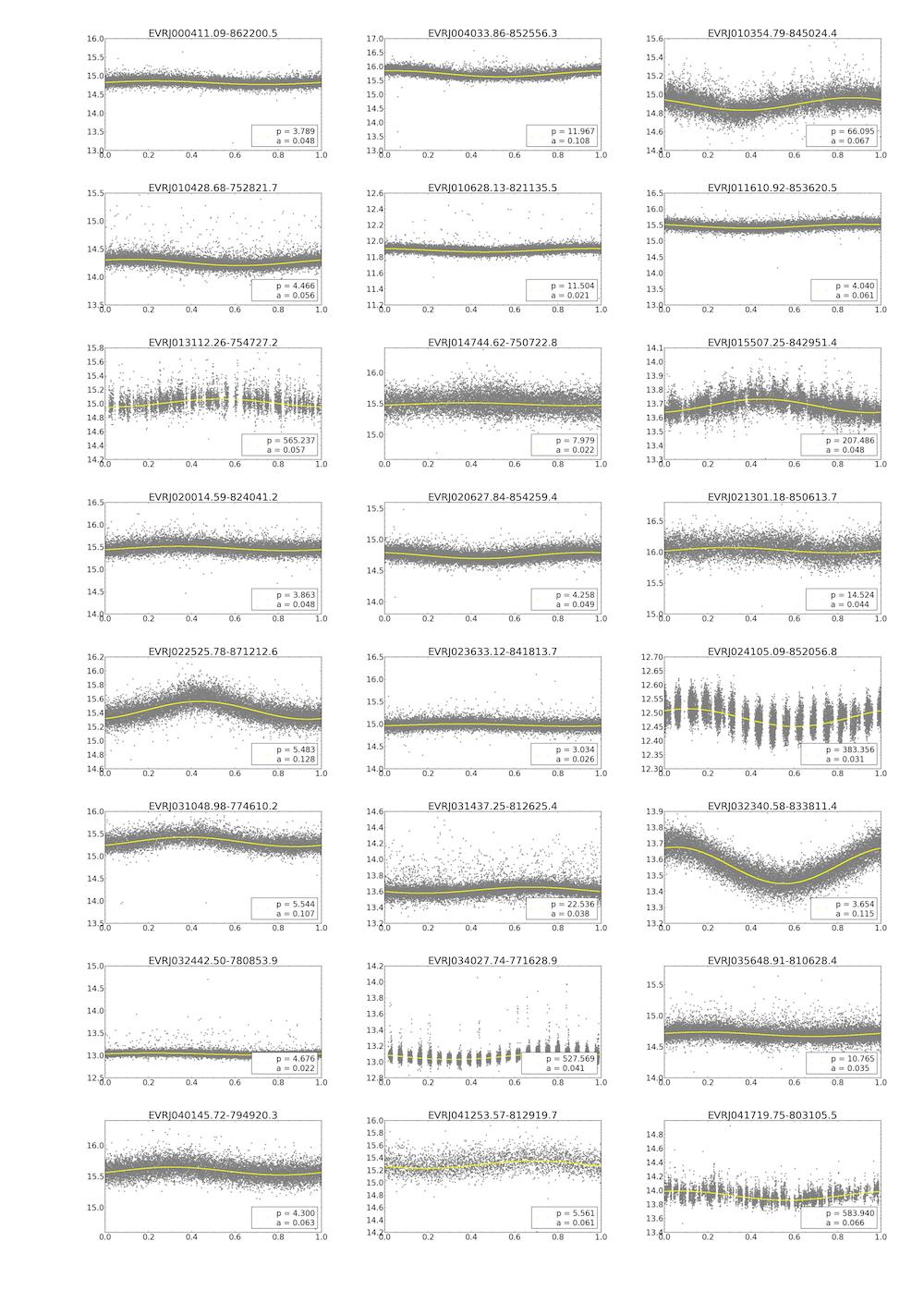}
\caption{Variable star discoveries. Y-axis is instrument magnitude, x-axis is the phase, p = period found in hours, a = amplitude change in magnitude. Gray points are two minute cadence, yellow is the best LS fit.}
\end{figure*}

\newpage
\begin{figure*}[ht]
\figurenum{16}
\epsscale{1}
\includegraphics[width=.95\textwidth]{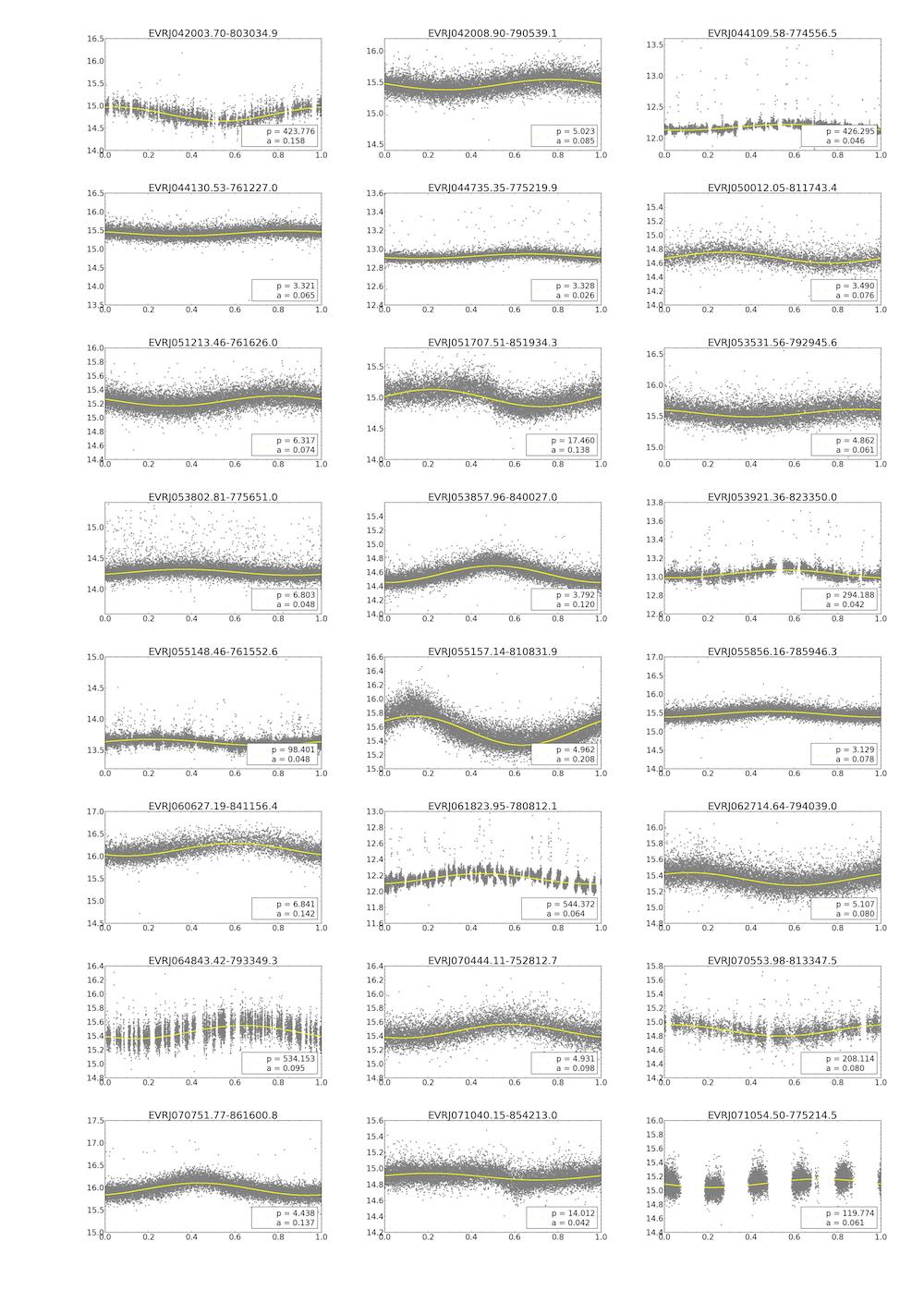}
\caption{Variable star discoveries continued. Y-axis is instrument magnitude, x-axis is the phase, p = period found in hours, a = amplitude change in magnitude. Gray points are two minute cadence, yellow is the best LS fit.}
\end{figure*}

\newpage
\begin{figure*}[ht]
\figurenum{16}
\epsscale{1}
\includegraphics[width=.95\textwidth]{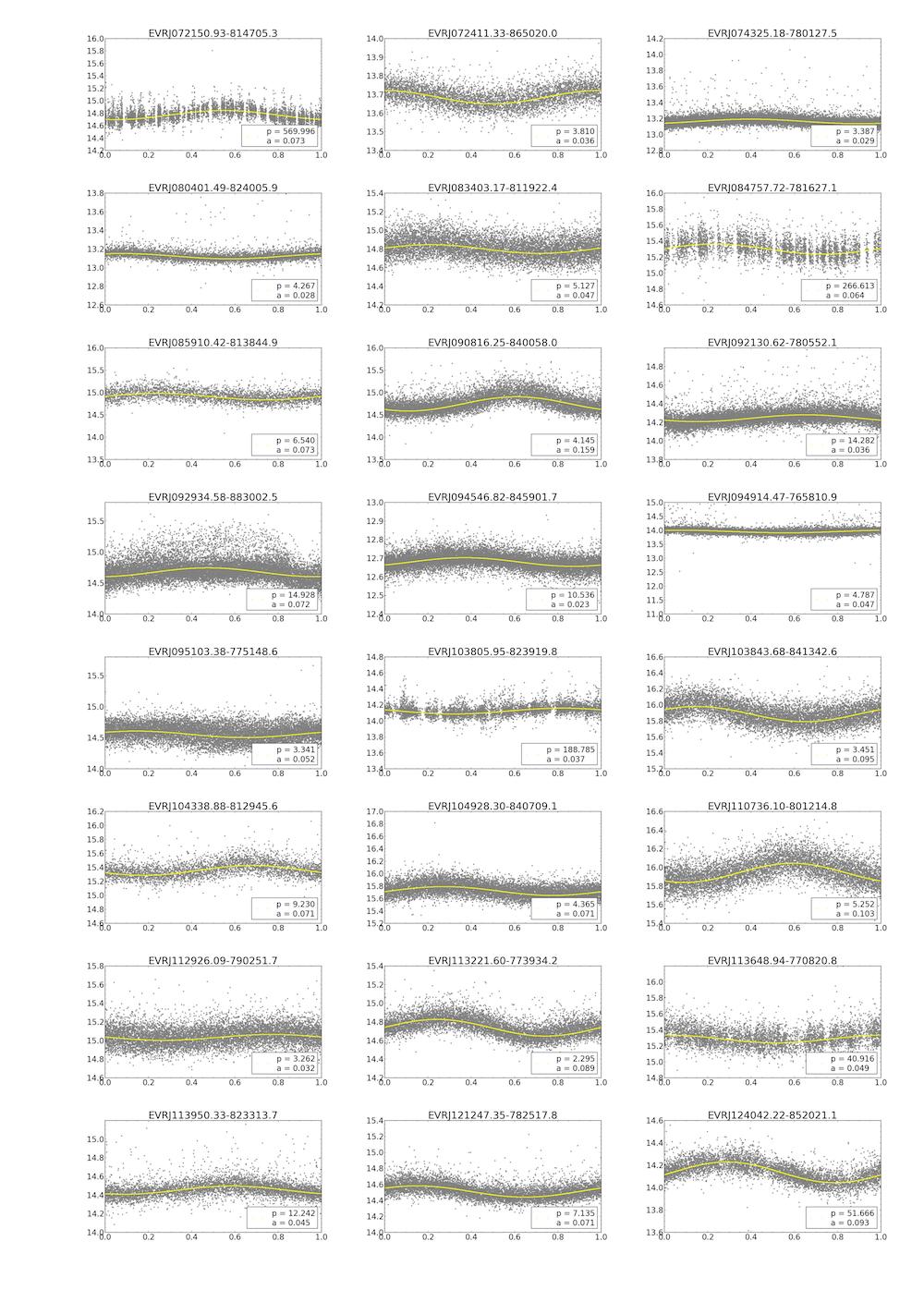}
\caption{Variable star discoveries continued. Y-axis is instrument magnitude, x-axis is the phase, p = period found in hours, a = amplitude change in magnitude. Gray points are two minute cadence, yellow is the best LS fit.}
\end{figure*}

\newpage
\begin{figure*}[ht]
\figurenum{16}
\epsscale{1}
\includegraphics[width=.95\textwidth]{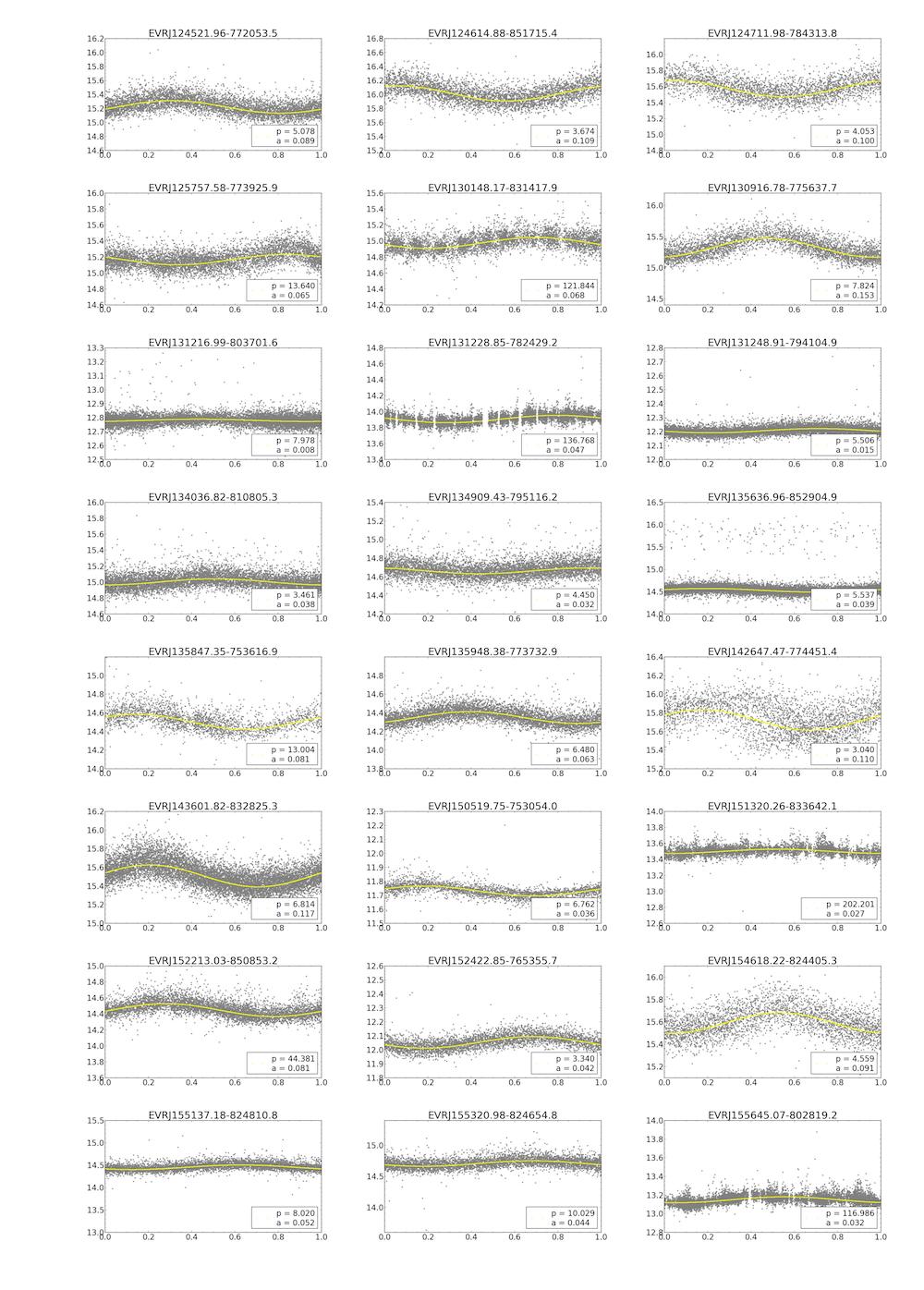}
\caption{Variable star discoveries continued. Y-axis is instrument magnitude, x-axis is the phase, p = period found in hours, a = amplitude change in magnitude. Gray points are two minute cadence, yellow is the best LS fit.}
\end{figure*}

\newpage
\begin{figure*}[ht]
\figurenum{16}
\epsscale{1}
\includegraphics[width=.95\textwidth]{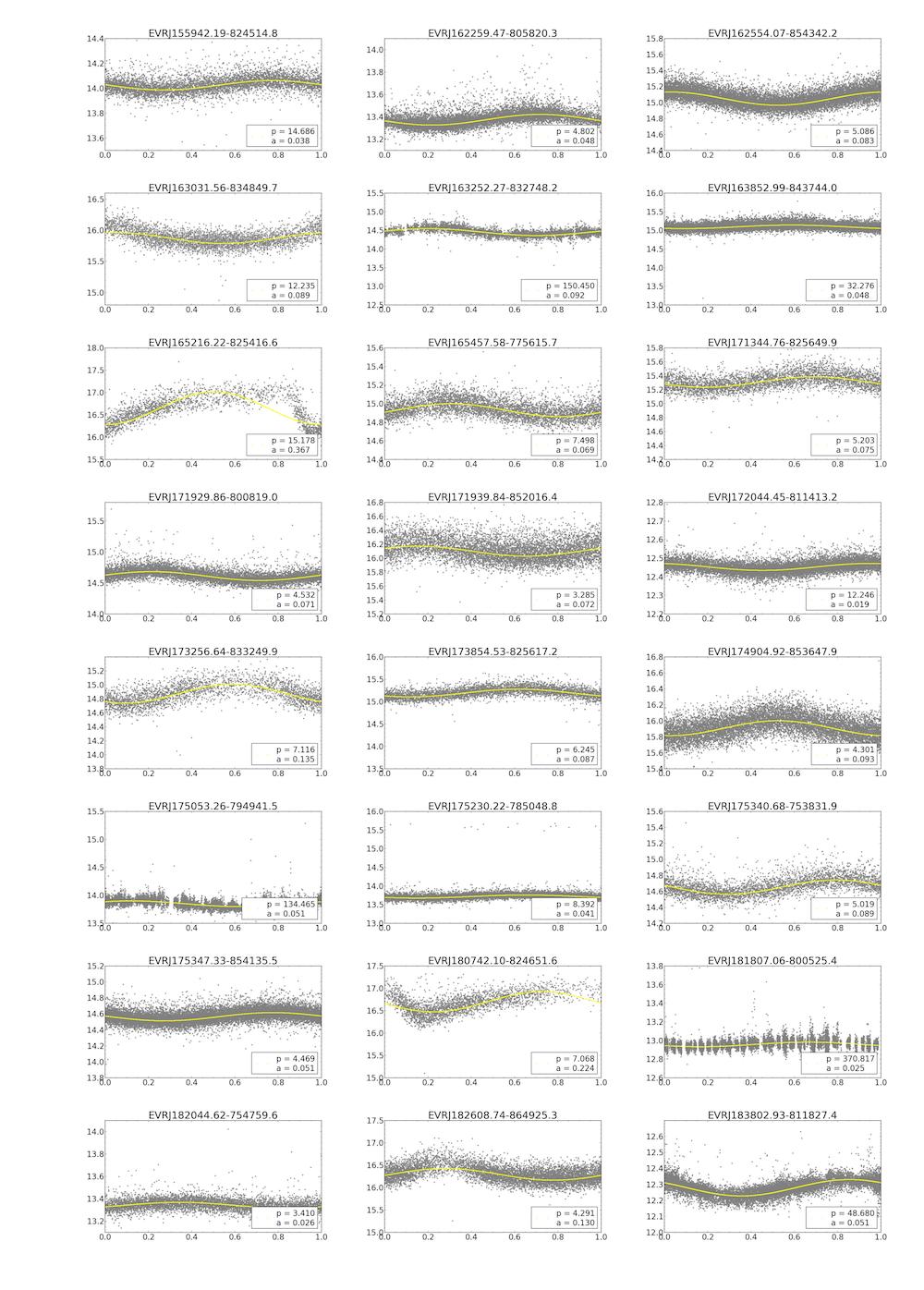}
\caption{Variable star discoveries continued. Y-axis is instrument magnitude, x-axis is the phase, p = period found in hours, a = amplitude change in magnitude. Gray points are two minute cadence, yellow is the best LS fit.}
\end{figure*}

\newpage
\begin{figure*}[ht]
\figurenum{16}
\epsscale{1}
\includegraphics[width=.95\textwidth]{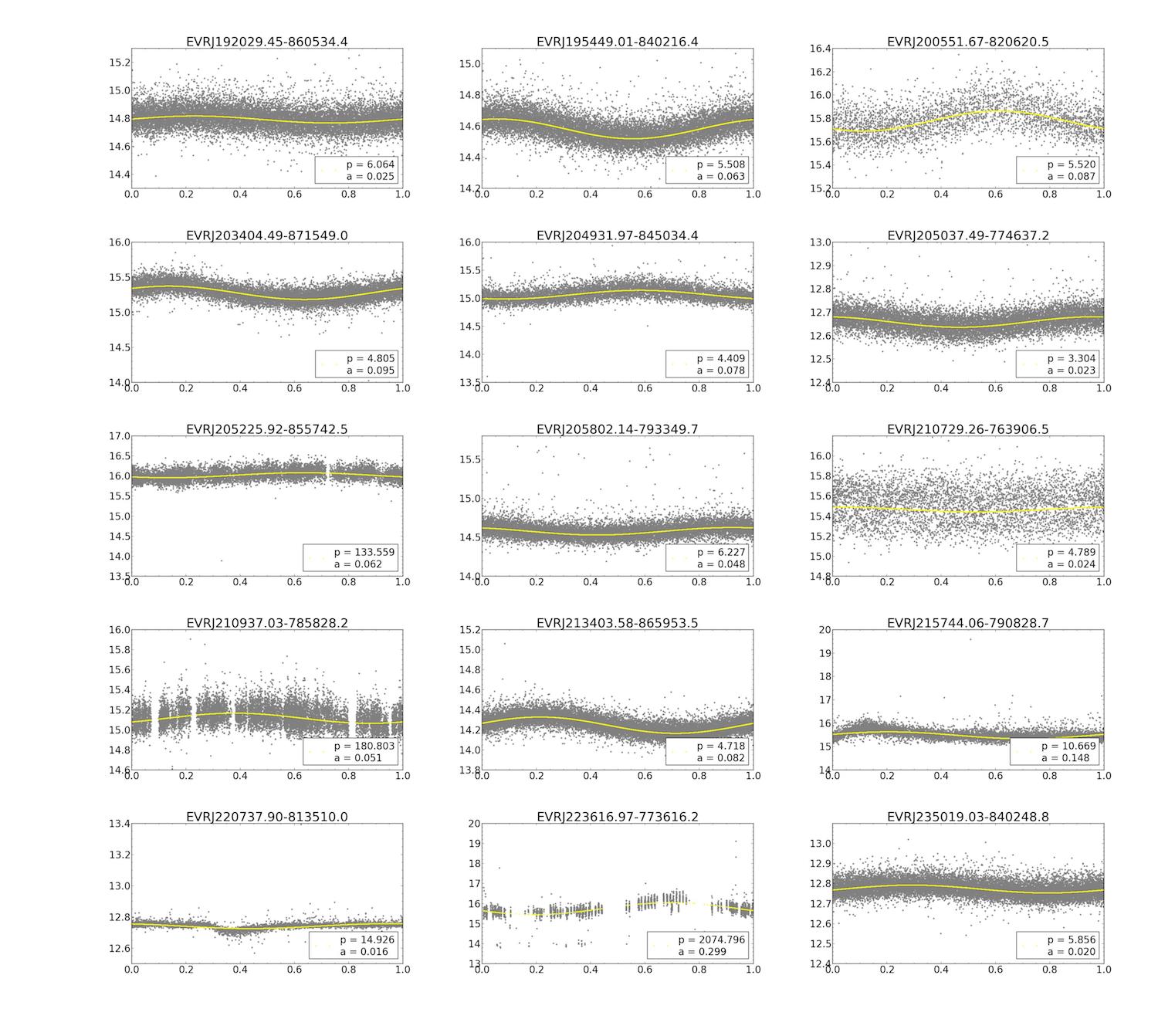}
\caption{Variable star discoveries continued. Y-axis is instrument magnitude, x-axis is the phase, p = period found in hours, a = amplitude change in magnitude. Gray points are two minute cadence, yellow is the best LS fit.}
\end{figure*}

\newpage
\begin{figure*}[ht]
\figurenum{17}
\epsscale{1}
\includegraphics[width=.95\textwidth]{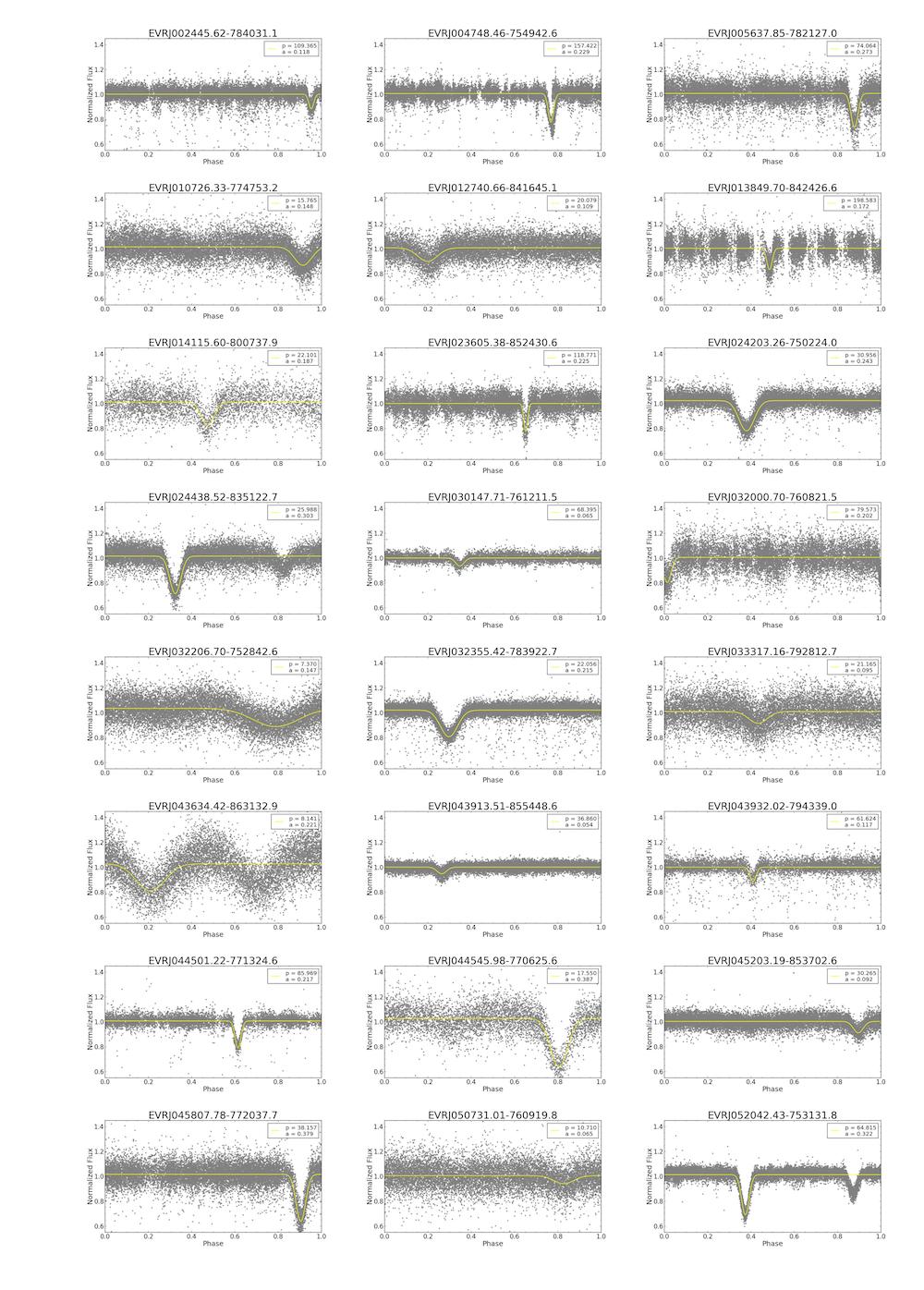}
\caption{Eclipsing Binary discoveries. Y-axis is normalized flux, x-axis is the phase, p = period found in hours, a = eclipse depth. Gray points are two minute cadence, yellow is the best fit.}
\end{figure*}

\newpage
\begin{figure*}[ht]
\figurenum{17}
\epsscale{1}
\includegraphics[width=.95\textwidth]{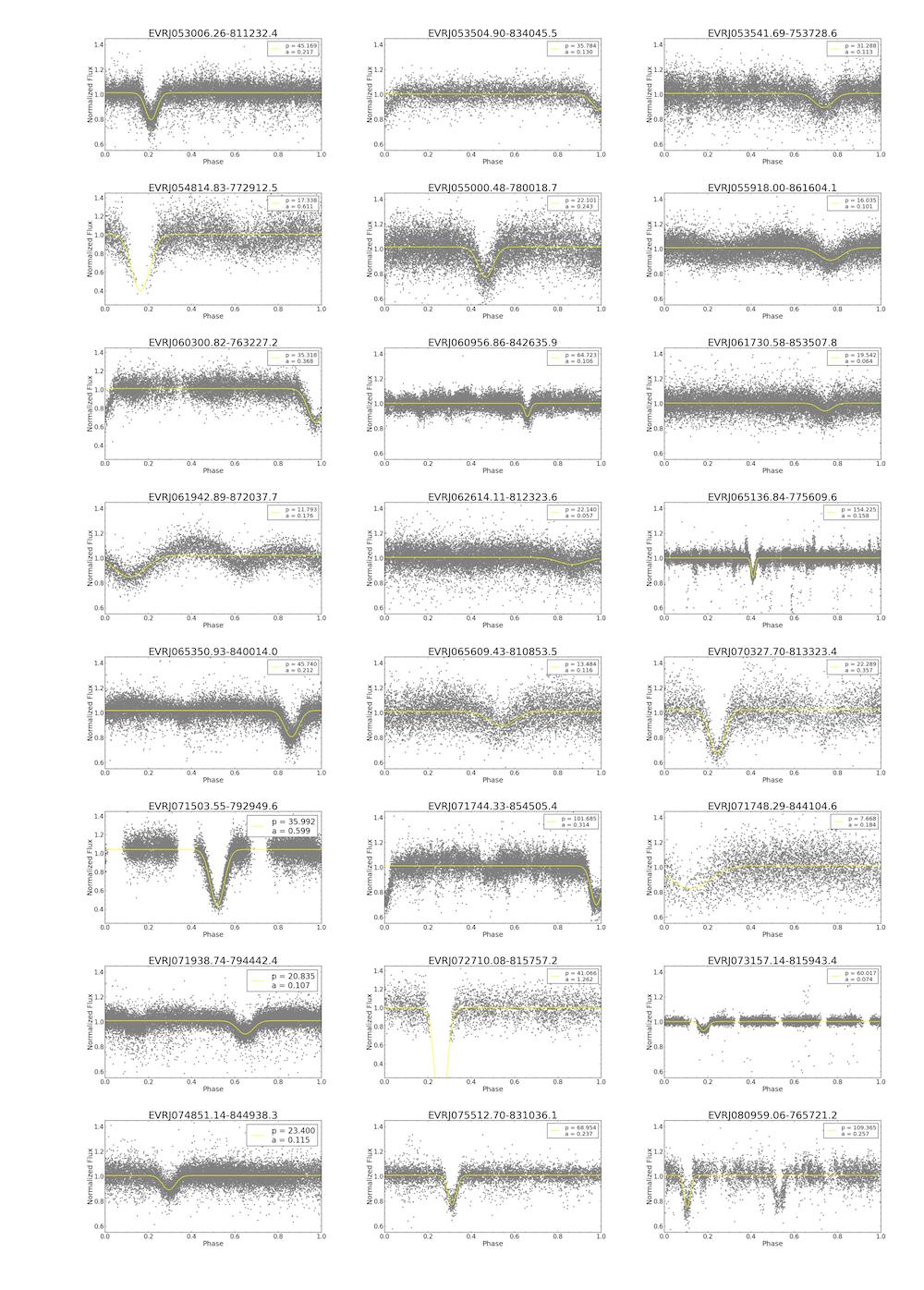}
\caption{Eclipsing Binary discoveries continued. Y-axis is normalized flux, x-axis is the phase, p = period found in hours, a = eclipse depth. Gray points are two minute cadence, yellow is the best fit.}
\end{figure*}

\newpage
\begin{figure*}[ht]
\figurenum{17}
\epsscale{1}
\includegraphics[width=.95\textwidth]{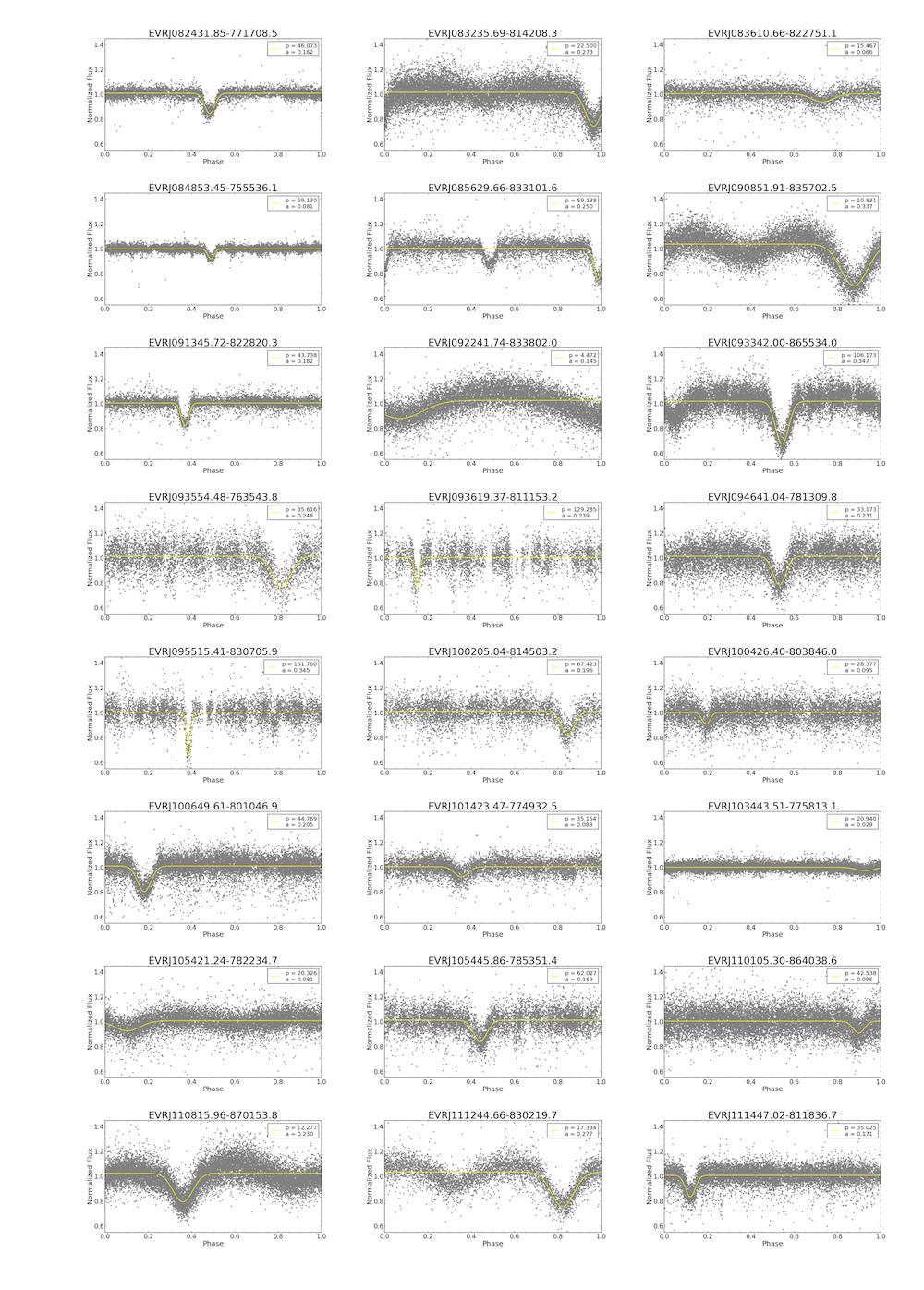}
\caption{Eclipsing Binary discoveries continued. Y-axis is normalized flux, x-axis is the phase, p = period found in hours, a = eclipse depth. Gray points are two minute cadence, yellow is the best fit.}
\end{figure*}

\newpage
\begin{figure*}[ht]
\figurenum{17}
\epsscale{1}
\includegraphics[width=.95\textwidth]{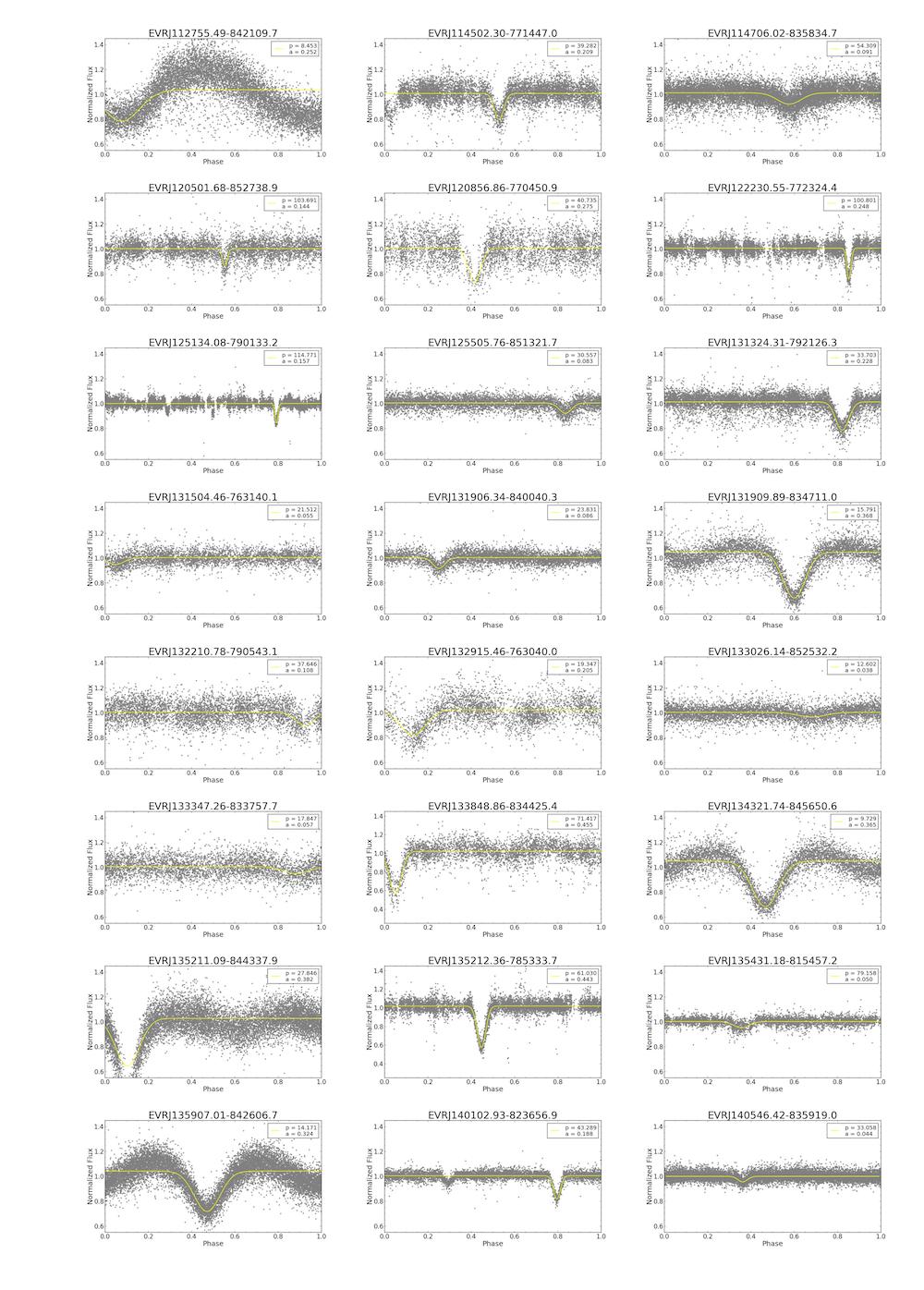}
\caption{Eclipsing Binary discoveries continued. Y-axis is normalized flux, x-axis is the phase, p = period found in hours, a = eclipse depth. Gray points are two minute cadence, yellow is the best fit.}
\end{figure*}

\newpage
\begin{figure*}[ht]
\figurenum{17}
\epsscale{1}
\includegraphics[width=.95\textwidth]{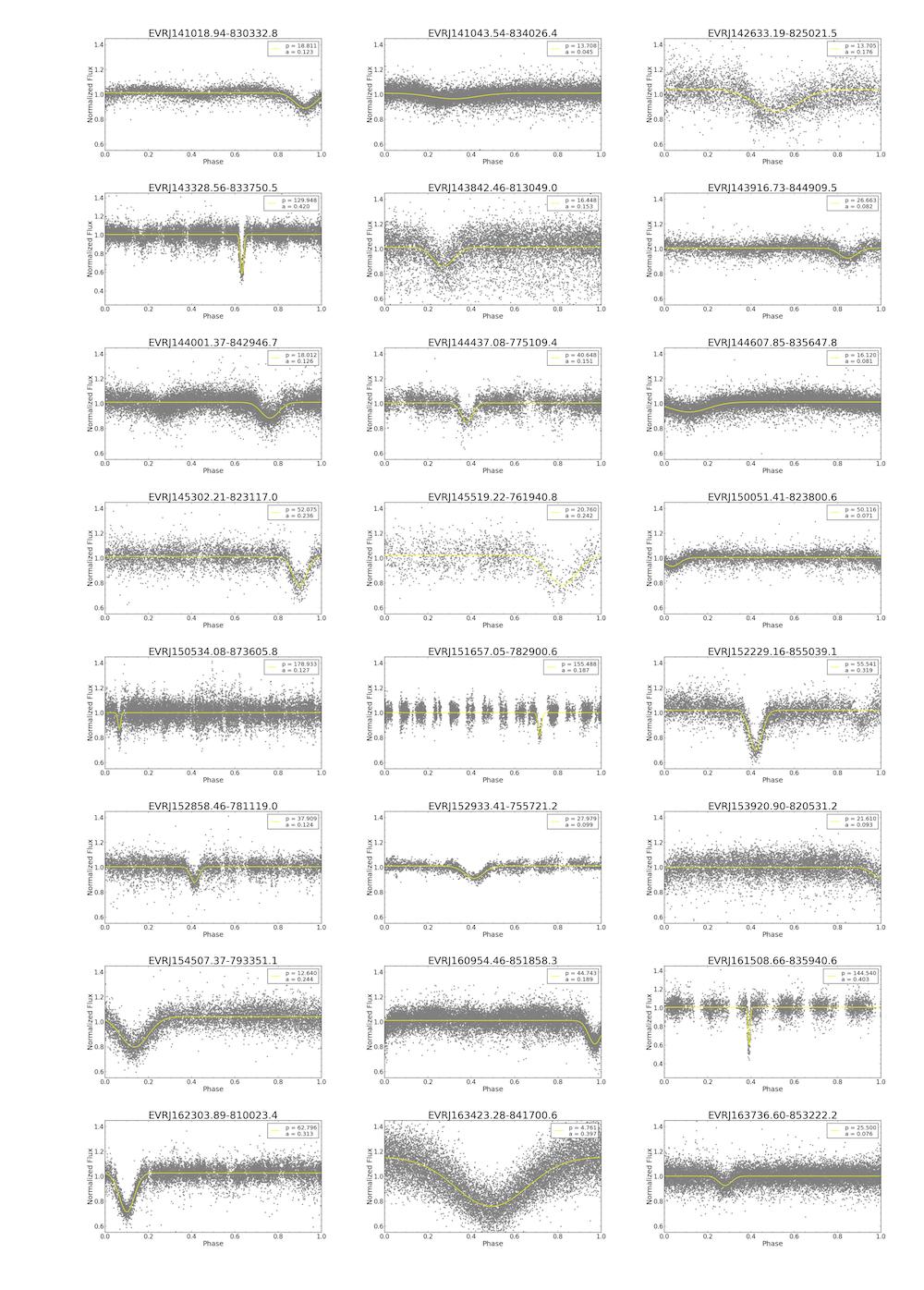}
\caption{Eclipsing Binary discoveries continued. Y-axis is normalized flux, x-axis is the phase, p = period found in hours, a = eclipse depth. Gray points are two minute cadence, yellow is the best fit.}
\end{figure*}

\newpage
\begin{figure*}[ht]
\figurenum{17}
\epsscale{1}
\includegraphics[width=.95\textwidth]{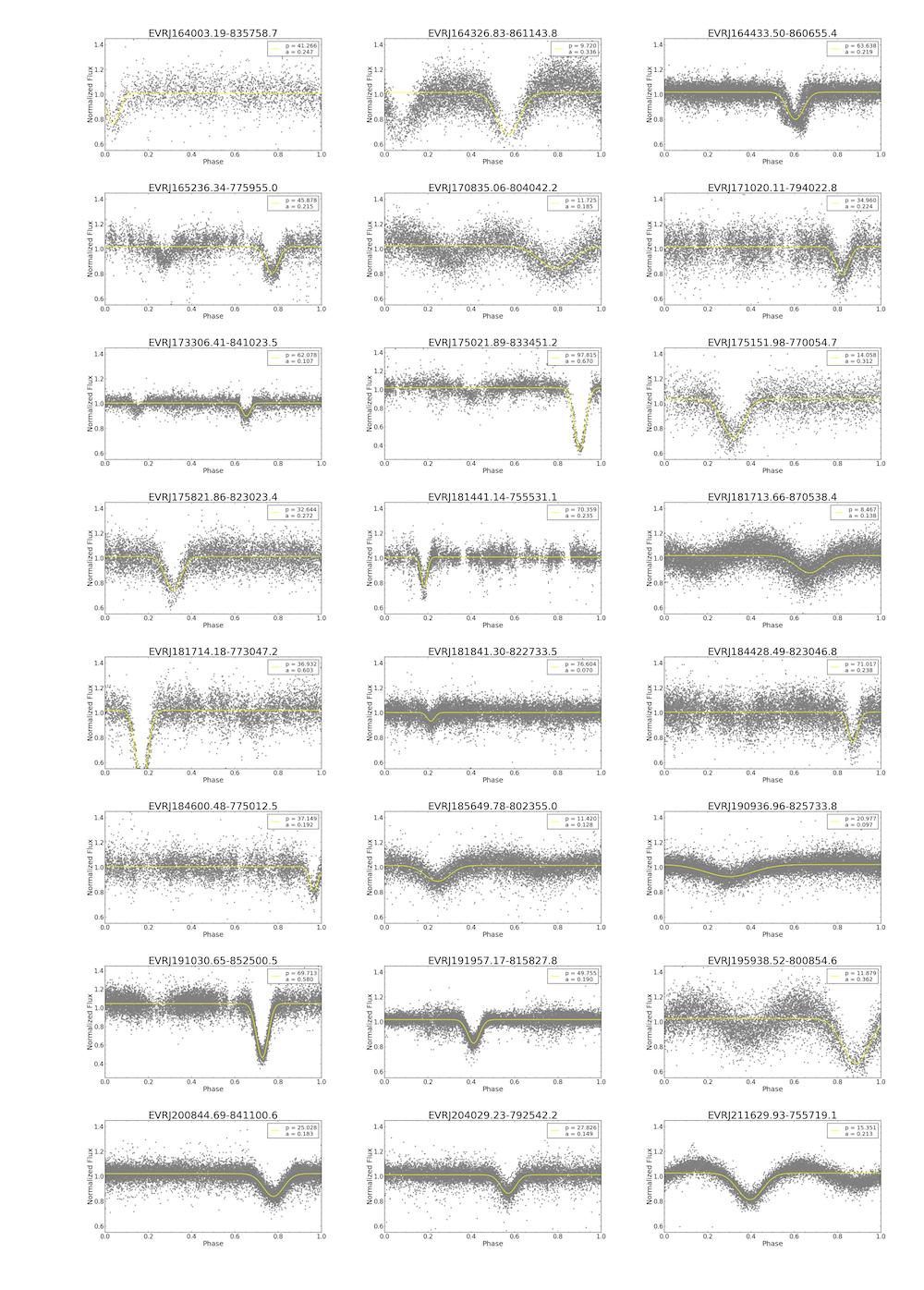}
\caption{Eclipsing Binary discoveries continued. Y-axis is normalized flux, x-axis is the phase, p = period found in hours, a = eclipse depth. Gray points are two minute cadence, yellow is the best fit.}
\end{figure*}

\newpage
\begin{figure*}[ht]
\figurenum{17}
\epsscale{1}
\includegraphics[width=.95\textwidth]{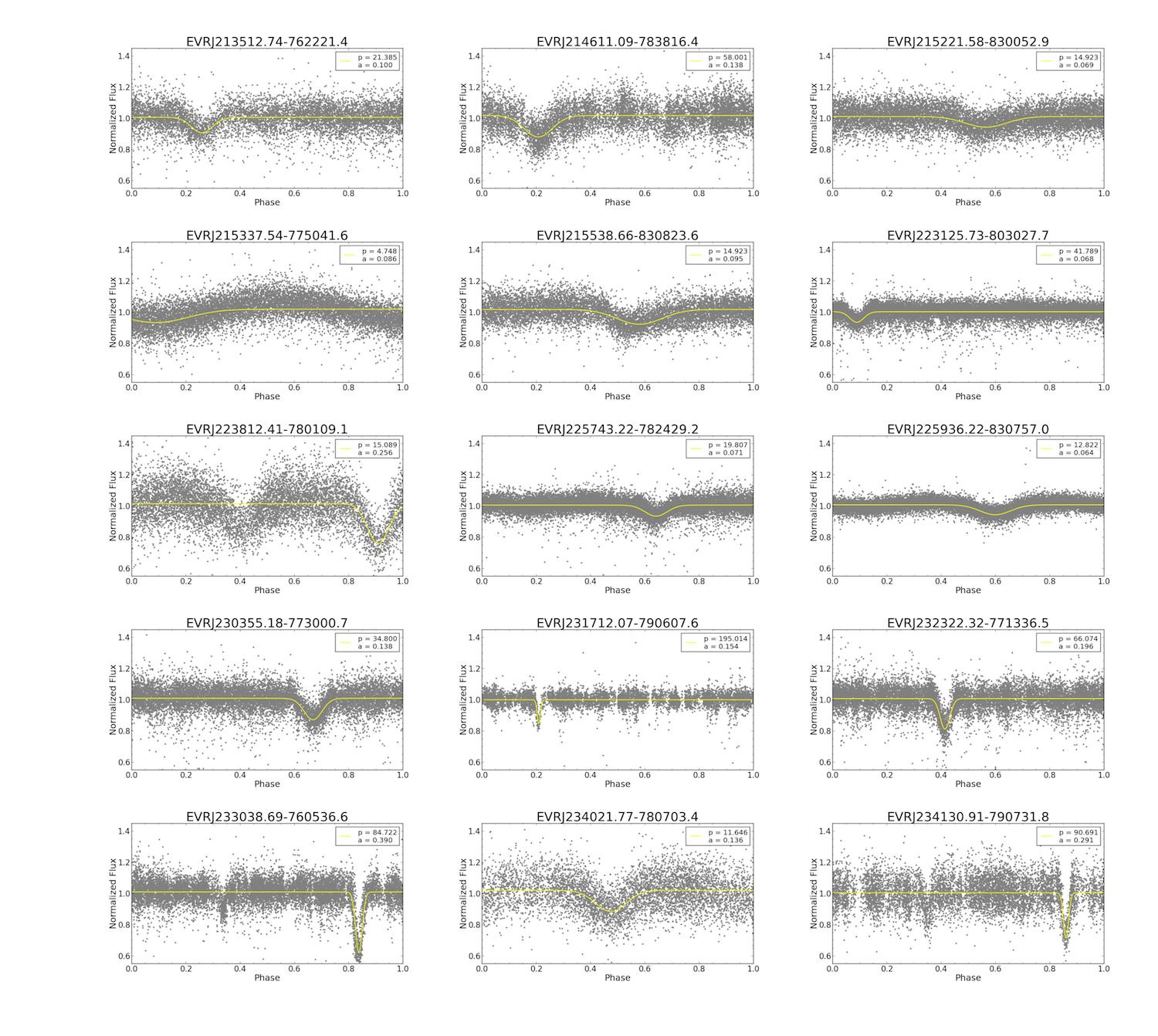}
\caption{Eclipsing Binary discoveries continued. Y-axis is normalized flux, x-axis is the phase, p = period found in hours, a = eclipse depth. Gray points are two minute cadence, yellow is the best fit.}
\end{figure*}

\end{document}